\documentstyle[preprint,prd,aps,eqsecnum]{revtex}
\begin{document}
\addtolength{\baselineskip}{-.3\baselineskip}
\preprint{{\vbox{\hbox{UR-1524} \hbox{TUIMP-TH-98/101} 
\hbox {Nov. 1998}
\hbox {revised Feb. 2000} }}}
\draft
%
%
%
\title{Supersymmetry without R-parity :
\newline
\phantom{aaaaaaaaaaaaa}
Constraints from
Leptonic Phenomenology}
\author{\bf Mike Bisset$^{1,2}$, Otto C.~W.~Kong$^2$, Cosmin Macesanu$^2$, 
and Lynne H.~Orr$^2$
\footnote{E-mail: bisset@urhepf.rochester.edu\ ; kong@pas.rochester.edu
\ ;\\  \hspace*{.7in} mcos@pas.rochester.edu \ ; orr@pas.rochester.edu \
.}
}
\address{$^1$Department of Physics, Tsinghua University,
             Beijing 100084, China\\
$^2$Department of Physics and Astronomy,\\
University of Rochester, Rochester NY 14627-0171}
\maketitle
\vskip -0.75cm
\begin{abstract}
R-parity conservation is an {\it ad hoc} assumption in the
most popular version of the supersymmetric standard model.
Most studies of models which do allow for R-parity violation
have been restricted to various limiting scenarios. 
The single-VEV parametrization used in this paper
provides a workable framework to analyze phenomenology
of the most general theory of SUSY without R-parity.  We perform a 
comprehensive study  of leptonic phenomenology at tree-level.
Experimental constraints on various processes are studied individually
and then combined to yield regions of admissible parameter space.
In particular, we show that large R-parity violating bilinear
couplings are not ruled out, especially for large $\tan\!\beta$.
\end{abstract}
\newpage
%
%
\section{Introduction}

The Minimal Supersymmetric Standard Model (MSSM) incorporates all of the 
Standard Model (SM) gauge symmetries  in its Lagrangian
(see \cite{SUSYrev,Kanebook} for reviews). 
The MSSM Lagrangian is also constrained by 
a new discrete symmetry, R-parity, defined by 
\begin{eqnarray}
{\cal R} = (-1)^{3B+L+2S}
\end{eqnarray}
where $B, L,$ and $S$ are respectively  baryon number, lepton number,
and spin.  Imposing R-parity conservation prohibits baryon
number and lepton number violating terms which
could otherwise lead to superparticle-mediated proton
decays on a weak interaction time scale, in stark disagreement with
observations (for a review and references, see
\cite{Dreiner}). 
Because R-parity distinguishes ordinary particles from their
supersymmetric partners, the minimal model gives rise to a
distinctive phenomenology.  Supersymmetric particles can be
produced only in pairs, and the lightest supersymmetric partner 
cannot decay.  These features drive many, if not most, 
supersymmetry (SUSY) search strategies. 

There is, however, no compelling reason to require R-parity conservation.
Less restrictive symmetries --- conservation of baryon number alone, for 
example --- can be imposed to prohibit unwanted proton decay.  
Furthermore, R-parity is not gauged or required by dynamics in the SM
or MSSM, and
hence there is no theoretical justification for requiring its
conservation.  
It is of course possible to devise extensions of the MSSM in
which R-parity is naturally conserved \cite{Rnat}, but such models
remain largely {\it ad hoc}.

Recently there has been a surge of interest in SUSY theories without
R-parity.  It is clear that their phenomenology 
can differ dramatically from that of the MSSM, and must
therefore be taken into account in SUSY searches.    
Introducing R-parity violating terms into the superpotentials
of these theories complicates any analysis enormously.  For that reason,
most studies make simplifying but otherwise unmotivated assumptions
that preclude general application of the results.  We have adopted a purely
phenomenological approach to supersymmetric theories without R-parity 
that provides the framework necessary to greatly simplify analyses without
necessitating so many {\it a priori} assumptions.  We introduced
our approach in \cite{PapI} and we elaborate on it here.

The most general renormalizable superpotential without R-parity for
a supersymmetric model with the minimal particle content may be written as
\begin{eqnarray}
     W \; = \;
     \varepsilon_{ab} [ h^u_{ij}\hat{Q}^a_i\hat{H}^b_u\hat{U}^C_j  \, + \, 
     &h^d_{ij}\hat{Q}^a_i\hat{H}^b_d\hat{D}^C_j \,\; + \,
     h^e_{ij}\hat{L}^a_i\hat{H}^b_d\hat{E}^C_j \,\; + \,
     {\mu}_{\scriptscriptstyle 0} \hat{H}^a_d \hat{H}^b_u   ] 
     \hbox{\phantom{aaaaaaaaaaaaaaaaaaa}}
     \nonumber \\
      & \hbox{\phantom{aaaaaaaaaaaaaaaaaaaaaaaaaaaaaaaaaa}}
     \, + \, 
     \lambda^{\prime\prime}_{ijk} \hat{D}^C_i\hat{D}^C_j\hat{U}^C_k 
     \nonumber \\
      \, + \, 
     \varepsilon_{ab} 
     [&\lambda^{\prime}_{ijk}\hat{Q}^a_i\hat{L}^b_j\hat{D}^C_k \, + \,
     \lambda_{ijk}\hat{L}^a_i\hat{L}^b_j\hat{E}^C_k
     \, + \, {\mu}_k \hat{L}^a_k\hat{H}^b_u  ] 
     \hbox{\phantom{aaaaaaaaaaaaaaaaa}} ,
     \label{super1}
\end{eqnarray}
where $i$, $j$, and $k$ are family (flavor) indices.
The coefficients
$\lambda$ and $\lambda^{\prime\prime}$ are antisymmetric in
the first two indices as required by
$SU(2)_{\hbox{\smash{\lower 0.25ex \hbox{${\scriptstyle L}$}}}}$
and $SU(3)_c$ product rules, respectively.  The
$SU(2)_{\hbox{\smash{\lower 0.25ex \hbox{${\scriptstyle L}$}}}}$
indices $a$ and $b$ are shown explicitly contracted with
the antisymmetric tensor ${\epsilon}_{ab}$,
with ${\epsilon}_{12} = -{\epsilon}_{21} = -1$, to generate
$SU(2)_{\hbox{\smash{\lower 0.25ex \hbox{${\scriptstyle L}$}}}}$
singlets.  $R$-parity conservation corresponds to setting ${\mu}_k = 0$
and all $\lambda = \lambda^\prime = \lambda^{\prime\prime} = 0$;
for baryon-number conservation, only $\lambda^{\prime\prime} = 0$ is
required.
 
We can obtain a more compact form for the superpotential $W$ by
noting that the
$SU(2)_{\hbox{\smash{\lower 0.25ex \hbox{${\scriptstyle L}$}}}}$
doublets $\hat{L}_k$ transform under the same SM gauge group 
representations as $\hat{H}_d$,
({$SU(3)_c$}, 
{$SU(2)_{\hbox{\smash{\lower 0.25ex \hbox{${\scriptstyle L}$}}}}$},
{$U(1)_{\hbox{\smash{\lower 0.25ex \hbox{${\scriptstyle Y}$}}}}$})
= ({\bf 1}, {\bf 2}, $\frac{1}{2}$).  (Here the hypercharge $Y$ is
normalized such that $Q = T_3 + Y$.)
In fact, without the assumption of lepton number
conservation, there is nothing to distinguish the $\hat{H}_d$
superfield from the $\hat{L}_k$ superfields, and they can mix.  
Their separate treatment in Eqn.$\!$ (\ref{super1}) is an artifact 
of starting with the MSSM superpotential and adding R-parity violating
terms.  Therefore a more appropriate form for $W$, with 
extended flavor indices $\alpha$ and $\beta$ running from $0$ to $3$, is      
\begin{eqnarray}
     W \; = \;
     \varepsilon_{ab} [h^u_{ij}\hat{Q}^a_i\hat{H}^b_u\hat{U}^C_j \, + \,
     {\lambda^\prime}_{i{\alpha}j}\hat{Q}^a_i\hat{L}^b_{\alpha}
     \hat{D}^C_j \, + \,
     {\lambda}_{{\alpha\beta}j}\hat{L}^a_{\alpha}\hat{L}^b_{\beta}
     \hat{E}^C_j \, + \,
     {\mu}_\alpha \hat{L}^a_\alpha \hat{H}^b_u] \, + \,
     \lambda^{\prime\prime}_{ijk} \hat{D}^C_i\hat{D}^C_j\hat{U}^C_k \; .
\label{superp2}
\end{eqnarray}
$U(4)$ flavor rotations can be used to transform between bases of the
four superfield doublets.  
      
It is in principle possible for the neutral scalar component in
{\em each} of these doublets to acquire a non-zero vacuum expectation
value (VEV).  The key to our approach, the 
single-VEV parametrization \cite{PapI,Nardi1}, is to rotate 
the doublets into a basis in which $\hat{L}_0$
alone bears a non-zero VEV.
The remaining admissible leptonic flavor rotations are
sufficient to diagonalize the lepton Yukawas 
$h^e_{ik} = 2\lambda_{i0k} = -2\lambda_{0ik}$  which
are then given by
$\frac{\sqrt{2}}{v_0}\rm{diag}\{ m_1, m_2, m_3 \}$.
There is then no additional freedom to  set
the $\mu_k$ bilinear coefficients equal to zero;  to maintain
complete generality they must be left arbitary.  

As implied above and discussed in \cite{PapI,ChFe}, care must be taken
to specify what choice of flavor basis (if any) is implied when
a specific set of of RPV parameters is given.
For example, if  the 
sneutrino VEV's and $\mu_k$ bilinear terms are left arbitrary,
then they are not truly physically independent because 
of the freedom to rotate between bases.
In another extreme, setting all sneutrino VEV's and  $\mu_k$ bilinears 
to zero results in a loss of generality, while still not being sufficient to 
uniquely determine the flavor basis.  
The common approach of using $R$-parity conserving MSSM
particle states with the {\it ad hoc} addition of a few
RPV trilinear couplings results in ill-defined RPV
parameters (since such parameters are in general basis-dependent
and the flavor basis is not specified).
In principle, that renders the  analysis internally inconsistent.  
In practice, this approach can reasonably approximate some regions 
of parameter space, but an inherent ambiguity remains.
One way out of this quandary would be to construct
``basis-independent'' observables.  This has been explored  with 
strong {\it a priori} restrictions 
placed on the sneutrino VEV's \cite{DavEl}.  However, the
observables were found to be phenomenologically ``messy'' and
impractical for most experimental situations even in this
limiting case.  The alternative resolution is to carefully choose
a convenient basis that renders the experimental analysis as simple
as possible --- the path taken in this work with the 
single-VEV parametrization.

As we described in \cite{PapI},
this parametrization has the advantage that
the tree-level mass matrices of {\em all}$\,$ fermions
in the theory are independent of all trilinear RPV couplings.
In particular, R-parity-violating contributions to leptonic phenomenology
at tree level are almost entirely determined by the  $\mu_k$.
In this paper, we examine this tree-level leptonic phenomenology in detail
within the single VEV parametrization, focusing on the
constraints that can be obtained from existing experimental results. 
The rich variety of relevant data include precision measurements at the 
$Z^0$ pole, charged lepton and pion decay rates and branching ratios,  
 neutrino mass bounds, lepton-neutrino scattering cross sections, and 
limits on neutrinoless double beta decay.  
A complete list of the measurements we will use can be found in Table I. 

In the present analysis, we assume that these processes are mediated by
gauge bosons, {\it i.e.}, on-shell or off-shell $Z^0$'s or $W^{\pm}$'s.
Scalar intermediaries are also possible for various processes, and
these would re-introduce dependence on the trilinear RPV couplings
(and other RPV parameters from the scalar sector)
even at tree-level.  We assume that contributions from such scalar modes
are negligible  due to kinematic supression
from the presumably much heavier (relative to the gauge bosons)
scalars.  It is certainly possible in principle  that  relative
strengths of the couplings involved and not-quite-so-heavy scalars
may conspire to invalidate our assumption. However, for these
to affect our experimental bounds (or for that matter, to be larger 
than the loop corrections we omit), one has to invoke very special and/or
unlikely combinations of parameters. Including
the additional parameters needed to accommodate these tiny corners of
parameter space needlessly complicates the analysis and will be avoided here.
In what follows we will point out neglected scalar modes where
appropriate.

We begin our analysis in Section II, where we examine the
neutral current interactions of the charged fermions (charged leptons
and charginos);  in Section III we do the same for the neutral
fermions.  We then consider charged current interactions in Section IV.
Having considered the constraints separately, we then combine them in
section V, where we determine exclusion regions in the
parameter space from both charged and neutral current
processes.  Finally, we present a summary
of our results and our conclusions in Section VI.

 \section{Color-Singlet Charged Fermions}

The color-singlet charged fermions, the charged leptons and the 
charginos, are linear combinations of two-component Weyl spinor 
charged winos (written as ${\lambda}_{\pm}$), the spinors of the 
$\hat{E}_i$ superfields, and the spinors for the charged superfields 
of the $\hat{L}_{\alpha}$ and $\hat{H}_u$
$SU(2)_{\hbox{\smash{\lower 0.25ex \hbox{${\scriptstyle L}$}}}}$ 
superdoublets.  In terms of Dirac spinors for a theory with spontaneous
symmetry breaking, denoting spinors corresponding to the superfields in 
Eqn.$\!$ (\ref{superp2}) by the corresponding lower-case letters  
(and where  $\tilde{\lambda}$ is the wino and $\tilde{\psi}$ is the
``higgsino'' built from $l_0$ and $\tilde{h}_u$), 
the contributing Lagrangian mass terms are given by     
\begin{eqnarray}
{\cal L}   \ni  - \left( \begin{array}{ccccc}
 i\bar{\tilde{\lambda}} & \bar{\tilde{\psi}} &
 \bar{{\ell}_1} & \bar{{\ell}_2} & \bar{{\ell}_3} 
  \end{array} \right)  
 \left[
  {\cal M}_{\scriptscriptstyle {\cal C}}^T P_{\scriptscriptstyle L} +
  { {\cal M}_{\scriptscriptstyle {\cal C}} } P_{\scriptscriptstyle R}
 \right]
   \left( \begin{array}{ccccc}
 -i\tilde{\lambda} & \tilde{\psi} & {\ell}_1 & {\ell}_2 & {\ell}_3
  \end{array} \right)^T \;\; ,
\label{Diracnot}
\\
\hbox{with} \;\;\;
  {\cal M}_{\scriptscriptstyle {\cal C}} = \left( \begin{array}{ccccc}
  M_{\scriptscriptstyle 2}  & \frac{gv_u}{\sqrt{2}} & 0   & 0   & 0   \\
  \frac{gv_d}{\sqrt{2}} & {\mu}_{\scriptscriptstyle 0} & 0   & 0   & 0   \\
  0 & {\mu}_{\scriptscriptstyle 1} & m_{\scriptscriptstyle 1} & 0   & 0   \\
  0 & {\mu}_{\scriptscriptstyle 2} & 0   & m_{\scriptscriptstyle 2} & 0   \\
  0 & {\mu}_{\scriptscriptstyle 3} & 0   & 0   & m_{\scriptscriptstyle 3}
  \end{array} \right) \;\; .
\hbox{\phantom{aaaaaaaaaaaaaaaaaaaa}}
\label{MatrixC}
\end{eqnarray}
     Here the two vacuum expectation values are in general complex,
     as are the ${\mu}_{\alpha}$'s and the soft gaugino mass 
     $M_{\scriptscriptstyle 2}$.
     As is customary, all of the 
     parameters entering the fermion mass matrices
     will henceforth be assumed to be real.  Potentially very
     interesting $CP$-violating effects related to the possible
     complex nature of these parameters will be relegated to a
     future paper.  On the other hand, it must be strongly
     emphasized that there is no loss of generality concerning the
     trilinear RPV terms --- {\em these need not be set to zero}.  
     They simply do not appear at tree level
     in the fermionic mass matrices in the single-VEV parametrization.

We now wish to find the mass eigenstates.  In the MSSM, the 
Yukawa mass entries,
$m_1$, $m_2$, and $m_3$ ($m_i = h^e_{ii}v_d$ in the single-VEV basis),
are identical to the known physical charged lepton masses.
Without R-parity conservation this correspondence is spoiled 
by the presence of the $\mu_i$ in ${\cal M_C}$.  
Masses of the well-known charged leptons (and in fact the
eigenvectors and mass eigenvalues for all five physical states) 
now result from diagonalizing the $5 \times 5$ chiral mass matrix 
${\cal M}_{\scriptscriptstyle {\cal C}}$: 
 \begin{eqnarray}
  U_L^{\dagger} {\cal M}_{\scriptscriptstyle {\cal C}} U_R = {\rm diag} \{
                               \overline{M}_{c\scriptscriptstyle{ 1}},
                               \overline{M}_{c{\scriptscriptstyle 2}},
                               \overline{m}_1,
                               \overline{m}_2,
                               \overline{m}_3
                                    \} \; ,
\label{barMcidef}
\end{eqnarray}
where $U_L$ and $U_R$ are unitary matrices which diagonalize
${\cal M}_{\scriptscriptstyle {\cal C}}$, and
$m_e = \overline{m}_1$,
$m_{\mu} = \overline{m}_2$, and
$m_{\tau} = \overline{m}_3$.
In general, 
the $\overline{m}_i$ depend on the $m_i$ {\em and} the $\mu_i$.

Therefore for non-zero ${\mu}_i$ the input Yukawa parameters for 
${\cal M_C}$ need to be determined, {\it i.e.}\/   for a fixed set of 
$\mu_i$, we need to find the values of $m_i$ that give the correct 
physical masses $\overline{m}_i$.
This is done by writing a system of linear
differential equations for the infinitesimal change in the
$m_i$'s due to an infinitesimal change in the $\mu_i$'s.
Beginning with all ${\mu}_i=0$ (where the solution of the
system of equations is known), an acceptable solution for
a chosen set of ${\mu}_i$'s is then obtainable via numerical
integration \cite{caveat}.
Note that only the three `lepton' masses need to be fixed in this way.
The heavier so-called `chargino' masses 
$\overline{M}_{c {\scriptscriptstyle 1}}$ and
$\overline{M}_{c {\scriptscriptstyle 2}}$
also depend on the ${\mu}_i$, but we do not
(yet) have experimental constraints telling us what the
physical chargino masses should be.

We now consider interactions with the $Z^0$ boson, following Ref.\cite{NovPil} 
but using the single-VEV parametrization.
The couplings of the five mass eigenstates to the $Z^0$ boson are given
by
\begin{eqnarray}
{\cal L}_{{\chi}^+ {\chi}^- Z^0} =
\frac{g_2}{2\cos{\theta}_w} {\chi}_a^+ {\gamma}_{\mu}
\left(
P_{\scriptscriptstyle L} \tilde{A}^{\scriptscriptstyle L}_{ab}
\; + \; P_{\scriptscriptstyle R} \tilde{A}^{\scriptscriptstyle R}_{ab}
\right)
{\chi}_b^- Z^{\mu} \; ,
\label{Zlag}
\end{eqnarray}
where $\bar{{\chi}}_a^+ = {\chi}_a^- = ( \chi_1, \chi_2, \ell_e,
\ell_{\mu}, \ell_{\tau} )$,
$P_{\scriptscriptstyle R,L} = \frac{1}{2}(1 \pm {\gamma}_5)$, and
\begin{eqnarray}
\tilde{A}^{\scriptscriptstyle L}_{ab} & = & U_L^{1a} U_L^{1b}
                 \; + \; {\delta}_{ab}(1 - 2{\sin}^2\!\!\;{\theta}_w)
   \; \equiv \; \delta\!\tilde{A}^{\scriptscriptstyle L}_{ab}
                 \; + \; {\delta}_{ab}(1 - 2{\sin}^2\!\!\;{\theta}_w)
   \;\; ,
\label{ULAL}
    \\
\tilde{A}^{\scriptscriptstyle R}_{ab} & = & 
                             2 U_R^{1a} U_R^{1b} + U_R^{2a} U_R^{2b}
                      \; + \; 2{\delta}_{ab}{\sin}^2\!\!\;{\theta}_w
   \; \equiv \; \delta\!\tilde{A}^{\scriptscriptstyle R}_{ab}
                      \; + \; 2{\delta}_{ab}{\sin}^2\!\!\;{\theta}_w
\label{URAR}
\end{eqnarray}
(notation here follows that of \cite{NovPil}
except for the ordering of basis vectors).
The ${\delta}_{ab}$ terms are the
SM expressions, and  deviations from SM expectations originate from
non-zero $\delta\!\tilde{A}^{\scriptscriptstyle L}_{ab}$ and
$\delta\!\tilde{A}^{\scriptscriptstyle R}_{ab}$.
The anomalous coupling of any two charged fermions to the $Z^0$ can
thus be
determined in terms of the $U_L$ and $U_R$ matrices found numerically
from diagonalizing  ${\cal M}_{\scriptscriptstyle {\cal C}}$,
and this is precisely what we do to obtain the parameter space plots
presented later in this paper.
The exact analytic expressions for the eigenvalues and
eigenvectors of the $5\, \times \, 5$ 
${\cal M}_{\scriptscriptstyle {\cal C}}$
prove cumbersome and not very illuminating and we do not reproduce 
them here.  However,
to improve our understanding of the results of the exact
numerical analysis, it is very useful to consider a couple of
interesting  analytic approximations.

First, we treat the R-parity violation as a perturbation, taking  the
limit in which the ${\mu}_i$ are small.  If
the ${\mu}_i$ were zero, the (now MSSM) chargino sector, which is the
upper $2 \times 2$ portion of ${\cal M}_{\scriptscriptstyle {\cal C}}$, 
would be orthogonal to the already diagonal SM charged lepton sector.  
So first we introduce new $2 \times 2$ rotation matrices
$R_{\scriptscriptstyle L}( \theta_{\!\scriptscriptstyle L} )$ and
$R_{\scriptscriptstyle R}( \theta_{\!\scriptscriptstyle R} )$
such that
\begin{eqnarray}
R_L^{\dagger} 
  \left( \begin{array}{cc}
  M_{\scriptscriptstyle 2}  & \frac{1}{\sqrt{2}}gv_u  \\
  \frac{1}{\sqrt{2}}gv_d & {\mu}_{\scriptscriptstyle 0}
  \end{array} \right) R_R
  \; = \; {\rm diag} 
  \{ M_{c{\scriptscriptstyle 1}}, M_{c{\scriptscriptstyle 2}} \} \;\;\; .
\label{Mcidef}
\end{eqnarray}
$M_{c{\scriptscriptstyle 1}}$ and $M_{c{\scriptscriptstyle 2}}$ 
are the chargino masses in the ${\mu}_i = 0$ MSSM limit.  
Applying this rotation to
${\cal M}_{\scriptscriptstyle {\cal C}}$ we find
\begin{eqnarray}
{{\cal M}_{\scriptscriptstyle {\cal C}}}^{\prime} \; \equiv \;
{R_{\scriptscriptstyle L}}^{\dagger} 
{\cal M}_{\scriptscriptstyle {\cal C}} R_{\scriptscriptstyle R} \; = \;
 \left( \begin{array}{ccccc}
 M_{c{\scriptscriptstyle 1}} & 0 & 0   & 0 & 0     \\
 0 & M_{c{\scriptscriptstyle 2}} & 0   & 0 & 0   \\
 {\mu}_{\scriptscriptstyle 1} \sin\!\theta_{\!\scriptscriptstyle R} 
  & {\mu}_{\scriptscriptstyle 1} \cos\!\theta_{\!\scriptscriptstyle R}
  & m_1 & 0 & 0  \\
 {\mu}_{\scriptscriptstyle 2} \sin\!\theta_{\!\scriptscriptstyle R} 
  & {\mu}_{\scriptscriptstyle 2} \cos\!\theta_{\!\scriptscriptstyle R}
  & 0   & m_2 & 0   \\
 {\mu}_{\scriptscriptstyle 3} \sin\!\theta_{\!\scriptscriptstyle R} 
  & {\mu}_{\scriptscriptstyle 3} \cos\!\theta_{\!\scriptscriptstyle R}
  & 0   & 0   & m_3 \\
 \end{array} \right) \;\; .
\label{Mcidef2}
\end{eqnarray}
To obtain the general $5 \times 5$ $U_L$ and $U_R$ matrices
from the MSSM's $2 \times 2$ $R_{\scriptscriptstyle L}$ and 
$R_{\scriptscriptstyle R}$ matrices, we
treat the $3 \times 2$ off-diagonal block containing the
${\mu}_i$'s as a perturbation.  Then
\begin{eqnarray}
U_{L,R}^{\dagger} \; = \;
  \left( \begin{array}{cc}
  R_{{\scriptscriptstyle L,R}}^{\dagger}
  & -R_{{\scriptscriptstyle L,R}}^{\dagger} V_{L,R}^{\dagger} \\
  V_{L,R}              & I_{3 \times 3}                   
  \end{array} \right) \; ,
\end{eqnarray}
where the elements of the $3 \times 2$ $V_L$ and $V_R$ matrices
are given by
\begin{eqnarray}
\begin{tabular}{rlrl}
$V_L^{i1} =$ & $\sqrt{2}\frac{\mu_i M_{\scriptscriptstyle W} \cos\!\beta}
{M_{\scriptscriptstyle 0}^2}$
& $\;\;\;\;\;\;\;\;\;\;$ $V_R^{i1} =$ & $\sqrt{2}\frac{m_i}
{M_{\scriptscriptstyle 0}}
 \frac{\mu_i M_{\scriptscriptstyle W} ( M_{\scriptscriptstyle 2}\sin\!\beta
   + {\mu}_{\scriptscriptstyle 0} \cos\!\beta )}{M_{\scriptscriptstyle 0}^3}$
\\
$V_L^{i2} =$ & 
 $-\frac{\mu_i M_{\scriptscriptstyle 2}}{M_{\scriptscriptstyle 0}^2}$
& $\;\;\;\;\;\;\;\;\;\;$ $V_R^{i2} =$ & $-\frac{m_i}{M_{\scriptscriptstyle 0}}
           \frac{\mu_i (M_{\scriptscriptstyle 2}^2
  + 2M_{\scriptscriptstyle W}^2{\cos}^2\!\!\>{\beta} )}
     {M_{\scriptscriptstyle 0}^3}
           \;\;\;\;\;\;\;\; ,$
\end{tabular}
\label{vvalues}
\end{eqnarray}
and where
\vskip -1.4cm
\begin{eqnarray}
M_{\scriptscriptstyle 0}^2 \; \equiv \;
{\mu}_{\scriptscriptstyle 0} M_{\scriptscriptstyle 2} - 
M_{\scriptscriptstyle W}^2\sin\! 2\beta
\end{eqnarray}
which is in fact the determinant of the $2 \times 2$ MSSM section
of ${\cal M}_{\scriptscriptstyle {\cal C}}$.  In the limit of large 
$\tan\!\beta$
and/or ${\mu}_{\scriptscriptstyle 0}M_{\scriptscriptstyle 2} \gg 
M_{\scriptscriptstyle W}^2$,
$M_{\scriptscriptstyle 0}^2$ tends toward 
${\mu}_{\scriptscriptstyle 0}M_{\scriptscriptstyle 2}$.

In terms of the $V_L$ and $V_R$ matrix elements, we can now write
the deviations of the fermions' coupling coefficients to the $Z^0$ boson
from the SM case as
\begin{eqnarray}
\delta\!\tilde{A}^{\scriptscriptstyle L}_{ab}
& = & V_L^{i1}V_L^{j1}
\; = \; 2\frac{ \mu_i \mu_j }{ M_{\scriptscriptstyle 0}^2 }
         \frac{ M_{\scriptscriptstyle W}^2 {\cos}^2\!\!\>{\beta} }
              { M_{\scriptscriptstyle 0}^2 }
\\
\noalign{\hbox{and}}
\delta\!\tilde{A}^{\scriptscriptstyle R}_{ab}
& = & 2V_R^{i1}V_R^{j1} \; + \; V_R^{i2}V_R^{j2}
                      \nonumber \\
& = &
\frac{m_im_j}{M_{\scriptscriptstyle 0}^2}\frac{\mu_i\mu_j}
{M_{\scriptscriptstyle 0}^2}
\frac{1}{M_{\scriptscriptstyle 0}^4}
[M_{\scriptscriptstyle 2}^4 + 
4M_{\scriptscriptstyle W}^2(M_{\scriptscriptstyle 2}^2
+ {\mu}_{\scriptscriptstyle 0}M_{\scriptscriptstyle 2}\sin\! 2\beta
+ {\mu}_{\scriptscriptstyle 0}^2{\cos}^2\!\!\>{\beta}
          + M_{\scriptscriptstyle W}^2 {\cos}^4\!\!\>{\beta}) ] \; ,
\end{eqnarray}
for $\{ a,i \}, \{ b,j \} = \{e,1\}, \{\mu,2\},
\{ \tau, 3\}$.\footnote{
In the formul\ae\ above and those that follow,
index labels $a$ and $b$ are reserved for the physical mass eigenstates,
$a,b = e,\mu,\tau$ (or the heavier charginos).
Index labels $i$ and $j$ are used for basis state parameters in the
single-VEV parametrization, $i,j = 1,2,\, \hbox{or}\, 3$.  {\em In the
small-${\mu}_i$ approximation}, $a$ and $i$ (or $b$ and $j$) have a
simple one-to-one correspondence.  
Hence, equations could be written with just one pair of index labels.
However, both pairs will be kept to make clear that this is not true 
in general.
}
This provides us with simple, quantitative expressions of how
the gaugino and higgsino contents of the charged lepton mass
eigenstates ({\it i.e.}, of the $e$, the $\mu$, and the $\tau$) 
affect the $Z^0$ leptonic
decay widths.  These effects, present when the $\mu_i$ are non-zero, 
can lead to non-universality among the $Z^0$
leptonic branching ratios or to mixed-flavor $Z^0$ leptonic decays.

Notice that the deviations of the $Z^0$ coupling coefficients are
in fact proportional to $\frac{\mu_i\mu_j}{M_{\scriptscriptstyle 0}^2}$, 
which must be
small in this perturbative approximation (this tells us quantitatively
what it means for the $\mu_i$ to be ``small'').  In addition, the
$\delta\!\tilde{A}^{\scriptscriptstyle R}_{ab}$ are suppressed by the factor
$\frac{m_im_j}{M_{\scriptscriptstyle 0}^2}$.  Thus one might expect that 
the effects of the
$\delta\!\tilde{A}^{\scriptscriptstyle L}_{ab}$ will dominate  those of the
$\delta\!\tilde{A}^{\scriptscriptstyle R}_{ab}$.  However,  the
$\delta\!\tilde{A}^{\scriptscriptstyle L}_{ab}$ are themselves proportional to
${\cos}\!^2\!\beta$ which will strongly suppress their values in the
large $\tan\!\beta$ regime (for example, ${\cos}^2\!\beta \sim 10^{-3}$
for $\tan\!\beta = 45$).  Neglecting terms proportional to $\cos\!\beta$
in the $\delta\!\tilde{A}^{\scriptscriptstyle R}_{ab}$
(recall $\sin\! 2\beta$ is small when
$\cos\!\beta$ is small), we find
\begin{eqnarray}
\frac{
\delta\!\tilde{A}^{\scriptscriptstyle R}_{ab} }
{ \delta\!\tilde{A}^{\scriptscriptstyle L}_{ab}
}
& \simeq &
\frac{1}{2}(1 + {\tan}\!^2\!\beta)
\frac{m_im_j}{M_{\scriptscriptstyle 0}^2}\frac{M_{\scriptscriptstyle 2}^2
[(M_{\scriptscriptstyle 2}/M_{\scriptscriptstyle W})^2 + 4]}
{M_{\scriptscriptstyle 0}^2}
\nonumber \\
& \simeq &
\frac{1}{2}(1 + {\tan}\!^2\!\beta)
\frac{m_im_j}{{\mu}_{\scriptscriptstyle 0}^2}
[(M_{\scriptscriptstyle 2}/M_{\scriptscriptstyle W})^2 + 4]
\;\;\;\;\;\;\;\;
(\hbox{large}\; \tan\!\beta \; \hbox{or} \;
{\mu}_{\scriptscriptstyle 0}M_{\scriptscriptstyle 2}\gg 
M_{\scriptscriptstyle W}^2) \;\;\; .
\end{eqnarray}
Thus if $M_{\scriptscriptstyle 2}$ is large and 
$|{\mu}_{\scriptscriptstyle 0}|$ is small,
the effect of deviations from the SM for the right-hand component of the
$Z^0$-$\tau$-$\tau$ coupling (and even the $Z^0$-$\tau$-$\mu$
coupling) can be as significant as those for the left-hand component.

The $Z^0$ branching fraction into any pair of colorless
charged fermions (with $M_{\scriptscriptstyle Z} > 
M_{ {\chi}_a } + M_{ {\chi}_b }$) 
is given by
\begin{eqnarray}
\!\!\!\!
Br( Z^0 \rightarrow {\chi}^+_a {\chi}^-_b) =
\frac{\smash{\lower 0.6ex \hbox{
${\alpha}_2 \, {\lambda}^{\scriptscriptstyle \!\!\!\!
\smash{\raise 0.9ex \hbox{ $\frac{1}{2}$ }}} $
}}
\!\!\!\!\!
\left(
{\scriptstyle
 1, \frac{ M_{ {\chi}_a }^2 }{ M_{\scriptscriptstyle Z}^2 },
 \frac{ M_{ {\chi}_b }^2 }{ M_{\scriptscriptstyle Z}^2 }
}
 \right)
}{ 24{\cos}^2\!\!\;{\theta}_w }
\frac{M_{\scriptscriptstyle Z}}{ {\Gamma}_{\scriptscriptstyle Z} }
\left[
\smash{\lower 0.4ex \hbox{$1$}}
{\scriptstyle
 \, - \, \frac{ M_{ {\chi}_a }^2 + M_{ {\chi}_b }^2 }
{ 2M_{\scriptscriptstyle Z}^2 }
\, - \, \frac{ (M_{ {\chi}_a }^2 - M_{ {\chi}_b }^2)^2 }
{ 2M_{\scriptscriptstyle Z}^4 }
\, + \, \frac{ 6M_{ {\chi}_a } M_{ {\chi}_b } }{ M_{\scriptscriptstyle Z}^2 }
}
\frac{ \tilde{A}^{\scriptscriptstyle L}_{ab}
\tilde{A}^{\scriptscriptstyle R}_{ab} }
     { {{\tilde{A}}^2}_{ab} }
\right]
{{\tilde{A}}^2}_{ab}
\!\!\!\!\!\!\!\!\!\!\!\!\!\!\!\!\!\!\!\!\!\!\!\!\!\!\!\!\!\!
\nonumber \\
\label{Zbr}
\end{eqnarray}
\vskip -0.5cm
where ${\alpha}_2 \equiv g_2^2/4\pi$ and
\begin{eqnarray}
{{\tilde{A}}^2}_{ab} \; \equiv \; | \tilde{A}^{\scriptscriptstyle L}_{ab} |^2
                          \, + \, | \tilde{A}^{\scriptscriptstyle R}_{ab} |^2
\;\;\; .
\end{eqnarray}
The experimentally determined total decay width of the
$Z^0$ is ${\Gamma}_{\scriptscriptstyle Z} = (2.4948 \pm 0.0075) \, \hbox{GeV}$
\cite{LEP1}.
The kinematic $\lambda$-function is
$\lambda (a,b,c) = (a - b - c)^2 - 4bc$. 
In the small ${\mu}_i$ perturbative approximation, and for $\tan\!\beta$ not
too large, we have 
for the leptonic $Z^0$ decays,
\begin{eqnarray}
{{\tilde{A}}^2}_{ii} &\simeq & (1 - 2{\sin}^2\!\!\;{\theta}_w)^2
+ (4{\sin}^4\!\!\;{\theta}_w)
+ 2(1 - 2{\sin}^2\!\!\;{\theta}_w)
\delta\!\tilde{A}^{\scriptscriptstyle L}_{ii}
 \nonumber \\
&=& .5027 + 2.148\,
\frac{M_{\scriptscriptstyle W}^2
\cos^2\!\!\beta}{M_{\scriptscriptstyle 0}^2}
\left(\frac{\mu_i}{M_{\scriptscriptstyle 0}}\right)^{\!\! 2}
\; , 
\label{Aiisq} \\
\hbox{and} \;\;\;\;\;\;\;\; & &
\nonumber \\
{{\tilde{A}}^2}_{ab} & = & |\delta\!\tilde{A}^{\scriptscriptstyle L}_{ab}|^2
\; = \; \frac{4 M_{\scriptscriptstyle W}^4 \cos^4\!\!\beta }
{M_{\scriptscriptstyle 0}^4}
\left(\frac{\mu_i}{M_{\scriptscriptstyle 0}}\right)^{\!\! 2}
\left(\frac{\mu_j}{M_{\scriptscriptstyle 0}}\right)^{\!\! 2}
 \quad (a,i \ne b,j) \; . 
\label{Aijsq}
\end{eqnarray}
If $\tan\!\beta$ is large, then the right-hand component
cannot be neglected, as noted above, and
the ${{\tilde{A}}^2}_{\tau\tau}$ and
${{\tilde{A}}^2}_{\mu\tau}$ formul\ae\
should be modified to
\begin{eqnarray}
{{\tilde{A}}^2}_{\tau\tau} &\simeq & (1 - 2{\sin}^2\!\!\;{\theta}_w)^2
+ (4{\sin}^4\!\!\;{\theta}_w) + 2(1 - 2{\sin}^2\!\!\;{\theta}_w)
\delta\!\tilde{A}^{\scriptscriptstyle L}_{\tau\tau}
-  4{\sin}^2\!\!\;{\theta}_w\delta\!\tilde{A}^{\scriptscriptstyle R}_{\tau\tau}
\nonumber \\
&=& .5027
  + \left[ 2.148 \frac{ M_{\scriptscriptstyle W}^2{\cos}^2\!\!\>{\beta} }
                     { M_{\scriptscriptstyle 0}^2 }
 -.926 \frac{ m_{\scriptscriptstyle 3}^2 
(M_{\scriptscriptstyle 2}^2 + 4M_{\scriptscriptstyle W}^2) 
M_{\scriptscriptstyle 2}^2 }
            { M_{\scriptscriptstyle 0}^6 } \right]
\left(\frac{{\mu}_{\scriptscriptstyle 3}}
{M_{\scriptscriptstyle 0}}\right)^{\!\! 2} \; ,
\label{A33sq} \\
\noalign{\hbox{and}}
{{\tilde{A}}^2}_{\mu\tau} &=& 
 |\delta\!\tilde{A}^{\scriptscriptstyle L}_{\mu\tau}|^2
 + |\delta\!\tilde{A}^{\scriptscriptstyle R}_{\mu\tau}|^2
\nonumber \\
&=& \left[ \frac{ 4M_{\scriptscriptstyle W}^4{\cos}^4\!\!\>{\beta} }
                { M_{\scriptscriptstyle 0}^4 }
 + \frac{ m_{\scriptscriptstyle 2}^2 m_{\scriptscriptstyle 3}^2
         (M_{\scriptscriptstyle 2}^2
+ 4M_{\scriptscriptstyle W}^2)^2 M_{\scriptscriptstyle 2}^4 }
         { M_{\scriptscriptstyle 0}^{12} }
    \right]
\left(\frac{ {\mu}_{\scriptscriptstyle 2} }
{ M_{\scriptscriptstyle 0} }\right)^{\!\! 2}
\left(\frac{ {\mu}_{\scriptscriptstyle 3} }
{ M_{\scriptscriptstyle 0} }\right)^{\!\! 2} \; .
\label{A23sq}
\end{eqnarray}
Note here that the numerical value ${\sin}^2\!\!\;{\theta}_w = 0.2315$
\cite{LEP1} has been used.  This effective value for
${\sin}^2\!\!\;{\theta}_w$ absorbs the SM radiative corrections to the
$Z^0 {\ell}_i {\ell}_i$ couplings.
Note also that the two terms inside the bracket in Eqn.$\!$ (\ref{A33sq})
enter with opposite signs, implying that in some region of the
parameter space, the deviation of
$\Gamma (Z^0 \rightarrow {\tau}^+{\tau}^- )$
from the SM prediction could be suppressed by cancellation between
these two terms even if ${\mu}_{\scriptscriptstyle 3}$ is quite substantial.

We now apply these results for couplings and branching ratios to 
specific processes in order to obtain limits on the $\mu_i$.

\subsection{Mixed-flavor leptonic $Z^0$ decays}

For non-zero ${\mu}_i$'s, the Lagrangian of Eqn.$\!$ (\ref{Zlag}) leads
to the tree-level flavor-violating $Z^0$ decays,
$Z^0 \rightarrow e\mu$, $e\tau$, and $\mu\tau$.  The predicted
branching ratios for these decay modes are given by Eqn.$\!$ (\ref{Zbr}).
The experimental bounds from LEP on these processes are shown in
the third column of Table I.
In the small-${\mu}_i$ approximation,
Eqn.$\!$ (\ref{Aijsq}) translates these  constraints into
the following bounds:
\begin{eqnarray}
\frac{ |{\mu}_i {\mu}_j | }{ M_{\scriptscriptstyle 0}^2 } & \le &
{\cal K}_{ij} (1 + {\tan}\!^2\!\beta ) 
\frac{M_{\scriptscriptstyle 0}^2}{M_{\scriptscriptstyle W}^2}
\label{cijform}
\end{eqnarray}
\begin{eqnarray}
{\cal K}_{12} & = & 1.8 \times 10^{-3}  
   \;\; \hbox{from} \; Z^0 \rightarrow e^{\pm} {\mu}^{\mp}
\nonumber \\
{\cal K}_{13} & = & 4.3 \times 10^{-3}  
   \;\; \hbox{from} \; Z^0 \rightarrow e^{\pm} {\tau}^{\mp}
\nonumber \\
{\cal K}_{23} & = & 4.7 \times 10^{-3}  
   \;\; \hbox{from} \; Z^0 \rightarrow {\mu}^{\pm} {\tau}^{\mp}
\;\;\; .
\nonumber 
\end{eqnarray}
Note that bounds apply to the products $|{\mu}_i {\mu_j}|$.  That the
constraints can be cast in such a simple form and in terms of so few
RPV input parameters is a key strength of the single-VEV parametrization.
Note also that the constraints become weaker as $\tan\!\beta$ increases.

\subsection{Flavor-violating charged lepton decays}

The $Z^0 {\ell}_i {\ell}_j$ couplings can also produce the tree-level
FCNC decays of $\mu$ and $\tau$ via a virtual $Z^{0}$.
  These branching ratios are given by 
\begin{eqnarray}
Br({\ell}^-_a \rightarrow {\ell}^-_b {\ell}^+_c {\ell}^-_c )
& = & \frac{ {\alpha}_2^2 {\cos}\!^4\!\beta }{ 1536 \pi }
\left( \frac{ m_{ {\ell}_a } }{ M_{\scriptscriptstyle Z} } \right)^4
\frac{ m_{ {\ell}_a } }{ {\Gamma}_{ {\ell}_a } }
{{\tilde{A}}^2}_{ab} {{\tilde{A}}^2}_{cc} 
\label{ldecss}
\\
Br({\ell}^-_a \rightarrow {\ell}^+_b {\ell}^-_c {\ell}^-_c )
& = & \frac{ {\alpha}_2^2 {\cos}\!^4\!\beta }{ 1536 \pi }
\left( \frac{ m_{ {\ell}_a } }{ M_{\scriptscriptstyle Z} } \right)^4
\frac{ m_{ {\ell}_a } }{ {\Gamma}_{ {\ell}_a } }
\left( {{\tilde{A}}^2}_{ac} {{\tilde{A}}^2}_{bc} \, + \,
| \tilde{A}^{\scriptscriptstyle L}_{ac} |^2
| \tilde{A}^{\scriptscriptstyle R}_{bc} |^2 \, + \,
| \tilde{A}^{\scriptscriptstyle R}_{ac} |^2
| \tilde{A}^{\scriptscriptstyle L}_{bc} |^2 \right) \;\;\; ,
\label{ldecos}
\end{eqnarray}
where ${\Gamma}_{ {\ell}_a }$ is the total decay width and
$m_{ {\ell}_a }$ the mass of the decaying ${\ell}_a$
(masses of the daughter leptons are neglected).
It is assumed the virtual intermediate is an off-shell $Z^0$;
possible slepton/Higgs intermediates are assumed to be heavy and
therefore to yield negligible contributions.
Note from Eqn.$\!$ (\ref{ldecos}) that when the odd-flavored
daughter lepton ${\ell}_b$ has the charge opposite to the parent lepton, 
the branching ratio 
is suppressed by an extra factor of 
${{\tilde{A}}^2}_{ab}$ $(a \ne b)$ compared to the case where the
odd-flavor daughter lepton has the same sign as the parent.  
Actually, the latter case also has a contribution
analogous to Eqn.$\!$ (\ref{ldecos}); however, this is insignificant
relative to the Eqn.$\!$ (\ref{ldecss}) contribution.

The experimental limits on these decays are again given in
the third column of Table I.
Using Eqn.$\!$ (\ref{Aijsq}) along with the lowest order
value for ${{\tilde{A}}^2}_{cc}$ ($= \; 0.5027$) from
Eqn.$\!$ (\ref{Aiisq}), bounds can be cast in the same form as in
Eqn.$\!$ (\ref{cijform}), with :
\begin{eqnarray}
{\cal K}_{12} & = & 1.4 \times 10^{-6}  
   \;\; \hbox{from} \; {\mu}^- \rightarrow e^-e^+e^-
\nonumber \\
{\cal K}_{13} & = & 4.5 \times 10^{-3}  
   \;\; \hbox{from} \; {\tau}^- \rightarrow e^- {\mu}^+ {\mu}^- 
\nonumber \\
{\cal K}_{12} & = & 4.3 \times 10^{-3}  
   \;\; \hbox{from} \; {\tau}^- \rightarrow {\mu}^- e^+e^-
\;\;\; .
\nonumber 
\end{eqnarray}
We see that while the $\tau$ decay constraints are comparable
to the $Z^0 \rightarrow \tau \ell$ ($\ell = e \, \hbox{or} \, \mu$)
constraints, the $\mu \rightarrow eee$ constraint is much more
strict than that for $Z^0 \rightarrow e\mu$ due to the much
stronger experimental bound on $\mu \rightarrow eee$.
In Fig.$\!$ 1, predicted values from the exact numerical
calculation of this branching ratio are shown as a function of the
$\mu_i$ for the MSSM parameters
$M_{\scriptscriptstyle 2} = 
{\mu}_{\scriptscriptstyle 0} = 200\, \hbox{GeV}$ and $\tan\!\beta = 2,45$.
The constraint is very stringent for small $\tan\!\beta$ and remains
relevant, although much weaker, even for large $\tan\!\beta$ (unlike
the other constraints
discussed so far).
Plots for $Br(Z^0 \to e \mu)$ would look very similar except with
a much weaker experimental bound.
Note that in Fig.~1, the useful RPV parameter
\begin{eqnarray}
{\mu}_{\scriptscriptstyle 5} \, \equiv \,
\sqrt{ {\mu}_{\scriptscriptstyle 1}^2
     + {\mu}_{\scriptscriptstyle 2}^2
     + {\mu}_{\scriptscriptstyle 3}^2 }
\label{mu5def}
\end{eqnarray}
is introduced to permit different
${\mu}_{\scriptscriptstyle 1}:{\mu}_{\scriptscriptstyle 2}$ ratios
to be plotted simultaneously.  This parameter will appear repeatedly
in subsequent discussions.  Interpreted as a constraint on 
${\mu}_{\scriptscriptstyle 5}$, Fig.$\!$ 1 shows that the 
$\mu \rightarrow eee$ constraint
can be evaded by supposing a strong hierarchy among the ${\mu}_i$'s
({\it i.e.}, ${\mu}_{\scriptscriptstyle 1}
              \ll {\mu}_{\scriptscriptstyle 2}
              \ll {\mu}_{\scriptscriptstyle 3}$), as can also be clearly
seen from the approximate Eqn.$\!$ (\ref{cijform}).

Fig.$\!$ 2 shows 
$Br({\tau}^- \rightarrow e^- {\mu}^+ {\mu}^-)$ for $\tan\!\beta=2$,
again based on exact numerical calculations.  There is no
meaningful constraint for $\tan\!\beta = 45$ due to the much weaker
experimental bound.  We need not explicitly show
results for the remaining processes:  
results for $Br({\tau}^- \rightarrow e^- e^+ e^-)$
and $Br({\tau}^- \rightarrow {\mu}^- {\mu}^+ {\mu}^-)$
are almost identical to Fig.~2, as are those for
$Br({\tau}^- \rightarrow {\mu}^- e^+ e^-)$
and $Br({\tau}^- \rightarrow {\mu}^- {\mu}^+ {\mu}^-)$ if the roles
of ${\mu}_{\scriptscriptstyle 1}$ and ${\mu}_{\scriptscriptstyle 2}$
are interchanged.  
Plots for $Br(Z^0 \rightarrow e\tau)$ and $Br(Z^0 \rightarrow \mu\tau)$
also yield very similar results. Note that in using
Eqn.$\!$ (\ref{cijform})
to obtain constraints on $\tau$ decays in the small ${\mu}_i$
approximation, extra contributions from 
$\tilde{A}^{\scriptscriptstyle R}_{\mu\tau}$
have been neglected. 
These $\tilde{A}^{\scriptscriptstyle R}_{\mu\tau}$ contributions are 
only significant at large $\tan\!\beta$, where constraints on $\mu_i$ 
from this process as given by Eqn.$\!$ (\ref{cijform}) 
are superseded by limits from other processes.
This is confirmed by exact numerical results.
Finally, note also that for $\tau$ decays
in which the odd-flavored daughter lepton has the flipped charge, the
result is proportional to ${{\tilde{A}}^2}_{e\mu}$ and thus very strongly
limited by constraints from
$Br({\ell}^-_a \rightarrow {\ell}^-_b {\ell}^+_c {\ell}^-_c )$ such that
$Br({\tau}^- \rightarrow {\ell}^+_b {\ell}^-_c {\ell}^-_c ) 
\raisebox{-.3em}{$\stackrel{\displaystyle <}{\sim}$}
10^{-18}$.

\subsection{Universality violations at the $Z^0$ peak}

Eqns.$\!$ (\ref{Aiisq}) and (\ref{A33sq}) also produce deviations from
SM predictions for $Br(Z^0 \rightarrow {\ell}_a {\ell}_a)$ which can
break lepton universality.  Fig.$\!$ 3 shows leptonic partial widths
as a function of ${\mu}_{\scriptscriptstyle 5}$ for several choices of
${\mu}_{\scriptscriptstyle 1}:
{\mu}_{\scriptscriptstyle 2}:
{\mu}_{\scriptscriptstyle 3}$, again using the
representative MSSM
parameter point $M_{\scriptscriptstyle 2} = 
{\mu}_{\scriptscriptstyle 0} = 200\, \hbox{GeV}$
and $\tan\!\beta =2$.
The experimental $3\sigma$ bounds are only exceeded for
${\mu}_{\scriptscriptstyle 5}$ values in excess of $50\, \hbox{GeV}$.
As the value of ${\mu}_{\scriptscriptstyle 5}$ increases, the
small-${\mu}_i$ approximation loses its validity.
The numerical results show that
for sufficiently large ${\mu}_i$ values, the partial decay widths
stop increasing as the ${\mu}_i$'s increase and in fact turn over
and decrease.
This behavior is  common to all the decay widths discussed
so far (although the maxima will in general occur at different
$\mu_5$ values, sometimes above the upper limits on the plots shown).
This behavior can be understood 
via the ``large ${\mu}_i$'' approximation employed in subsection E 
below to derive chargino masses.
Note that if, at some point after reaching its maximum,
the deviation in a partial width again drops below its experimental
limit, then the large $\mu_5$ values above this point are
again acceptable.
Thus it is possible that no upper bounds can be placed on 
$\mu_5$ from these processes, but
instead either only a finite range of $\mu_5$ values is excluded or
(if the maximum is too low)
no ${\mu}_{\scriptscriptstyle 5}$ values are excluded at all!
For further details, see \cite{MRST99}.
In practice,  other constraints will rule out arbitrarily large
${\mu}_{\scriptscriptstyle 5}$ values. 

Universality constraints on $Z^0 \rightarrow {\ell}_a {\ell}_b$ decays
are quantified via the observable $U_{br}^{({\ell}_a {\ell}_b)}$ 
\cite{BKPS}:
\begin{eqnarray}
U_{br}^{(\ell_a\ell_b )} \; \equiv \;
\frac{\Gamma(Z^0\to \ell_a^+\ell_a^-)
- \Gamma (Z^0\to \ell_b^+\ell_b^-)}
{\Gamma(Z^0\to \ell_a^+\ell_a^-)
+ \Gamma (Z^0\to \ell_b^+\ell_b^-)}
= \frac{{{\tilde{A}}^2}_{aa} - {{\tilde{A}}^2}_{bb}}
{{{\tilde{A}}^2}_{aa} + {{\tilde{A}}^2}_{bb}}
= 2.136\,
\frac{({\mu}_i^2 - {\mu}_j^2) M_{\scriptscriptstyle W}^2 \cos^2\!\!\beta}
{M_{\scriptscriptstyle 0}^4} \; ,
\end{eqnarray}
where the first equality follows from Eqn.$\!$ (\ref{Zlag})
and the second from Eqn.$\!$ (\ref{Aiisq}).
LEP experimental measurements on the partial widths can now be
translated into the restrictions on the $U_{br}^{({\ell}_a {\ell}_b)}$
listed in the third column of Table I.
These are all compatible with the SM prediction of
$U_{br}^{({\ell}_a {\ell}_b)} = 0$, and, neglecting the (small) nonzero
central values, translate into the following bounds on the ${\mu}_i$:
\begin{eqnarray}
 \frac{1}{M_{\scriptscriptstyle 0}^2} | {\mu}_i^2 - {\mu}_j^2 | & \le &
\overline{{\cal K}}_{ij} (1 + {\tan}\!^2\!\beta )
\frac{M_{\scriptscriptstyle 0}^2}{M_{\scriptscriptstyle W}^2}
\;\;\; .
\label{cbijform}
\end{eqnarray}
The $\overline{{\cal K}}_{ij}$'s have values of
$\overline{{\cal K}}_{12} = 2.05 \times 10^{-3}$,
$\overline{{\cal K}}_{13} = 2.33 \times 10^{-3}$, and
$\overline{{\cal K}}_{23} = 2.62 \times 10^{-3}$;
they are comparable to the 
${\cal K}_{ij}$'s of Eqn.$\!$ (\ref{cijform}) obtained from
flavor-violating $Z^0$-decays and $\tau$-decays.
For large $\tan\!\beta$, deviations from the SM are highly suppressed;
{\it i.e.}, very high values of ${\mu}_{\scriptscriptstyle 5}$ 
(well beyond the range of
validity of this approximation) are allowed.  Exact numerical studies
also confirm that these constraints vanish.  Using the above
formula from the small ${\mu}_i$ approximation, we find that ${\mu}_i$'s
(or more precisely their difference in magnitudes) as large
as $M_{\scriptscriptstyle 0}$ become allowable for $\tan\!\beta \sim 20$.
In fact, for large $\tan\!\beta$, Eqn.$\!$ (\ref{A33sq}) should be used
for the $Z^0 \rightarrow {\tau}^+{\tau}^-$ partial width.  Cancellation
among terms in this equation would further
weaken any surviving $U_{br}^{({\ell}_a \tau)}$ bound on
${\mu}_{\scriptscriptstyle 3}$.

\subsection{Leptonic left--right asymmetry}
 
Predictions for left-right asymmetries in $Z^0$ leptonic decays,
which are defined by
\begin{eqnarray}
{\cal A}_a \equiv \frac{ | \tilde{A}^{\scriptscriptstyle L}_{aa} |^2
                        - | \tilde{A}^{\scriptscriptstyle R}_{aa} |^2 }
                       { | \tilde{A}^{\scriptscriptstyle L}_{aa} |^2
                        + | \tilde{A}^{\scriptscriptstyle R}_{aa} |^2 } \;
\end{eqnarray}
follow immediately from Eqns.$\!$ (\ref{Aiisq}) and (\ref{A33sq}):
\begin{eqnarray}
{\cal A}_a = {\cal A}_\ell^{\scriptscriptstyle (SM)}  + 
4.273\,
\frac{M_{\scriptscriptstyle W}^2 \cos\!^2\!\beta}
{M_{\scriptscriptstyle 0}^2}
\left(\frac{\mu_j}{M_{\scriptscriptstyle 0}}\right)^{\!\! 2} \;\;\;
(a=e,j=1\; \hbox{or}\; a=\mu, j=2) \; ,
\label{asym12}
\end{eqnarray}
\begin{eqnarray}
{\cal A}_{\tau} = {\cal A}_\ell^{\scriptscriptstyle (SM)}  + \left[4.273\,
\frac{M_{\scriptscriptstyle W}^2
\cos^2\!\!\beta}{M_{\scriptscriptstyle 0}^2}
+ 1.842\, \frac{ m_{\scriptscriptstyle 3}^2
(M_{\scriptscriptstyle 2}^2
+ 4M_{\scriptscriptstyle W}^2) M_{\scriptscriptstyle 2}^2}
{M_{\scriptscriptstyle 0}^6} \right]
\left(\frac{{\mu}_{\scriptscriptstyle 3}}
{M_{\scriptscriptstyle 0}}\right)^{\!\! 2} \; .
\label{asym3}
\end{eqnarray}
First note that the $|{\mu}_i|$'s enter individually
rather than in products or differences of two distinct
$|{\mu}_i|$'s.
This makes the left-right asymmetries potentially
very useful in distinguishing effects from the three ${\mu}_i$'s.
Note also that non-zero ${\mu}_i$'s always increase the ${\cal A}_a$'s
from their SM values, and that the
$\delta\!\tilde{A}^{\scriptscriptstyle R}_{\tau\tau}$
contribution now reinforces that of
$\delta\!\tilde{A}^{\scriptscriptstyle L}_{\tau\tau}$.
This could be important if these contributions cancel in
${{\tilde{A}}^2}_{\tau\tau}$ which
enters into the previously-discussed RPV $\tau$ effects.
Another immediate consequence is that for the favored case of
${\mu}_{\scriptscriptstyle 3} > {\mu}_{\scriptscriptstyle 1}$,
${\cal A}_{\tau} > {\cal A}_e$.
  
Eqns.$\!$ (\ref{asym12}) and (\ref{asym3}) and conclusions drawn from them
are valid in the small-${\mu}_i$ approximation.  In Fig.$\!$ 4, the
asymmetries are shown using exact numerical calculations (again for
$M_{\scriptscriptstyle 2} = {\mu}_{\scriptscriptstyle 0} = 200\, \hbox{GeV}$).
For large values of ${\mu}_{\scriptscriptstyle 5}$, the ${\cal A}_a$'s
cease rising with increasing ${\mu}_{\scriptscriptstyle 5}$,
deviating from the approximate behavior of Eqns.$\!$ (\ref{asym12}) and
(\ref{asym3}).
As with the $Z^0$ leptonic partial decay widths, a maximum is
reached and then the slope turns negative.  For low $\tan\!\beta$,
this occurs for ${\mu}_{\scriptscriptstyle 5}$ values excluded for
other reasons.
But for high $\tan\!\beta$ it may occur in admissible regions of the
parameter space.  Also notice from Fig.$\!$ 4 for $\tan\!\beta = 45$
that ${\cal A}_{\tau}$ shows an increase relative to
${\cal A}_e$ and ${\cal A}_{\mu}$ arising from the right-handed
contribution.

Eqns.$\!$ (\ref{asym12}) and (\ref{asym3})  require
${\cal A}_{\ell}^{(SM)}$ as input --- this must be determined
independent of the leptonic asymmetry measurements.
Unfortunately,
${\cal A}_\ell^{\scriptscriptstyle (SM)}$
depends strongly on
${\sin}^2\!\!\;{\theta}_w$; for instance,
${\cal A}_{\ell}^{\scriptscriptstyle (SM)}$ decreases by more than 11\% when
${\sin}^2\!\!\;{\theta}_w$ is increased by 1\%.
The effective value of ${\sin}^2\!\!\;{\theta}_w$ for the
$Z^0\ell\ell$ coupling depends on radiative corrections
\cite{rcsw2}; ${\sin}^2\!\!\;{\theta}_w = 0.2315$ employed
here yields
${\cal A}_\ell^{\scriptscriptstyle (SM)}= 0.147$, but
includes only SM corrections and not additional
corrections depending upon SUSY parameters.  Thus
uncertainty in the effective value of ${\sin}^2\!\!\;{\theta}_w$
leads to even larger uncertainty in 
${\cal A}_{\ell}^{\scriptscriptstyle (SM)}$,
which also includes beyond-SM contributions not apparent
in the simple separations seen in the formul\ae\ above.
In an attempt to reduce such uncertainties, Ref.$\!$ \cite{NovPil}
(following the idea put forward in Ref.$\!$ \cite{BernPil}) suggests
using
\begin{eqnarray}
\Delta{\cal A}_{ab} \; \equiv \;
\frac{{\cal A}_a - {\cal A}_b}{{\cal A}_a + {\cal A}_{b}} 
\end{eqnarray}
instead of the individual ${\cal A}_a$'s.
$\Delta{\cal A}_{ab} = 0$ in the SM.  In the small ${\mu}_i$
approximation, where
${\cal A}_a - {\cal A}_\ell^{\scriptscriptstyle (SM)}$ is also small,
\begin{eqnarray}
\Delta{\cal A}_{ab} \; = \;
\left(\frac{1}{{\cal A}_\ell^{\scriptscriptstyle (SM)}}
- 1\right)\, U_{br}^{(\ell_a\ell_b)}
\label{compcond}
\end{eqnarray}
(as noted in Ref.$\!$ \cite{NovPil}), {\em if} 
$\delta\!\tilde{A}^{\scriptscriptstyle R}_{aa}$'s are neglected.
If this is permissible, then the
$\Delta{\cal A}_{ab}$ constraints have the same dependence on the
${\mu}_i$'s as the constraints for the $U_{br}^{(\ell_a\ell_b)}$, providing 
a compatibility condition for the RPV framework.  The
$\delta\!\tilde{A}^{\scriptscriptstyle R}_{aa}$'s may safely be
neglected for the first two generations, meaning the
allowed window for $\Delta{\cal A}_{e\mu}$ given in Table I follows
immediately from that for $U_{br}^{(e\mu)}$ (though perhaps a small
amount of extra widening to reflect the uncertainly in 
${\cal A}_\ell^{\scriptscriptstyle (SM)}$ should be included).
Unfortunately, the present experimental situation is far too
imprecise for any  incompatibility to be seen.
As noted previously,
$\delta\!\tilde{A}^{\scriptscriptstyle R}_{\tau\tau}$ may not be negligible,
especially for larger $\tan\!\beta$.  Adding these terms to
$\Delta{\cal A}_{m\tau}$ and $U_{br}^{({\ell}_m\tau)}$ means that
the value of one is no longer determinable given only the value
of the other.  Furthermore, deviations from Eqn.$\!$ (\ref{compcond}),
which predicts $| \Delta{\cal A}_{m\tau} | 
\raisebox{-.3em}{$\stackrel{\displaystyle <}{\sim}$} 0.03$,
indicate a substantial contribution from
$\delta\!\tilde{A}^{\scriptscriptstyle R}_{\tau\tau}$
(within the framework of the small-${\mu}_i$ approximation).

In the $\tau$ case, then, since $U_{br}^{({\ell}_a\tau)}$ experimental 
constraints
cannot  place bounds on the
$\Delta{\cal A}_{a\tau}$'s, direct ${\cal A}_{\ell}$ measurements
must be considered.  These are listed in Table II and used to
derive the $3\sigma$ bounds on the $\Delta{\cal A}_{a\tau}$'s shown in
Table I.  The ${\cal A}_e$ and ${\cal A}_{\tau}$ used in the  
$\Delta{\cal A}_{e\tau}$ bounds of Table I are obtained from
LEP measurements of $\tau$ polarization
(note however that the SLD group at SLAC \cite{SLD} measured
${\cal A}_e$ directly with their polarized electron beam and
obtained a substantially higher value for ${\cal A}_e$).
The $\Delta{\cal A}_{\mu\tau}$ bounds are obtained from
LEP forward-backward asymmetry measurements (which provide the best
${\cal A}_{\mu}$ value but larger uncertainty in ${\cal A}_{\tau}$).
Note that the bounds are considerably weaker than the $0.03$
value given above, meaning that substantial deviation of
${\cal A}_{\tau}$ from ${\cal A}_\ell^{\scriptscriptstyle (SM)}$
including an important contribution from
$\delta\!\tilde{A}^{\scriptscriptstyle R}_{\tau\tau}$ is possible.

\subsection{Limits on charginos}

We now turn to constraints associated with the remaining color-singlet
charged fermions, the charginos.
The term ``chargino'' is here applied to the two (heaviest)
mass eigenstates remaining after the other three eigenstates
in Eqn.$\!$ (\ref{barMcidef}) are fixed to give the ``leptons'' with
well-known experimentally observed masses
($\overline{m}_1 = m_e$,
$\overline{m}_2 = m_{\mu}$,
$\overline{m}_3 = m_{\tau}$).
In the MSSM, the charginos have masses given by
$M_{c{\scriptscriptstyle 1}}$ and $M_{c{\scriptscriptstyle 2}}$ in 
Eqn.$\!$ (\ref{Mcidef}).
For nonzero values of the ${\mu}_i$, the chargino masses are modified to
$\overline{M}_{c{\scriptscriptstyle 1}}$ and
$\overline{M}_{c{\scriptscriptstyle 2}}$ in Eqn.$\!$ (\ref{barMcidef}).
If the ${\mu}_i$ are ``small,'' then
$\overline{M}_{c{\scriptscriptstyle 1}} \simeq M_{c{\scriptscriptstyle 1}}$ 
and
$\overline{M}_{c{\scriptscriptstyle 2}} \simeq M_{c{\scriptscriptstyle 2}}$; 
however,
larger values of the ${\mu}_i$'s can produce more pronounced effects. From 
Eqn.$\!$ (\ref{Mcidef2}), 
\begin{eqnarray}
{{\cal M}_{\scriptscriptstyle {\cal C}}}
({{\cal M}_{\scriptscriptstyle {\cal C}}})^{\dagger}
\; = \;
\left[
\begin{array}{ccccc}
\frac{g^2 v_u^2}{2} + M_{\scriptscriptstyle 2}^2
& \frac{g v_d}{\sqrt{2}}M_{\scriptscriptstyle 2} + 
\frac{g v_u}{\sqrt{2}}\mu_{\scriptscriptstyle 0} 
                   &  \frac{gv_u}{\sqrt{2}} \mu_{\scriptscriptstyle 1}  
                   &  \frac{gv_u}{\sqrt{2}} \mu_{\scriptscriptstyle 2} 
                   &  \frac{gv_u}{\sqrt{2}} \mu_{\scriptscriptstyle 3} \\
\frac{gv_d}{\sqrt{2}}M_{\scriptscriptstyle 2} 
+ \frac{gv_u}{\sqrt{2}}\mu_{\scriptscriptstyle 0} 
                & \frac{g^2 v_d^2}{2} + \mu_{\scriptscriptstyle 0}^2
                & \mu_{\scriptscriptstyle 0} \mu_{\scriptscriptstyle 1} 
                & \mu_{\scriptscriptstyle 0} \mu_{\scriptscriptstyle 2}
                & \mu_{\scriptscriptstyle 0} \mu_{\scriptscriptstyle 3} \\
\frac{g v_u}{\sqrt{2}} \mu_{\scriptscriptstyle 1} 
        & \mu_{\scriptscriptstyle 0} \mu_{\scriptscriptstyle 1} 
        & m_1^2+\mu_{\scriptscriptstyle 1}^2
        & \mu_{\scriptscriptstyle 1} \mu_{\scriptscriptstyle 2} 
        & \mu_{\scriptscriptstyle 1} \mu_{\scriptscriptstyle 3} \\
\frac{g v_u}{\sqrt{2}} \mu_{\scriptscriptstyle 2} 
        & \mu_{\scriptscriptstyle 0} \mu_{\scriptscriptstyle 2} 
        & \mu_{\scriptscriptstyle 1} \mu_{\scriptscriptstyle 2}
        & m_2^2+ \mu_{\scriptscriptstyle 2}^2 
        & \mu_{\scriptscriptstyle 2} \mu_{\scriptscriptstyle 3} \\
\frac{g v_u}{\sqrt{2}} \mu_{\scriptscriptstyle 3}
        & \mu_{\scriptscriptstyle 0} \mu_{\scriptscriptstyle 3} 
        & \mu_{\scriptscriptstyle 1} \mu_{\scriptscriptstyle 3}
        & \mu_{\scriptscriptstyle 2} \mu_{\scriptscriptstyle 3} 
        & m_3^2+ \mu_{\scriptscriptstyle 3}^2 \\
\end{array}
\right]  \; .
\label{McMc}
\end{eqnarray}
Now by applying the rotation
\begin{eqnarray}
 \left( \begin{array}{cc}
  I_{2 \times 2} & 0             \\
  0              & R_5
 \end{array} \right)
 \nonumber
\;\;\;\;\;\; \hbox{where} \;\;\;
R_5^{\dagger} \pmatrix{ {\mu}_{\scriptscriptstyle 1} \cr
              {\mu}_{\scriptscriptstyle 2} \cr
              {\mu}_{\scriptscriptstyle 3} \cr} =
\pmatrix{ {\mu}_{\scriptscriptstyle 5} \cr 0 \cr 0 \cr}
\end{eqnarray}
and ${\mu}_{\scriptscriptstyle 5}$ is given by Eqn.$\!$ (\ref{mu5def}),
${{\cal M}_{\scriptscriptstyle {\cal C}}}
({{\cal M}_{\scriptscriptstyle {\cal C}}})^{\dagger}$ can be rotated into
the form
\begin{eqnarray}
\left[
\begin{array}{ccccc}
\frac{g^2 v_u^2}{2} + M_{\scriptscriptstyle 2}^2  
        & \frac{gv_d}{\sqrt{2}}M_{\scriptscriptstyle 2}
        + \frac{gv_u}{\sqrt{2}}\mu_{\scriptscriptstyle 0}
        & \frac{gv_u}{\sqrt{2}} \mu_{\scriptscriptstyle 5} & 0 & 0 \\
\frac{gv_d}{\sqrt{2}}M_{\scriptscriptstyle 2} + 
\frac{gv_u}{\sqrt{2}}\mu_{\scriptscriptstyle 0} 
        & \frac{g^2 v_d^2}{2} + \mu_{\scriptscriptstyle 0}^2
        & \mu_{\scriptscriptstyle 0} \mu_{\scriptscriptstyle 5} & 0 & 0  \\
\frac{g v_u}{\sqrt{2}} \mu_{\scriptscriptstyle 5} 
        & \mu_{\scriptscriptstyle 0} \mu_{\scriptscriptstyle 5}
        & \mu_{\scriptscriptstyle 5}^2 & 0 & 0 \\
0 & 0 & 0  & 0 & 0 \\
0 & 0 & 0  & 0 & 0 
\end{array} \right]
\, + \,
\left[
\begin{array}{rc}   
 & \\
\smash{\raise 0.8ex \hbox{$\hbox{\huge{ 0}}_{2 \times 2}$}} &
\smash{\raise 1.1ex \hbox{$0$}}
 \\
\smash{\lower 4.7ex \hbox{$0\;\;\;\;\;$}}
& \smash{\lower 4.7ex \hbox{$
 R_5^{\dagger}
 \pmatrix{ m_1^2 & 0     &     0 \cr
           0     & m_2^2 & 0     \cr
           0     & 0     & m_3^2 \cr}
 R_5
$}} \\
 & \\
 &
\end{array} \right] \, .
\label{largemu}
\end{eqnarray}
 The matrix $R_5$ will appear again in the next section when the 
 color-singlet neutral fermions are considered.
In the limit where
${\mu}_{\scriptscriptstyle 5} \; \gg \; m_3 \simeq m_{\tau}$
(referred to henceforth as the large ${\mu}_i$ approximation\footnote{
Note that this is {\em not} the converse of the small $\mu_i$ 
approximation, which requires $\mu_i \ll M_0$.
}),
the second matrix of expression (\ref{largemu}) that is proportional
to the $m_i$'s may be dropped.  This leads to simple
analytic formul\ae\ for the chargino mass eigenvalues: 
\begin{eqnarray}
\overline{M}_{c {\scriptscriptstyle 1},c {\scriptscriptstyle 2}}^2
& = & \frac{1}{2} \left[  M_{\scriptscriptstyle 2}^2 
+ 2 M_{\scriptscriptstyle W}^2 +
{\mu}_{\scriptscriptstyle 0}^2 + {\mu}_{\scriptscriptstyle 5}^2 \right]
\nonumber \\
& & \;\;\;\;\; \pm 
\; \frac{1}{2} \left[ \left( {\mu}_{\scriptscriptstyle 0}^2
 + {\mu}_{\scriptscriptstyle 5}^2 - M_{\scriptscriptstyle 2}^2
- 2 M_{\scriptscriptstyle W}^2 \cos\!2\beta  \right)^2
+ 8 M_{\scriptscriptstyle W}^2 \left( M_{\scriptscriptstyle 2} \sin\!\beta
+ {\mu}_{\scriptscriptstyle 0}\cos\!\beta \right)^2
\right]^{1/2}
\label{mgino-a} \\
& = & 
\frac{1}{2} \left( \bar{\alpha_{\scriptscriptstyle 1}} +
\bar{\alpha_{\scriptscriptstyle 2}} \right)
\pm \frac{1}{2}
\sqrt{ (\bar{\alpha_{\scriptscriptstyle 1}} - 
\bar{\alpha_{\scriptscriptstyle 2}})^2 +
        2(g v_u M_{\scriptscriptstyle 2} + 
g v_d \mu_{\scriptscriptstyle 0})^2}
\label{mgino-b} \\
& = &
\frac{1}{2} \left[ M_{c {\scriptscriptstyle 1}}^2 + 
       M_{c {\scriptscriptstyle 2}}^2 +
       {\mu}_{\scriptscriptstyle 5}^2 \right]
 \pm \frac{1}{2}
\left[ \left( M_{c {\scriptscriptstyle 2}}^2 - 
              M_{c {\scriptscriptstyle 1}}^2 -
              {\mu}_{\scriptscriptstyle 5}^2 \right)^2
- 4 {\mu}_{\scriptscriptstyle 5}^2 {\cos}^2{\theta}_{\!\scriptscriptstyle R}
\left( M_{c {\scriptscriptstyle 2}}^2 - 
M_{c {\scriptscriptstyle 1}}^2 \right)^2
\right]^{1/2}
\label{mgino-c}
\end{eqnarray}
where
\begin{eqnarray}
\bar{\alpha_{\scriptscriptstyle 1}} \; \equiv \;
\frac{g v_u^2}{2} + \mu_{\scriptscriptstyle 0}^2 + 
\mu_{\scriptscriptstyle 5}^2
\;\;\; \hbox{and} \;\;\;
\bar{\alpha_{\scriptscriptstyle 2}} \; \equiv \; \frac{g v_d^2}{2} + 
M_{\scriptscriptstyle 2}^2 \;\;\; .
\label{alphadef}
\end{eqnarray}
Within this approximation the chargino masses only depend
on the ${\mu}_i$ through the single parameter 
${\mu}_{\scriptscriptstyle 5}$.
The roles of ${\mu}_{\scriptscriptstyle 0}$ and 
${\mu}_{\scriptscriptstyle 5}$ in the mass
formul\ae\ are very similar.  Note from Eqn.$\!$ (\ref{mgino-c})
that the chargino masses reduce to the MSSM chargino masses when
${\mu}_{\scriptscriptstyle 5} = 0$.  Also note that a non-zero value
for ${\mu}_{\scriptscriptstyle 5}$ {\em increases} the lighter chargino mass;
hence, some region of the 
$M_{\scriptscriptstyle 2}$~--~${\mu}_{\scriptscriptstyle 0}$ parameter space
that is ruled out in the MSSM by the chargino mass bound \cite{LEPrev}
can be re-instated when ${\mu}_{\scriptscriptstyle 5} \ne 0$.  However,
correct chargino mass bounds for RPV scenarios require analysis of
the decay modes.  One such analysis was recently performed for a
specific bilinear RPV model \cite{ADV}; a general analysis
in the single-VEV parametrization, including contributions from
both bilinear and trilinear RPV couplings, is currently in progress.

It is possible to simultaneously have
${\mu}_{\scriptscriptstyle 5} \gg m_{\tau}$ and ${\mu}_i 
\ll M_{\scriptscriptstyle 0}$,
meaning that both the ``large-${\mu}_i$'' and the ``small-${\mu}_i$''
approximations will yield quite accurate results.  In fact, results from
the two methods agree well with each other and with the exact numerical
results in most relevant regions of the parameter space.  
Using the exact numerical calculation, the region in MSSM parameter
space that would be excluded by a $90\, \hbox{GeV}$ chargino mass bound 
is plotted in Fig.$\!$ 5, from which it is clear that constraints based on
such an analysis will be significant if $M_{\scriptscriptstyle 2}$ and/or
${\mu}_{\scriptscriptstyle 0}$ are small.  

Bounds from $Z^0$ decays into two charginos or a chargino and
a lepton follow from the general formula for $Z^0$ decays into
a pair of charged fermions, Eqn.$\!$ (\ref{Zlag}).  
However, determining the exact values for these bounds
requires detailed information about chargino decays from detector
simulations which is beyond the scope of this work.  Therefore, only
a conservative bound is applied:  $Z^0$ decays involving at least
one (on-shell) chargino are required to have branching ratios of
less than $10^{-5}$ (as noted in Table I).

Turning to the lepton sector, we see that in the limit where $m_i^2=0$, 
the matrix ${{\cal M}_{\scriptscriptstyle {\cal C}}}
({{\cal M}_{\scriptscriptstyle {\cal C}}})^{\dagger}$
 of Eqn.$\!$ (\ref{McMc})
[or Eqn.$\!$ (\ref{largemu})] has three zero eigenvalues.
Thus, in this approximation, the masses of the physical leptons are
zero.  Despite this degeneracy, the correct eigenvectors can be found 
by requiring that they do not differ significantly from the exact 
massive eigenstates
obtained when the non-zero $m_i$ are retained 
{\sl if} ${\mu}_{\scriptscriptstyle 5}^2$ {\sl is large}. 
 Three such massless
eigenvectors for
${{\cal M}_{\scriptscriptstyle {\cal C}}}
({{\cal M}_{\scriptscriptstyle {\cal C}}})^{\dagger}$ with $m_i^2=0$ are
given by:
\begin{eqnarray}
\begin{array}{ccccccccc}
 |\ell^{\prime}_1> & = &  (\frac{g v_d}{\sqrt{2}}\mu_{\scriptscriptstyle 1} ,
             & \ - M_{\scriptscriptstyle 2}\mu_{\scriptscriptstyle 1} ,
             & M_{\scriptscriptstyle 0}^2 , & 0 , & \ 0 & )^T & /\Delta_1
 \\
 |\ell^{\prime}_2> & = &  (\frac{g v_d}{\sqrt{2}}\mu_{\scriptscriptstyle 2} ,
             & \ - M_{\scriptscriptstyle 2}\mu_{\scriptscriptstyle 2} ,
             & 0 , & M_{\scriptscriptstyle 0}^2 , & \ 0 & )^T & /\Delta_2
 \\
 |\ell^{\prime}_3> & = &  (\frac{g v_d}{\sqrt{2}}\mu_{\scriptscriptstyle 3} ,
             & \ - M_{\scriptscriptstyle 2}\mu_{\scriptscriptstyle 3} ,
             & 0 , & 0 , & \ M_{\scriptscriptstyle 0}^2 & )^T & /\Delta_3 
\end{array}
\label{lepvec}
\end{eqnarray}
(the $\Delta_i \, \equiv \,
       \sqrt{M_{\scriptscriptstyle 0}^4 + 
     \bar{\alpha_{\scriptscriptstyle 2}}{\mu}_i^2}$ are
normalization constants).
Since these are degenerate states, any linear
combination of the three will be a massless eigenstate.
The natural hierarchy of the
$m_i$ values ($m_3 > m_2 > m_1$)  requires  the following
choice of orthogonal eigenvectors:
\begin{eqnarray}
\begin{tabular}{ccccccccc}
$|\ell_e>$    & $ \; \propto \; $ &
\multicolumn{4}{l}{$|\ell^{\prime}_1>$} & & &
 \\
       & $ \; = \; $ &  $(\frac{g v_d}{\sqrt{2}}\mu_{\scriptscriptstyle 1}$ ,
               & $\ - M_{\scriptscriptstyle 2}\mu_{\scriptscriptstyle 1}$ ,
               & $M_{\scriptscriptstyle 0}^2$ , & $0$ , & $\ 0$ & $)^T$
       & $/\Delta_e$
           \\
$|\ell_\mu>$  & $ \; \propto \; $ &
\multicolumn{4}{l}{$|\ell^{\prime}_2>
- |\ell^{\prime}_1><\ell^{\prime}_1|\ell^{\prime}_2>$} & & &
 \\
     & $ \; = \; $ &  $(\frac{g v_d}{\sqrt{2}}\mu_{\scriptscriptstyle 2}$ ,
         & $\ - M_{\scriptscriptstyle 2}\mu_{\scriptscriptstyle 2}$  ,
         & $\ - \bar{\alpha_{\scriptscriptstyle 2}}\mu_{\scriptscriptstyle 2}
           \mu_{\scriptscriptstyle 1}/M_{\scriptscriptstyle 0}^2$ ,
           & $\Delta_e^2/M_{\scriptscriptstyle 0}^2$ , & $\ 0$ & $)^T$
              & $M_{\scriptscriptstyle 0}^2/(\Delta_e \Delta_\mu)$
           \\
$|\ell_\tau>$ & $ \; \propto \; $ &
\multicolumn{4}{l}{$|\ell^{\prime}_3>
- |\ell^{\prime}_1><\ell^{\prime}_1|\ell^{\prime}_3>
-  |\ell^{\prime}_2><\ell^{\prime}_2|\ell^{\prime}_3>$} & & &
 \\
    & $ \; = \; $ & $(\frac{g v_d}{\sqrt{2}}\mu_{\scriptscriptstyle 3}$ ,
        & $\ - M_{\scriptscriptstyle 2}\mu_{\scriptscriptstyle 3}$ ,
        & $\ - \bar{\alpha_{\scriptscriptstyle 2}}\mu_{\scriptscriptstyle 3}
        \mu_{\scriptscriptstyle 1}/M_{\scriptscriptstyle 0}^2$ ,
        & $\ - \bar{\alpha_{\scriptscriptstyle 2}}\mu_{\scriptscriptstyle 3}
        \mu_{\scriptscriptstyle 2}/M_{\scriptscriptstyle 0}^2$ ,
            & $\ \Delta_\mu^2/M_{\scriptscriptstyle 0}^2$ & $)^T$
            & $M_{\scriptscriptstyle 0}^2/(\Delta_\mu \Delta_\tau)$
\end{tabular}
\label{cos-in}
\end{eqnarray}
where the normalization constants are:
\begin{eqnarray}
\Delta_e \, \equiv \,
             \sqrt{M_{\scriptscriptstyle 0}^4 + 
 \bar{\alpha_{\scriptscriptstyle 2}}\mu_{\scriptscriptstyle 1}^2}
             \, , \;
\Delta_{\mu} \, \equiv \,
    \sqrt{M_{\scriptscriptstyle 0}^4 + \bar{\alpha_{\scriptscriptstyle 2}}
    (\mu_{\scriptscriptstyle 1}^2 +\mu_{\scriptscriptstyle 2}^2)} \, , \;
\Delta_{\tau} \, \equiv \,
     \sqrt{M_{\scriptscriptstyle 0}^4 + \bar{\alpha_{\scriptscriptstyle 2}}
      (\mu_{\scriptscriptstyle 1}^2 +
       \mu_{\scriptscriptstyle 2}^2 +
       \mu_{\scriptscriptstyle 3}^2)} \, .
\end{eqnarray}

The first components of these three vectors are the elements
$U_L^{a1}$ of the left rotation matrix in Eqn.$\!$ (\ref{barMcidef}):
\begin{eqnarray}
 U_L^{e1}=\frac{g v_d}{\sqrt{2}}\  
 \frac{\mu_{\scriptscriptstyle 1}}{\Delta_e} \ ; \ \
 U_L^{\mu 1}=\frac{g v_d}{\sqrt{2}}\ 
 \frac{\mu_{\scriptscriptstyle 2} M_0^2}{\Delta_e \Delta_{\mu}} \ ; \ \
U_L^{\tau 1}=\frac{g v_d}{\sqrt{2}}\
 \frac{\mu_{\scriptscriptstyle 3} M_0^2}{\Delta_{\mu} \Delta_{\tau}} 
\;\;\; .
\end{eqnarray} 
These quantities appear in the expression 
for the anomalous coupling coefficients
$\delta \tilde{A}^L_{ab}$ of Eqn.$\!$ (\ref{ULAL}).
Note that in the limit $\mu_i \rightarrow 0$,
$\Delta_a \rightarrow M_0^2$ and
 we recover the ``small $\mu_i$'' approximation results given by
Eqns.$\!$ (\ref{vvalues}).
The dependence of the normalization constants (the $\Delta_a$'s) on the
$\mu_i$'s indicates that the decay widths will {\em decrease}
(for $\mu$ and $\tau$) as the ${\mu}_i$'s increase, and exposes the
reason behind the maxima seen in Figs.$\!$ 3 and 4. Thus if 
an experimental bound actually allows the maximum value
(for instance at large $\tan\!\beta$ where there is strong suppression),
then that bound may provide no constraint on ${\mu}_5$
(or equivalently on the ${\mu}_i$'s) at all.
Even if the maximum is excluded, a window of
large (possibly {\em very} large) ${\mu}_i$'s may be allowed.
However, this region of the RPV parameter space is largely ruled out
by limits on neutrino masses and bounds from neutrino scattering 
and charged current processes.  Therefore  we do not consider 
very large ${\mu}_i$ values further at present and instead
proceed to a discussion of these additional constraints.

\section{Color-Singlet Neutral Fermions}

     In the single-VEV parametrization, the Lagrangian terms
     contributing to the color-singlet neutral fermion (neutrino
     and neutralino) masses may be written as
\begin{eqnarray}
{\cal L}  & \ni &
 - \! \left( \! \begin{array}{ccccccc}
 i{\lambda}_0^C & i{\lambda}_3^C
 & {\tilde{h_u}}^{0\,C} & {\tilde{h_d}}^{0\,C}
 & {\nu}_1^C & {\nu}_2^C & {\nu}_3^C
  \end{array} \! \right) 
  \! {\cal M}_{\scriptscriptstyle {\cal N}} \!
  \left( \! \begin{array}{ccccccc}
 -i{\lambda}_0 & -i{\lambda}_3 &
 {\tilde{h_u}}^0 & {\tilde{h_d}}^0 &
 {\nu}_1 & {\nu}_2 & {\nu}_3
  \end{array} \! \right)^T \!\! + \, \hbox{c.c.} \!
\label{nLag}
\end{eqnarray}
 where
 ${\tilde{h_u}}^0$, ${\tilde{h_d}}^0$, and ${\nu}_i$ ($i=1$-$3$)
 are the
 Dirac spinors
 associated with the neutral superfields in
 $\hat{H}_u$, $\hat{L}_0$, and $\hat{L}_i$, respectively, of
 Eqn.$\!$ (\ref{superp2})
 and $-i{\lambda}_0$ and $-i{\lambda}_3$ are respectively the bino and
 wino components.  $C$ refers to charge conjugation acting
 on each spinor.
 Since neutrinos are assumed to have zero Yukawa masses,
 the same $R_5$ rotation that was used when the Yukawa masses
 for the charged leptons were neglected is also applicable
 to ${\cal M}_{\scriptscriptstyle {\cal N}}$.  Now there is no approximation,
 no Yukawa part to throw away; two massless neutrino states
 decouple\cite{JN1}, leaving a $5\times 5$ matrix:
\begin{eqnarray}
{\cal M}_{\scriptscriptstyle {\cal N}} =
{\textstyle
\left( \begin{array}{ccccccc}
  M_{\scriptscriptstyle 1}
  & 0
  & \frac{g^{\prime}v_u}{2} \!\!
  & {\scriptstyle -}\frac{g^{\prime}v_d}{2} \!\!
  & 0 & 0 & 0 \\
  0
  & M_{\scriptscriptstyle 2}
  & {\scriptstyle -}\frac{gv_u}{2} \!\!
  &  \frac{gv_d}{2}  \!\!
  & 0 & 0 & 0  \\
  \frac{g^{\prime}v_u}{2} \!\!
  & {\scriptstyle -}\frac{gv_u}{2} \!\!
  & \; 0
  & {\scriptstyle -}{\mu}_{\scriptscriptstyle 0} \!\!
  & {\scriptstyle -}{\mu}_{\scriptscriptstyle 1} \!\!
  & {\scriptstyle -}{\mu}_{\scriptscriptstyle 2} \!\!
  & {\scriptstyle -}{\mu}_{\scriptscriptstyle 3} \!\!  \\
  {\scriptstyle -}\frac{g^{\prime}v_d}{2} \!\!
  & \frac{gv_d}{2} \!\!
  & {\scriptstyle -}{\mu}_{\scriptscriptstyle 0} \!\!
  & 0 & 0 & 0 & 0 \\
  0
  & 0
  & {\scriptstyle -}{\mu}_{\scriptscriptstyle 1} \!\!\!
  & 0 & 0 & 0 & 0 \\
  0
  & 0
  & {\scriptstyle -}{\mu}_{\scriptscriptstyle 2} \!\!\!
  & 0 & 0 & 0 & 0 \\
  0
  & 0
  & {\scriptstyle -}{\mu}_{\scriptscriptstyle 3} \!\!\!
  & 0 & 0 & 0 & 0 
  \end{array} \right)
}
\;
\stackrel
{ \smash{\raise 0.5ex \hbox{$R_5$}}}{\Rightarrow}
\;
\left( \begin{array}{ccccc}
  M_{\scriptscriptstyle 1}  & 0  & \frac{g^{\prime}v_u}{2}
  & -\frac{g^{\prime}v_d}{2} & 0 \\
  0 & M_{\scriptscriptstyle 2}  & -\frac{gv_u}{2}
  &  \frac{gv_d}{2}          & 0  \\
  \frac{g^{\prime}v_u}{2} & -\frac{gv_u}{2} & 0
  & -{\mu}_{\scriptscriptstyle 0} & -{\mu}_{\scriptscriptstyle 5} \!\! \\
 -\frac{g^{\prime}v_d}{2} & \frac{gv_d}{2} & -{\mu}_{\scriptscriptstyle 0}
  & 0 & 0 \\
  0 & 0 & -{\mu}_{\scriptscriptstyle 5}
  & 0 & 0 \\
  \end{array} \right) \;\; .
\label{MatrixN}
\end{eqnarray}

\subsection{ Neutrino mass (tree-level)}

The single massive neutrino
that results from diagonalizing this
tree-level mixing matrix was discussed in the first paper
in this series\cite{PapI}.  Here again we denote the massive state by 
$ \left|{\nu}_{\scriptscriptstyle 5}\right\rangle \, = \,
\frac{{\mu}_{\scriptscriptstyle 1}}{{\mu}_{\scriptscriptstyle 5}}
\left| {\nu}_{\scriptscriptstyle 1} \right\rangle
+
\frac{{\mu}_{\scriptscriptstyle 2}}{{\mu}_{\scriptscriptstyle 5}}
\left| {\nu}_{\scriptscriptstyle 2} \right\rangle
+
\frac{{\mu}_{\scriptscriptstyle 3}}{{\mu}_{\scriptscriptstyle 5}}
\left| {\nu}_{\scriptscriptstyle 3} \right\rangle
$.
Approximate analytic formul\ae\ for
its mass were found to be:
\begin{eqnarray}
m_{\nu_{\scriptscriptstyle 5}} =
 -\frac{1}{2}
\frac{ {\mu}_{\scriptscriptstyle 5}^{2} v^2 \cos^2\!\!\beta 
\left( xg^2 + {g}^{\prime 2} \right) }
{\mu_{\scriptscriptstyle 0} 
 \left[ 2xM_{\scriptscriptstyle 2} \mu_{\scriptscriptstyle 0} -
 \left( xg^2+{g}^{\prime 2} \right) 
 v^2 \sin\!\beta\cos\!\beta \right] }
\label{seesaw}
\end{eqnarray}
(where $v^2 \equiv v_u^2 + v_d^2$ and 
$M_{\scriptscriptstyle 1} = xM_{\scriptscriptstyle 2}$;
$x = \frac{5}{3}\tan\!\!^2{\theta}_w$ assuming gaugino unification,
as is done in all numerical calculations)
from a ``seesaw'' approximation in which ${\mu}_{\scriptscriptstyle 5}$
is taken to be small and 
\begin{eqnarray} 
m_{\nu_{\scriptscriptstyle 5}} = - \frac{1}{4} 
\frac{{\mu}_{\scriptscriptstyle 5}^{2} {v}^{2} {\cos}^2\!\!\>{\beta} 
\left(xg^2 + {g}^{\prime 2} \right)}
{ \left( \mu_{\scriptscriptstyle 0}^{2} + \mu_{\scriptscriptstyle 5}^{2} 
\right) xM_{\scriptscriptstyle 2} }
\label{numass}
\end{eqnarray}
from a perturbative treatment in which the electroweak
symmetry breaking terms are regarded as small but the
magnitude of ${\mu}_{\scriptscriptstyle 5}$ is not restricted.  Key
features to note are:
\begin{itemize}
\item
The mass has a simple dependence on only one RPV parameter, 
${\mu}_{\scriptscriptstyle 5}$.
This is to be contrasted with results found without using the
single-VEV parametrization \cite{other}.
The formul\ae\ above are also quite general.
In particular, the trilinear terms have {\em not} been set to zero
--- they simply do not contribute to the tree-level mass formul\ae.
They will reappear at the one-loop level; however, loop effects
are expected to be small compared to those at tree-level \cite{loops}. 

\item
The approximate expressions, Eqns.$\!$ (\ref{seesaw}) and (\ref{numass}),
indicate that the neutrino mass is proportional to 
${\mu}_{\scriptscriptstyle 5}^2$ if
${\mu}_{\scriptscriptstyle 5} \ll {\mu}_{\scriptscriptstyle 0}$.
Eqn.$\!$ (\ref{numass}) also shows
that for ${\mu}_{\scriptscriptstyle 5} \gg {\mu}_{\scriptscriptstyle 0}$, 
the neutrino mass approaches a constant asymptotic value.
The features of these approximate formul\ae\
are confirmed by exact numerical calculations, as shown in Fig.$\!$ 6 
for the MSSM parameters
$M_{\scriptscriptstyle 2} = 
{\mu}_{\scriptscriptstyle 0} = 200\, \hbox{GeV}$ and $\tan\!\beta = 2,45$.
Note in particular the  linear rise in 
$\log m_{{\nu}_{\scriptscriptstyle 5}}$ seen in the log-log inserts for low
$\log {\mu}_{\scriptscriptstyle 5}$
values, as expected if 
$m_{{\nu}_{\scriptscriptstyle 5}} \propto {\mu}_{\scriptscriptstyle 5}^2$.
Note also that there is no decrease in
$m_{\nu_{\scriptscriptstyle 5}}$ for high ${\mu}_{\scriptscriptstyle 5}$ 
values, in contrast to the case of several constraints in the previous 
section.  Therefore, unless the asymptotic value falls below the 
experimental bound, the ${\mu}_{\scriptscriptstyle 5}$ upper limit limit 
from $m_{\nu_{\scriptscriptstyle 5}}$ will close any 
large-${\mu}_{\scriptscriptstyle 5}$ window.
\item
There is strong suppression both at high $\tan\!\beta$ due to the
${\cos}^2\!\!\>{\beta}$ factor\footnote{This factor was also obtained 
in \cite{Hemp}, but additional assumptions led
to compensating factors that canceled the suppression.} 
and at high $M_{\scriptscriptstyle 2}$.
Inverting Eqn.$\!$ (\ref{numass}) yields a bound on 
${\mu}_{\scriptscriptstyle 5}$:
\begin{eqnarray}
\mu_{\scriptscriptstyle 5}^2
< \frac {4 {x}{\mu}_{\scriptscriptstyle 0}^{2}M_{\scriptscriptstyle 2} 
m_{\nu_{({\text b\!o\!u\!n\!d})}}}{{v}^{2} {\cos}^2\!\!\;{\beta}
\left( xg^2 + g^{\prime 2} \right) - 4{x}M_{\scriptscriptstyle 2}  
m_{\nu_{({\text b\!o\!u\!n\!d})}}} \; ,
\end{eqnarray}
clearly demonstrating the high $\tan\!\beta$ suppression.
The exact numerical results in both Figs.$\!$ 6a and 6b 
and Figs.$\!$ 7b and 7d also
illustrate that the asymptotic limit is much lower for large
$\tan\!\beta$.
In addition, as $M_{\scriptscriptstyle 2}$ increases, the denominator 
goes to zero,
beyond which $\mu_{\scriptscriptstyle 5}$ is unconstrained.
\item
If we assume $\mu_{\scriptscriptstyle 5} = \mu_{\scriptscriptstyle 3}
\Leftrightarrow (\mu_{\scriptscriptstyle 1}:
                 \mu_{\scriptscriptstyle 2}:
                 \mu_{\scriptscriptstyle 3}) = (0:0:1)$,
then the recently-improved mass bound on ${\nu}_{\tau}$ from LEP,
${\nu}_{\tau} < 18.2\, \hbox{MeV}$ \cite{18p2}
(shown by solid horizonal lines in Figs.$\!$ 6ab),
implies that, for $\tan\!\beta = 45$, there is no
$m_{{\nu}_{\tau}}$ constraint on ${\mu}_{\scriptscriptstyle 5}$ 
when $M_{\scriptscriptstyle 2} 
\raisebox{-.3em}{$\stackrel{\displaystyle >}{\sim}$} 280\, \hbox{GeV}$ 
(Fig.~7b).
Analogous arguments can be made for 
$\mu_{\scriptscriptstyle 5} = \mu_{\scriptscriptstyle 1}$ and
$\mu_{\scriptscriptstyle 5} = \mu_{\scriptscriptstyle 2}$ using
$m_{\scriptscriptstyle \nu_{e}} < 5\, \hbox{eV}$ \cite{PDG} and
$m_{\scriptscriptstyle \nu_{\mu}} < 170\, \hbox{keV}$ \cite{mnumu}, 
respectively.
Due to the much tighter bounds these constraints exclude all interesting
regions of parameter space {\em with such}
$({\mu}_1:{\mu}_2:{\mu}_3)$ {\em ratios}.
\item
The negative sign for the neutrino mass in
Eqns.$\!$ (\ref{seesaw}) and (\ref{numass})
can be removed by redefining the fields in Eqn.$\!$ (\ref{nLag})
(see \cite{oldTat} for more details).
\end{itemize}

If more than one of the ${\mu}_i$'s is non-zero, then the massive
neutrino will be an admixture of the three neutrino basis states
plus (especially for large ${\mu}_i$'s) the two gaugino and two
higgsino states.
Using the matrix $U_0$ to diagonalize 
${\cal M}_{\scriptscriptstyle {\cal N}}$, the eigenvalues are
\begin{eqnarray} 
U_0^{\dag} {\cal M}_{\scriptscriptstyle {\cal N}} U_0 = 
\mbox{diag}\{M_{n{\scriptscriptstyle 1}}, M_{n{\scriptscriptstyle 2}},
M_{n{\scriptscriptstyle 3}}, M_{n{\scriptscriptstyle 4}}, 
0, 0, m_{\nu_{\scriptscriptstyle 5}}\} \; ,
\label{neutev}
\end{eqnarray}
where the massive neutrino is defined to be the lightest of the
massive states and the four heavier states are termed
neutralinos.\footnote{Going beyond tree level will give small
masses to the two zero mass neutrino eigenstates found here.
Trilinear RPV terms will contribute to these corrections.
Attempts to fit sub-eV mass neutrinos such as suggested by results
of the Super-Kamiokande experiment \cite{superK}
into the RPV framework will require knowledge of these corrections.
This is beyond the scope of the present work.
See \cite{Otto1} for more details.}
Further, note that there is no reason to expect alignment
between the neutrino eigenstates and the charged lepton
mass eigenstates ($e$, $\mu$, and $\tau$).
Neutrino mass bounds in this more general case 
 are more complicated, and
better constraints are in fact obtainable from analysis of charged current
processes (to be discussed in detail in the next section)
than from direct mass bounds on ${\nu}_{\tau}$ (or ${\nu}_{\mu}$)
\cite{CCneut}.
Based on such an analysis, Bottini {\it et al.}
\cite{BFKM} gave a general neutrino mass bound of $149\, \hbox{MeV}$
for a massive neutrino that was an admixture of ${\nu}_e$,
${\nu}_{\mu}$ and ${\nu}_{\tau}$.  The present case differs from
theirs since in the RPV framework the massive neutrino can also have
gaugino and higgsino contributions.  Nonetheless, the $149\, \hbox{MeV}$
mass bound is in fact applicable as will be justified in the following
section treating charged current interactions.
This bound is shown in
Fig.$\!$ 6a as the upper horizontal line, but not shown for
the $\tan\!\beta = 45$ case depicted in Fig.$\!$ 6b since the bound is 
never reached.

Cosmological neutrino mass bounds also exist.  These are usually far
more stringent than the neutrino mass bounds discussed thus far, and 
upper limits of 
$m_{\nu} 
\raisebox{-.3em}{$\stackrel{\displaystyle <}{\sim}$} 35\, \hbox{eV}$ 
have been given\cite{cosbnd}.
However, additional assumptions about
cosmology enter when determining these values, which are also sensitive
to the decay modes of the massive neutrino, which  
is  expected to be unstable.  Due to these loopholes,
a MeV neutrino
is not cosmologically taboo\cite{cosmo}.

Upper bounds on ${\mu}_5$
(obtained from exact numerical calculations)
throughout the ${\mu}_0$-$M_{\scriptscriptstyle 2}$ MSSM parameter
space are shown for $\tan\!\beta = 2,45$ in Fig.$\!$ 7. 
Figs.$\!$ 7ab show results
assuming a mass bound of $18.2\, \hbox{MeV}$ --- {\it i.e.}, assuming
$(\mu_{\scriptscriptstyle 1}:
  \mu_{\scriptscriptstyle 2}:
  \mu_{\scriptscriptstyle 3}) = (0:0:1)$
(these update the plots of \cite{PapI} which assumed a mass bound of
$24\, \hbox{MeV}$ for ${\nu}_{\tau}$);
and Figs.$\!$ 7cd show results with the more general bound of
$149\, \hbox{MeV}$.  Bounds on ${\mu}_{\scriptscriptstyle 5}$ weaken as 
either $M_{\scriptscriptstyle 2}$ and/or $|{\mu}_{\scriptscriptstyle 0}|$ 
increase.  As noted in \cite{PapI}, for high $\tan\!\beta$, 
${\mu}_{\scriptscriptstyle 5}$ values in the hundreds of GeV are permitted 
by the ${\nu}_{\tau}$ mass bound. 

\subsection{Invisible {\protect\boldmath $Z^0$}-width}
The couplings of the seven Majorana mass eigenstates to the $Z^0$ boson
are given by
\begin{eqnarray}
{\cal L}_{{\chi}^0 {\chi}^0 Z^0} =
-\frac{g_2}{4\cos{\theta}_w} {\chi}_c^0 {\gamma}_{\mu}
\Big(
i{\Im}m \tilde{C}_{cd} 
\; - \; {\gamma}_{\scriptscriptstyle 5} {\Re}e \tilde{C}_{cd}
\Big)
{\chi}_d^0 Z^{\mu} \; ,
\label{Znnlag}
\end{eqnarray}
where
$\tilde{C}_{cd} = \tilde{C}^{\dagger}_{cd} = (U_0 T^Z U_0^{\dagger})_{cd}$
and $T^Z = \hbox{diag}(0,\, 0,\, -1,\, 1,\, 1,\, 1,\, 1)$
(again adopting the notation of \cite{NovPil}).
The partial $Z^0$ decay width into a pair of neutral fermions 
(with $M_{\scriptscriptstyle Z} > M_{ {\chi}_c } + M_{ {\chi}_d }$) is
then given by
\begin{eqnarray}
\Gamma(Z^0 \rightarrow \bar{\chi}^0_c \chi^0_d) =
\frac{\smash{\lower 0.6ex \hbox{
${\alpha}_2 \, {\lambda}^{\scriptscriptstyle \!\!\!\!
\smash{\raise 0.9ex \hbox{ $\frac{1}{2}$ }}} $
}}
\!\!\!\!\!
\left(
{\scriptstyle
 1, \frac{ M_{ {\chi}_c }^2 }{ M_{\scriptscriptstyle Z}^2 },
 \frac{ M_{ {\chi}_d }^2 }{ M_{\scriptscriptstyle Z}^2 }
}
 \right)
}{ 12{\cos}^2\!\!\;{\theta}_w }
\left[
\smash{\lower 0.4ex \hbox{$1$}}
{\scriptstyle
 \, - \, \frac{ M_{ {\chi}_c }^2 + 
M_{ {\chi}_d }^2 }{ 2M_{\scriptscriptstyle Z}^2 } \, - \, 
\frac{ (M_{ {\chi}_c }^2 - M_{ {\chi}_d }^2)^2 }
     { 2M_{\scriptscriptstyle Z}^4 }
\, - \, \frac{ 3M_{ {\chi}_c } M_{ {\chi}_d } }
             { M_{\scriptscriptstyle Z}^2 }
}
\right]
\left|\tilde{C}_{cd}\right|^2 \;\;\; .
\label{nZdw}
\end{eqnarray}

In the SM, the invisible $Z^0$ width,
${\Gamma}^{\scriptstyle S\!M}_{
\!\scriptstyle Z_{\scriptscriptstyle i\!n\!v}}$,
is given by the sum of the partial decay widths of the $Z^0$ into the
massless neutrinos.  From this width, the number of SM neutrino
flavors has been measured to be $3.09 \pm 0.13$ \cite{PDG}.
In the $R$-parity conserving MSSM, the
decay of the $Z^0$ into a pair of the stable lightest neutralinos
(taken as the LSP's) should also be 
included\footnote{Other contributions may also have to be included ---
such as $Z^0$ decays to the second lightest neutralino which in
turn decays to the lightest neutralino and a pair of neutrinos; or
$Z^0$ decays to  sparticles which are close in mass
to the LSP and thus produce decay products too soft to detect when
they in turn decay to the LSP.  These effects will always {\em increase}
${\Gamma}^{\scriptstyle M\!S\!S\!M}_{
\!\scriptstyle Z_{\scriptscriptstyle i\!n\!v}}$.
}, meaning
that
${\Gamma}^{\scriptstyle M\!S\!S\!M}_{
\!\scriptstyle Z_{\scriptscriptstyle i\!n\!v}} \ge
{\Gamma}^{\scriptstyle S\!M}_{
\!\scriptstyle Z_{\scriptscriptstyle i\!n\!v}}$.
When $R$-parity violation is permitted, the situation becomes more
complicated.  
$\Gamma (Z^0 \rightarrow {\nu}_{\scriptscriptstyle 5} 
                         {\nu}_{\scriptscriptstyle 5})$
is suppressed by kinematic factors for the now massive neutrino.
The $Z^0$ coupling may also change due to gaugino and higgsino contributions
to ${\nu}_{\scriptscriptstyle 5}$.  
Further, whether or not the partial decay width to the massive neutrino or 
to any of the more massive neutralinos should be included in 
${\Gamma}^{\scriptstyle R\!P\!V}_{
\!\scriptstyle Z_{\scriptscriptstyle i\!n\!v}}$
will depend on how these particles
decay.\footnote{Light scalar states might also contribute to the
${\Gamma}^{\scriptstyle R\!P\!V}_{
\!\scriptstyle Z_{\scriptscriptstyle i\!n\!v}}$,
either directly or as virtual propagators.  This further complicates
the situation due to the larger number of free parameters in the RPV
scalar sector.  We make the reasonable assumption here that such scalar
states are too heavy to make meaningful contributions.}
As a result,
${\Gamma}^{\scriptstyle R\!P\!V}_{
\!\scriptstyle Z_{\scriptscriptstyle i\!n\!v}}$
need not be larger than
${\Gamma}^{\scriptstyle S\!M}_{
\!\scriptstyle Z_{\scriptscriptstyle i\!n\!v}}$.
An example of this is shown in Fig.$\!$ 8, where
${\Gamma}^{\scriptstyle R\!P\!V}_{
\!\scriptstyle Z_{\scriptscriptstyle i\!n\!v}}$
is plotted for the generic parameter space point
$M_{\scriptscriptstyle 2} = 
{\mu}_{\scriptscriptstyle 0} = 200\, \hbox{GeV}$ and $\tan\!\beta = 2$
assuming contributions from the two massless neutrinos and
${\nu}_{\scriptscriptstyle 5}$ (dotted curve) and from the two massless 
neutrinos, ${\nu}_{\scriptscriptstyle 5}$, and the lightest of the 
neutralinos (dashed curve).  Since the LEP measurement of
${\Gamma}_{\!\scriptstyle Z_{\scriptscriptstyle i\!n\!v}}$ given in 
Table I is slightly below the SM value, the RPV framework can actually 
give better agreement.
The solid horizontal lines in Fig.$\!$ 8 give the $3\sigma$ bounds for
the LEP measurement.  These do not pose a strong constraint on 
$\mu_{\scriptscriptstyle 5}$, allowing values up to
${\sim}100\,\hbox{GeV}$ in this case, as compared to the neutrino mass
constraint which from Fig.$\!$ 7c demands
$\mu_{\scriptscriptstyle 5} 
\raisebox{-.3em}{$\stackrel{\displaystyle <}{\sim}$} 21.5\, \hbox{GeV}$ 
for $m_{\nu_{\scriptscriptstyle 5}} < 149\, \hbox{MeV}$.
A thorough study of the decay modes of the massive neutrino as well as
the other particle states in the model is necessary to make more definitive
predictions.  Such input will also be required if we wish to constrain this 
minimal RPV model using the full $Z^0$ width and the searches for
anomalous $Z^0$ decays,
such as to neutralinos 
($Z^0 \rightarrow \chi^{0}_i\chi^{0}_j,\chi^{0}_j\nu; j \ne 1$).

\section{Charged Current Interactions}
Up to this point, color-singlet fermion interactions with on- and off-shell
$Z^0$ bosons have been analyzed.
Important constraints can also be obtained from charged-current processes
such as the decays of pions, leptons, and heavy neutrinos.
The relevant  $W^{\pm}$ interactions 
may be written (again following notation from \cite{NovPil}) as
\begin{eqnarray}
{\cal L}^{\scriptscriptstyle W\!\chi^-\!\chi^0}_{\scriptscriptstyle int} 
\equiv -\frac{g_{\scriptscriptstyle 2}}{\sqrt{2}} W^{-\mu}
\bar{\chi}^-_a \gamma_\mu \left( P_{\scriptscriptstyle L} 
\tilde{B}^{\scriptscriptstyle L}_{\!ac}
\, + \, P_{\scriptscriptstyle R} \tilde{B}^{\scriptscriptstyle R}_{\!ac}
\right)\chi^0_c\ \, + \, \hbox{h.c.}\; ,
\label{Wlv}
\end{eqnarray}
in four-component mass eigenbasis notation. The 
$\tilde{B}^{\scriptscriptstyle L,R}$ matrices giving the effective coupling
strength among the mass eigenstates can be obtained from the
diagonalizing matrices of the charged and neutral fermions:
\begin{eqnarray} 
\tilde{B}^{\scriptscriptstyle L} = U^{\dag}_{\scriptscriptstyle L}
 T^L U_0\; \qquad \hbox{and} \qquad\;
\tilde{B}^{\scriptscriptstyle R} =  U^{\dag}_{\scriptscriptstyle R}
T^R U_0\; ,
\label{Bac}
\end{eqnarray}
where
\begin{eqnarray}
T^L=\left(\begin{array}{cccc}
0 &  \sqrt{2} & 0 & 0 \\
0 & 0 & 0 & I_{\scriptscriptstyle 4\times 4}  
\end{array}\right)\; \qquad \hbox{and} \qquad\; 
T^R =
\left(\begin{array}{cccc}
0 &  -\sqrt{2} & 0 & 0  \\
0 & 0 & 1 & 0_{\scriptscriptstyle 4\times 4}
\end{array}\right)\; .
\end{eqnarray}
Of particular interest are pion and lepton decays.
      Here discussion is limited to tree-level decays mediated by a
      virtual $W$ boson.  ($Z^0$ exchange in lepton decays is negligible
for the cases we consider below, the corresponding amplitude being
proportional to the product of two $\mu_i$'s.)
      An exhaustive discussion  would in principle also
      require consideration of possible virtual scalar intermediate
      states.  These (together with radiative corrections) would
      re-introduce trilinear RPV terms; however, as already noted,
      since supersymmetric scalar particles must be considerably
      more massive than the gauge bosons, their contributions can
      be expected to be very small.
Partial decay widths
for these processes are then given by
\begin{eqnarray}
&\Gamma&(\pi \to \ell \bar\nu_\ell) = 
\frac{G^2 f_\pi^2 |V^{ud}_{\scriptscriptstyle \!C\!K\!M}|^2 m_\pi^3}
{8 \pi} {\cal R}_{\pi \ell} 
\sum_{c=1}^3 \tilde{B}_{\ell \nu_c}^2 P_c^{\pi \ell}\;
\label{pidec}\\
\hbox{and} \;\;\;\;\;\;\;\;\;\;
&\Gamma&(\ell^{'} \to \ell \bar \nu_{\ell} \nu_{\ell'})  = 
\frac{G^2 m_{\ell'}^5}{192 \pi^3} {\cal R}_{\ell'}
\sum_{c,d=1}^3 \tilde{B}_{\ell'\nu_c}^2 \tilde{B}_{\ell\nu_d}^2
P_{cd}^{\ell' \ell}\; ,
\label{lnunu}
\end{eqnarray}
\begin{eqnarray} 
\hbox{where}\;\;\;\;\;\;\;\;
\tilde{B}^2_{\!ac} \equiv |\tilde{B}^{\scriptscriptstyle L}_{\!ac}|^2
+  |\tilde{B}^{\scriptscriptstyle R}_{\!ac}|^2 \; .
\label{B2}
\end{eqnarray}
Standard parameters are the Fermi constant
$G$ (see \cite{BFKM} for minor subtleties),
the pion decay constant $f_{\pi}$, and
$V^{ud}_{\scriptscriptstyle \!C\!K\!M} =$ 
the $ud$-component of the CKM quark-sector mixing matrix.  
${\cal R}_{\pi \ell}$ and ${\cal R}_{\ell'}$ are
leading radiative corrections to the processes.  The former depends on
the pion and charged lepton masses; the latter depends only on the mass 
of the decaying lepton (See \cite{PDG,MS} for more details.).
Finally, the functions $P_c^{\pi \ell}$ and $P_{cd}^{\ell'\ell}$
include the entire phase space factors for the decays as well as parts
of the matrix elements --- including all dependence on neutrino masses.
Explicit formul\ae\ from these functions are given in \cite{BFKM}; for 
use below, $P_c^{\pi \ell}$ is:
\begin{eqnarray}
P_c^{\pi \ell} & = &
\theta(m_\pi-m_\ell-m_{\nu_{\scriptscriptstyle c}})\,
[\delta_{\ell\pi}^2 + \delta_{{\nu_{\scriptscriptstyle c}}\pi}^2 -
(\delta_{\ell\pi}^2 - \delta_{{\nu_{\scriptscriptstyle c}}\pi}^2)^2]
\lambda^{1/2}
(1,\delta_{\ell\pi}^2,\delta_{{\nu_{\scriptscriptstyle c}}\pi}^2)\; ;
\\
& & \qquad\;\;\;\; \delta_{\ell\pi} = \frac{m_\ell}{m_\pi} \; ,
\delta_{{\nu_{\scriptscriptstyle c}}\pi} = 
\frac{m_{\nu_{\scriptscriptstyle c}}}{m_\pi} \; .
\nonumber 
\end{eqnarray}
As noted previously, Ref.$\!$ \cite{BFKM} analyzed three-neutrino mixing.
The present case differs since mixing of neutrinos with neutralinos
(and charged leptons with charginos) is also possible.  However, the
above formul\ae\ remain valid since the new states would be too
heavy to  contribute to these GeV or sub-GeV decays.

The experimental inputs are the following decay rates:
\[ \begin{array}{c}
\Gamma(\pi \rightarrow e \bar\nu_e)\; ,\quad
\Gamma(\pi  \rightarrow \mu \bar\nu_\mu)\; ,\\
\Gamma(\mu  \rightarrow e \bar\nu_e \nu_\mu\; )\; ,
\quad
\Gamma(\tau \rightarrow e \bar\nu_e \nu_\tau) \; ,\quad
\Gamma(\tau  \rightarrow \mu \bar\nu_\mu \nu_\tau)\; .
\end{array} \]
Here, by $\nu_e, \nu_\mu, \nu_\tau $ we mean the states produced 
alongside the $ e, \mu, \tau $ leptons in the corresponding decays.
To eliminate the uncertainty in some of the common factors and thus
better isolate the effects of non-SM leptonic masses and mixings,
it is preferable to work with ratios \cite{BFKM,NovPil}.
For pion decays, we will use
\begin{eqnarray}
R^{\pi e}_{\pi\mu} \; \equiv \;
\frac{\Gamma(\pi \rightarrow e \bar\nu_e)}
{\Gamma(\pi  \rightarrow \mu \bar\nu_\mu)} =
\frac{{\cal R}_{\pi e} \sum_i  \tilde{B}_{e\nu_i}^2 P_i^{\pi e}}
{{\cal R}_{\pi \mu} \sum_i  \tilde{B}_{\mu\nu_i}^2 P_i^{\pi \mu}}\; .
\label{pirate}
\end{eqnarray}
For the lepton decays, reduced decay widths are defined by
\begin{eqnarray}
\overline{\Gamma}^{\ell'\ell} \; \equiv \; 
\frac{192 \pi^3}{G^2 m_{\ell'}^5}
\Gamma(\ell^{'} \to \ell \bar \nu_{\ell} \nu_{\ell'})
\, = \,
{\cal R}_{\ell'} \sum_{c,d=1}^3  \tilde{B}_{\ell'\nu_c}^2
 \tilde{B}_{\ell\nu_d}^2 P_{cd}^{\ell'\ell}\; ,
\end{eqnarray}
which conveniently allows us to define the ratios:
\begin{eqnarray}
R^{\mu e}_{\tau e} \; \equiv \;
\frac{\overline\Gamma^{\mu e}}{\overline\Gamma^{\tau e}}=
\frac{{\cal R}_{\mu} \sum_{c,d}  \tilde{B}_{\mu\nu_c}^2 
 \tilde{B}_{e\nu_d}^2 P_{cd}^{\mu e}}
{{\cal R}_{\tau} \sum_{c,d}  \tilde{B}_{\tau\nu_c}^2
  \tilde{B}_{e\nu_d}^2 P_{cd}^{\tau e}} \;\;\; \hbox{and} \;\;\;
R^{\tau e}_{\tau \mu} \; \equiv \;
\frac{\overline\Gamma^{\tau e}}{\overline\Gamma^{\tau \mu}}=
\frac{ \sum_{c,d}  \tilde{B}_{\tau\nu_c}^2  \tilde{B}_{e\nu_d}^2 
P_{cd}^{\tau e}}{ \sum_{c,d}  \tilde{B}_{\tau\nu_c}^2 
 \tilde{B}_{\mu\nu_d}^2 P_{cd}^{\tau \mu}}\; . \label{Grll}
\end{eqnarray}

 Barring some miraculous alignment,
the physical neutrino mass eigenstates will not be the partners of
the charged leptons --- the $\nu_{\ell}$'s above.  Nor will they be
basis states in the single-VEV parametrization --- the ${\nu}_i$'s
of Eqn.$\!$ (\ref{nLag}).  The physical eigenstate which acquires a mass
at tree-level is denoted by ${\nu}_{\scriptscriptstyle 5}$; we will 
denote the other two 
massless degenerate states  by ${\nu}_{\scriptscriptstyle z1}$ and
${\nu}_{\scriptscriptstyle z2}$.  Note that these two
light eigenstates are not uniquely defined at tree-level;  any linear 
combination between them is another massless neutrino eigenstate.
This indeterminacy will not affect physical results;
moreover, the degeneracy will be lifted by radiative corrections.

In the rest of this section,  we will discuss implications of experimental
constraints on  the ratios (\ref{pirate}) and (\ref{Grll}). Exact numerical 
values for rotation matrices $U_L, U_R$, and $U_0$ will be used to 
compute the charged current couplings $\tilde{B}^L$ and $\tilde{B}^R$. 
However, in order to be able to understand the
qualitative features of our numerical results,
we find it worthwhile to 
derive and use approximate analytical expressions for these couplings.
To this end, we will use the following approximation.  We split
the diagonalizing rotation matrices into separate
``quasi-MSSM'' and ``quasi-SM'' blocks:
\[
U_{\scriptscriptstyle L} = 
 \hbox{diag}\{R_{\scriptscriptstyle L}, 
  I_{\scriptscriptstyle 3\times 3}\}\; , 
\qquad  U_{\scriptscriptstyle R} = 
 \hbox{diag}\{R_{\scriptscriptstyle R}, 
  I_{\scriptscriptstyle 3\times 3}\}\; , 
\qquad \hbox{and} \qquad 
U_0 = \hbox{diag}\{U_4, R_5\}\; , 
\]
where $U_4$ diagonalizes the first  $4\times 4$ block of 
${\cal M}_{\scriptscriptstyle {\cal N}}$ and $R_5$ is again the
$3\times 3$ CKM-like neutrino mixing matrix.
Note that, while this approximation is valid in the small $\mu_i$ limit,
it is not the same as the approximation used in section II. 
In that case, all first order terms in the $\mu_i$ were kept;
here, we keep only those terms
of the form $\mu_i/ \mu_5$.  This will
be the meaning of the ``small $\mu$ approximation" in this section.

The SM components of the ${\nu}_{\scriptscriptstyle 5}$ eigenstate are 
then given by
\begin{eqnarray}
R_5^{\scriptscriptstyle i3} \; = \; 
\frac{ {\mu}_i }{ {\mu}_{\scriptscriptstyle 5} } \; ,
\label{neuevc1}
\end{eqnarray}
and, by unitarity, the quasi-SM components of the other two eigenstates
obey the sum rule
\begin{eqnarray}
|R_5^{i1}|^2 + |R_5^{i2}|^2 = 1 - \left(
\frac{\mu_i}{\mu_{\scriptscriptstyle 5}}\right)^{\!\!2} \; .
\label{neuevc2}
\end{eqnarray}
Using Eqns.$\!$ (\ref{neuevc1}) and (\ref{neuevc2}) together
with the assumption of no right-handed neutrino fields,
the quasi-SM sector charged current couplings become
\begin{eqnarray}
\tilde{B}^{\scriptscriptstyle R}_{\ell_m\nu_n} = 0 \; , \;\;\;\;
\tilde{B}^{\scriptscriptstyle L}_{\ell_m\nu_n} = R_5^{mn} \;\;\;\;\;
\Longrightarrow \;\;\;
\tilde{B}_{\ell_m\nu_ n}^2 = |R_5^{mn}|^2 \; .
\label{Bln}
\end{eqnarray}

\subsection{Pion decays}

Inserting the small-${\mu}_i$ approximate expressions
from Eqns.$\!$ (\ref{neuevc1}), (\ref{neuevc2}) and
(\ref{Bln}) into
Eqn.$\!$ (\ref{pirate}), and disregarding $\tilde{B}_{\ell_m\chi_n}^2$
contributions involving the more massive quasi-MSSM neutral states
(as justified earlier), yields
(using $P_{\!\scriptscriptstyle z1}^{\pi \ell} = 
P_{\!\scriptscriptstyle z2}^{\pi \ell}$ for
the massless states):
\begin{eqnarray}
R^{\pi e}_{\pi\mu} &=&
\frac{{\cal R}_{\pi e}}{{\cal R}_{\pi \mu}}\; \frac{
\left[
1 - \left(\frac{\mu_{\scriptscriptstyle 1}}
{\mu_{\scriptscriptstyle 5}}\right)^{\!\!2}
\right]
P_{\!\scriptscriptstyle z1}^{\pi e}  +
\left( \frac{\mu_{\scriptscriptstyle 1}}
{\mu_{\scriptscriptstyle 5}}\right)^{\!\!2}\, 
P_{\!\scriptscriptstyle 5}^{\pi e} }
{ \left[
1 -\left( \frac{\mu_{\scriptscriptstyle 2}}
{\mu_{\scriptscriptstyle 5}}\right)^{\!\!2}
\right]
P_{\!\scriptscriptstyle z1}^{\pi \mu} + \left(
\frac{\mu_{\scriptscriptstyle 2}}
{\mu_{\scriptscriptstyle 5}}\right)^{\!\!2}\, 
P_{\!\scriptscriptstyle 5}^{\pi \mu} }
\; = \;
{\cal K}_{SM} \frac{1+ \left( \frac{\mu_{\scriptscriptstyle 1}}
{\mu_{\scriptscriptstyle 5}}\right)^{\!\!2} {\cal P}^{\pi e}}
{1+ \left( \frac{\mu_{\scriptscriptstyle 2}}
{\mu_{\scriptscriptstyle 5}}\right)^{\!\!2} {\cal P}^{\pi \mu}}
\; ,
\label{Rpiell}
\end{eqnarray}
where
\begin{eqnarray}
{\cal K}_{SM} = \frac{{\cal R}_{\pi e}}{{\cal R}_{\pi \mu}}
\frac{ P_{\!\scriptscriptstyle z1}^{\pi e} }
{ P_{\!\scriptscriptstyle z1}^{\pi \mu} }
\; = \;
\frac{{\cal R}_{\pi e}}{{\cal R}_{\pi \mu}}
\left( \frac{m_e}{m_{\mu}} \right)^2
\left[ \frac{m_{\pi}^2 - m_e^2}{m_{\pi}^2 - m_{\mu}^2} \right]^2
\; = \;
1.233 \times 10^{-4}
\end{eqnarray}
is the SM prediction
(with $\frac{{\cal R}_{\pi e}}{{\cal R}_{\pi \mu}} = 0.96103$),
and 
$
{\cal P}^{\pi \ell} = P_{\!\scriptscriptstyle 5}^{\pi \ell} /
P_{\!\scriptscriptstyle z1}^{\pi \ell} -1\; 
$.

The behavior of the $R^{\pi e}_{\pi\mu}$ ratio as a function of $\mu_5$
can be understood by analyzing
the dependence of the kinematic functions 
$P_{\!\scriptscriptstyle 5}^{\pi \ell}$
on the neutrino mass. Indeed, for a given set of $\mu_i$ ratios,
the only dependence on $\mu_5$ in $R^{\pi e}_{\pi\mu}$ comes
via $m_{\nu_5}$ in $P_{\!\scriptscriptstyle 5}^{\pi \ell}$.
Small changes in $m_{\nu_5}$ affect $P_{\!\scriptscriptstyle 5}^{\pi e}$ and 
$P_{\!\scriptscriptstyle 5}^{\pi \mu}$ differently.
$P_{\!\scriptscriptstyle 5}^{\pi e}$ increases quite rapidly with increasing 
$m_{\nu_5}$,
 as a consequence of matrix
element dependence on the masses of decay products in pion decay
[for $m_{\mu_5}< 80$ MeV, 
$P_{\!\scriptscriptstyle 5}^{\pi e} = (m_{\nu_5}/m_\pi)^2$
is a very good approximation].
In contrast,
$P_{\!\scriptscriptstyle 5}^{\pi \mu}$  decreases, albeit more slowly,
because of the phase space factor.
As a result, their ratio $R^{\pi e}_{\pi\mu}$   increases
with $m_{\nu_5}$, and this effect is further enhanced
the larger $\mu_1 / \mu_5$ is with respect to 
$\mu_2 / \mu_5$.

Looking in more detail at various mass ranges, we find that 
for  $m_{\nu_5} 
\raisebox{-.3em}{$\stackrel{\displaystyle <}{\sim}$} 20$ MeV, 
the kinematic function 
$P_{\!\scriptscriptstyle 5}^{\pi \mu}$ is nearly
the same as the constant
$P_{\!\scriptscriptstyle z1}^{\pi \mu}$ (${\cal P}^{\pi \mu} \simeq 0$). 
In this case, the experimental constraints on 
the ratio $R^{\pi e}_{\pi\mu}$ can be put in a simple form in 
terms of the RPV parameters. Using 
the $3\sigma$ bounds from Table I, we obtain:
\begin{equation}
\label{ineq1}
\hbox{~~~~ for ~~} m_{\nu_5} 
\raisebox{-.3em}{$\stackrel{\displaystyle <}{\sim}$} 20\ \hbox{MeV ~: ~~~~~}
\frac{\mu_1}{\mu_5} < \sqrt{\frac{1}{137}}\ \frac{m_e}{m_{\nu_5}} \;\; .
\end{equation}
Beyond the threshold value of $m_{\nu_5} = m_{\pi} - m_{\mu} = 33.91$ MeV, 
the $\pi$ decay into a muon and the massive neutrino can no longer
proceed; $P_{\!\scriptscriptstyle 5}^{\pi \mu}=0 $ and the denominator of 
(\ref{Rpiell}) becomes constant.
Then we have the following constraint on the $\mu_2/\mu_5$ ratio:
  \begin{equation}
   \label{ineq2}
   \hbox{~~~~ for ~~} 34\ \hbox{MeV } <  m_{\nu_5} < 139\ \hbox{MeV ~: ~~~~~}
   \frac{\mu_2}{\mu_5} < \sqrt{\frac{1}{137}}\
   \hbox{~~~~(} \mu_1=0 \hbox{).}
   \end{equation}
Upper limits on the $\mu_1/\mu_5$ ratio in this region are given in Fig.~9.
 
For values of $m_{\nu_5}$ of order 
100 MeV, the increase in $P_{\!\scriptscriptstyle 5}^{\pi e}$ due to the 
amplitude contribution are offset
by the decrease due to the phase space factor, and $R^{\pi e}_{\pi\mu}$ 
begins to  decrease. At the threshold
$m_{\nu_5} = m_{\pi} - m_{e} = 139.057$ MeV, the $\pi$ cannot
decay into $ e \nu_5 $ either. 
Above this threshold,
the $R^{\pi e}_{\pi\mu}$ ratio is constant (in the 
approximation used here), and the following constraints can be derived:    
\begin{equation}
\label{ineq3}
m_{\nu_5}>139\ \hbox{MeV ~: ~}
\left\{
\begin{array}{ccc}
\hbox{for~~} \mu_2 > \mu_1 \ : &
\frac{\mu_2^2}{\mu_5^2} - \frac{\mu_1^2}{\mu_5^2} < \frac{1}{137} &
\hbox{~ (upper experimental bound),} \\
\hbox{for~~} \mu_1 > \mu_2 \ : &
\frac{\mu_1^2}{\mu_5^2} - \frac{\mu_2^2}{\mu_5^2} < \frac{1}{\ 82\ } &
\hbox{~ (lower experimental bound).}
\end{array}
\right.
\end{equation}
It must be mentioned  that large neutrino masses of
${\cal O}(100\, \hbox{MeV})$ may be beyond
the range of validity of the small-${\mu}_i$ approximation unless
$\tan\!\beta$ is ``small''. For $\tan\!\beta = 2$ and for the
MSSM parameters $M_{\scriptscriptstyle 2} = 250\, \hbox{GeV}$ and
${\mu}_{\scriptscriptstyle 0} = 100\, \hbox{GeV}$, 
Eqn.$\!$ (\ref{numass}) gives
$m_{\nu_{\scriptscriptstyle 5}} = 149\, \hbox{MeV}$ when
${\mu}_{\scriptscriptstyle 5} = 13.9\, \hbox{GeV}$, in rough agreement 
with the see-saw prediction of
${\mu}_{\scriptscriptstyle 5} = 11.4\, \hbox{GeV}$ 
from Eqn.$\!$ (\ref{seesaw}),
and thus arguably still within the small ${\mu}_{\scriptscriptstyle 5}$ 
domain.
In addition, radiative corrections to this tree-level calculation may 
alter the numerical values given above.

Using exact numerical calculations, Fig.$\!$ 10  shows the dependence
of $R^{\pi e}_{\pi\mu}$ on $\mu_{\scriptscriptstyle 5}$ for some illustrative
ratios among the $\mu_i$'s and for the generic MSSM parameter point
$M_{\scriptscriptstyle 2} = {\mu}_{\scriptscriptstyle 0} = 200\, \hbox{GeV}$ 
and $\tan\!\beta =2,45$.
The solid horizonal line denotes a $3\sigma$ positive deviation from the
experimentally determined central value. The qualitative attributes of 
the curves agree well with the predictions made above on the basis of the 
analytic expression (\ref{Rpiell}). Note that for large tan$\beta$,
 high values of $m_{\nu_{\scriptscriptstyle 5}}$ are
unattainable no matter how large ${\mu}_{\scriptscriptstyle 5}$ is. In this 
case, values of ${\mu}_{\scriptscriptstyle 5}$ in the hundreds of
GeV are allowed.

\subsection{Decays of charged leptons}
Eqn.$\!$ (\ref{lnunu}) shows that $W^*$-mediated charged lepton decays
are proportional to two $\tilde{B}^2_{\ell_a\nu_c}$ factors.  This
along with the more complicated $P_{cd}^{\ell'\ell}$ function (which
gives the dependence on the neutrino mass) makes analytic expressions
unwieldy, even in the small-${\mu}_i$ approximation.  Hence only
results from the exact numerical calculations will be discussed.

$R^{\mu e}_{\tau e}$ of Eqns.$\!$ (\ref{Grll}) provides the most
interesting constraint, tightening the bound on 
$\frac{ {\mu}_{\scriptscriptstyle 2} }{ {\mu}_{\scriptscriptstyle 5} }$ 
in some regions where $R^{\pi e}_{\pi\mu}$ is less effective.
A similar conclusion was reached in \cite{BFKM}.
For $\tan\!\beta =2$ and ${\mu}_{\scriptscriptstyle 1}$ set to zero, exact
numerical results at the generic MSSM point
$M_{\scriptscriptstyle 2} = {\mu}_{\scriptscriptstyle 0} = 200\, \hbox{GeV}$
show $R^{\mu e}_{\tau e}$ to be more restrictive than
$R^{\pi e}_{\pi\mu}$ for
$\frac{ {\mu}_{\scriptscriptstyle 2} }
{ {\mu}_{\scriptscriptstyle 3} } \le \frac{1}{12}$ and 
$\frac{1}{4} \le \frac{ {\mu}_{\scriptscriptstyle 2} }
{ {\mu}_{\scriptscriptstyle 3} } \le 1$
(echoing Fig.$\!$ 6 of \cite{BFKM}).
Fig.$\!$ 11a depicts the actual behavior of $R^{\mu e}_{\tau e}$
{\it vs}. ${\mu}_{\scriptscriptstyle 5}$ at this 
$M_{\scriptscriptstyle 2}$-${\mu}_{\scriptscriptstyle 0}$ point
for $\tan\!\beta =2$ and assorted ${\mu}_i$ ratios.
The solid horizontal lines represent $3\sigma$ deviations above and
below the experimentally-determined central value (which is
consistent with the SM prediction of $1$).  For 
$ {\mu}_{\scriptscriptstyle 2} / {\mu}_{\scriptscriptstyle 3}$ closer to one
($\frac{1}{5} 
\raisebox{-.3em}{$\stackrel{\displaystyle <}{\sim}$} 
\frac{ {\mu}_{\scriptscriptstyle 2} }{ {\mu}_{\scriptscriptstyle 3} } < 1$)
and smaller values of ${\mu}_{\scriptscriptstyle 5}$
(${\mu}_{\scriptscriptstyle 5} 
\raisebox{-.3em}{$\stackrel{\displaystyle <}{\sim}$} 10\, \hbox{GeV}$), 
$R^{\mu e}_{\tau e}$ may dip below the experimentally-allowed band;
however, for any
$ {\mu}_{\scriptscriptstyle 2} / {\mu}_{\scriptscriptstyle 3} < 1$,
$R^{\mu e}_{\tau e}$ eventually goes above the acceptable band as
${\mu}_{\scriptscriptstyle 5}$ is increased.
The upper bound on ${\mu}_{\scriptscriptstyle 5}$ only runs from 
$19.5\, \hbox{GeV}$ down to $17.5\, \hbox{GeV}$ as
$\frac{ {\mu}_{\scriptscriptstyle 2} }{ {\mu}_{\scriptscriptstyle 3} }$ 
is changed from $\frac{1}{12}$ to $0$ (values below ${\sim}\!\frac{1}{50}$ 
are indistinguishable from $0$).  However, the neutrino decay constraints
from the WA66 and CHARM experiments described in the next subsection
may surpass the
constraint from $R^{\mu e}_{\tau e}$ for these somewhat larger
${\mu}_{\scriptscriptstyle 5}$ values.

For high $\tan\!\beta$, the behavior of $R^{\mu e}_{\tau e}$
{\it vs}. ${\mu}_{\scriptscriptstyle 5}$ is decidedly different, as seen in
Fig.$\!$ 11b for $\tan\!\beta = 45$.  Here the value of
$R^{\mu e}_{\tau e}$ always drops below the experimentally-allowed
region as ${\mu}_{\scriptscriptstyle 5}$ is increased irrespective of the
$ {\mu}_{\scriptscriptstyle 2} : {\mu}_{\scriptscriptstyle 3} $ ratio.  
The behavior is qualitatively reminiscent of that for the
$\frac{ {\mu}_{\scriptscriptstyle 2} }{ {\mu}_{\scriptscriptstyle 3} } = 1$ 
curve for $\tan\!\beta = 2$ even if now
$\frac{ {\mu}_{\scriptscriptstyle 2} }{ {\mu}_{\scriptscriptstyle 3} }$ is 
set to zero. Quantitatively though, the upper limit placed on 
${\mu}_{\scriptscriptstyle 5}$ becomes quite large as
$\frac{ {\mu}_{\scriptscriptstyle 2} }{ {\mu}_{\scriptscriptstyle 3} }$ 
drops to zero.
This bound is nevertheless still more stringent than that from
$R^{\pi e}_{\pi\mu}$ if ${\mu}_{\scriptscriptstyle 1} \simeq 0$, again 
irrespective of
$\frac{ {\mu}_{\scriptscriptstyle 2} }{ {\mu}_{\scriptscriptstyle 3} }$.

The experimentally-derived values for 
$\overline{\Gamma}^{\mu e}$, $\overline{\Gamma}^{\tau e}$, and
$\overline{\Gamma}^{\tau \mu}$ can also be applied individually
without taking ratios.  The $3\sigma$ bounds on these quantities are
also given in Table I.  However, these restrictions were always
found to be weaker than the constraints from the ratios
(in contrast to  Ref.$\!$ \cite{BFKM}).

\subsection{Decays of a massive neutrino}
Assuming the decay of the massive neutrino is mediated by
a virtual $W$-boson, the expected decay modes are 
${\nu}_5^{\!\!\!\!\!\! \scriptscriptstyle (-)} 
\rightarrow W^{* \pm} {\ell}^{\mp}$;
$W^{* \pm} \rightarrow {\ell}^{\prime \pm}
{\nu}_{ {\ell}^{\prime} }^{\!\!\!\!\!\! \scriptscriptstyle (-)},
(\bar{q}q^{\prime} \Rightarrow {\pi}^{\pm})$.

A crucial experiment restricting this process was performed by the
CERN WA66 Collaboration \cite{WA66} using the BEBC bubble chamber
placed in a neutrino beam resulting from dumping protons on a high
density target.   The  neutrino beam is  mainly
composed of `prompt' neutrinos from charmed meson decays
which can include massive neutrinos up to ${\sim}1.8\, \hbox{GeV}$.
WA66 is sensitive to massive neutrino decays into electrons, muons,
and pions.  From the absence
of any excess of such events, limits can be placed on mixing
of the massive neutrino state with either the ${\nu}_e$ or ${\nu}_{\mu}$
weak-flavor eigenstate.  
In the small-${\mu}_i$ approximation, $\nu_\ell \equiv \nu_i $ and
this mixing is simply
$|R_5^{mn}|^2 = \left( \frac{ {\mu}_i }
{ {\mu}_{\scriptscriptstyle 5} } \right)^2$.
Therefore, the WA66 results can be used to restrict these ratios
{\em if} the small-${\mu}_i$ approximation is valid; for
    $m_{{\nu}_{\scriptscriptstyle 5}} \sim {\cal O}(100\, \hbox{MeV})$,
 this would demand that $\tan\!\beta$ is small. 
 
The WA66 results can be summarized as follows
(results from CHARM\cite{CHARM} are similar in the parameter regions
of interest here). For neutrino mass values of order 100 MeV and above, 
the mixing parameter $\mu_1/\mu_5$ has to be smaller
than $10^{-3}$ ($ <10^{-4}$, for $m_{\nu_5} \sim 149$ MeV).
The constraint on $\mu_2/\mu_5$ is a little bit weaker:
at $m_{\nu_5} = 149$ MeV, $\mu_2/\mu_5 < 10^{-3}$. These limits 
as a function of $m_{\nu_5}$ are presented
in Fig.~9. 
Note that for $m_{\nu_5}$ less than $\cal{O}$(80-100)  MeV, the constraint
on $\mu_1/\mu_5$ coming from $\pi$ decay is stronger than that
from WA66, while for $m_{\nu_5} >  \cal{O}$(80-100)  MeV, the WA66 
constraint is stronger.
Thus, these two inputs play complementary roles
in setting limits on neutrino flavor-state mixings. 

The WA66 neutrino decay limits are also used in establishing for the
present RPV scenario the
$149\, \hbox{MeV}$ absolute upper mass bound on a neutrino
which is not a pure ${\nu}_{\tau}$.\footnote{This result from
Ref.$\!$ \cite{BFKM} is based on a charged current analysis
of $R^{\pi e}_{\pi \mu}$, $R^{\tau e}_{\tau \mu}$, $R^{\mu e}_{\tau e}$,
$\overline{\Gamma}^{\mu e}$, $\overline{\Gamma}^{\tau e}$, and
$\overline{\Gamma}^{\tau \mu}$.  As mentioned earlier, Ref.~\cite{BFKM} only
includes 3-flavor mixing, not possible mixing with gauginos and higgsinos.
It should also be noted that this value is given in \cite{BFKM} as
a $1\sigma$ bound whereas more conservative $3\sigma$ bounds are used
throughout the present work.}
As seen above, a neutrino with a mass of $149\, \hbox{MeV}$ is about 
99.9\% ${\nu}_{\scriptscriptstyle 3}$ which is now extremely well aligned
with the $\tau$ lepton.  In this case the tighter LEP bound of
$m_{\nu_{\tau}} < 18.2\, \hbox{MeV}$ is applicable; therefore,
the upper mass limit on a mixed
massive neutrino state is ${\sim}149\, \hbox{MeV}$ as adopted here.

There are other experiments on neutrino 
interactions which constrain the mixing among the three generations
(for example, the CHARM II experiment \cite{CHARMII} on $\nu_\mu - e$
scattering, and LAMPF \cite{LAMPF} on $\nu_e - e$ scattering).
 Some of the resulting constraints 
are stronger than the WA66 results. However, the interpretation of these 
data in our current framework
(constraints on the RPV parameters $\mu_i$) is straightforward only in 
the approximation used in this section; otherwise, mixing between the
neutrino and neutralino contributions should also be taken 
into account. For this reason, we have discussed only the WA66 results 
(for purposes of illustration), leaving a more detailed analysis to a 
future paper.

\section{Neutrinoless double-beta decay}

Neutrinoless double beta decay
($0\nu\beta\beta$) places very stringent bounds on trilinear RPV parameters,
as demonstrated in Ref.$\!$ \cite{HKK1}.
The bilinear ${\mu}_i$ RPV couplings can also mediate
$0\nu\beta\beta$; therefore, this process warrants serious
consideration here.  First, however, features unique to this process must be
carefully noted.  As with pion decay, this process involves
(valence) quarks rather than leptons in the initial state.  However,
unlike pion decay, for $0\nu\beta\beta$ the decay products are not
all colorless.  
Thus transitions can be mediated by $W$-bosons and
a Majorana neutrino with mixing to $\nu_e$
{\em or} by scalars (squarks and/or sleptons) and
sfermions (gluinos or neutralinos) \cite{MohaCo,HKK1}.  Tree-level
diagrams involving the former pair can derive the neutrino couplings
from the RPV bilinear ${\mu}_i$'s, while those
involving the latter pair will be proportional to
trilinear RPV couplings.  For leptonic decays, possible tree-level
trilinear RPV $\lambda$ coupling dependence due to scalar
intermediates was neglected on the grounds that this would be
kinematically suppressed since sleptons  and
  Higgs bosons are more massive than
the gauge bosons (and the trilinear couplings are expected to also
be small).  In contrast, with $0\nu\beta\beta$  there are
strongly-interacting sparticle intermediates which could perhaps
off-set the simple kinematic suppression; thus, trilinear RPV
couplings might play a significant role at tree-level for this process.
It should be stressed that this is not the case for the previous
sections --- there the trilinear RPV couplings have {\em not} been
set to zero by hand, they simply do not yield any significant
contributions to those processes at tree-level within the
single-VEV parametrization augmented by the reasonable assumption
  of sufficiently heavy scalars.
For $0\nu\beta\beta$, in order to concentrate on the bilinear
RPV ${\mu}_i$'s,
the ${\lambda}^{\prime}$ couplings can be {\em assumed} to be
negligible and/or the squark and gluino masses can be {\em assumed}
to be very large to kill these contributions.  Alternatively, to
set conservative bounds on the ${\mu}_i$'s it is sufficient to
{\em assume} that there is no destructive interference
between the two types of diagrams.  A more thorough analysis including
the scalar intermediates is underway.

With these caveats, the effective constraint from $0\nu\beta\beta$
becomes \cite{BBnu0Con}
\begin{eqnarray}
m_{\nu_{\scriptscriptstyle 5}}
|\tilde{B}^{\scriptscriptstyle L}_{e\nu_{\scriptscriptstyle 5}}|^2 < 
0.46 \, \hbox{eV}
\;\;\;\;\;\;\; \hbox{for}\;
m_{\nu_{\scriptscriptstyle 5}} < 10\, \hbox{MeV} \; .
\end{eqnarray}
Applying the small-${\mu}_i$ approximation, this translates into
     \begin{eqnarray}
     \frac{\mu_{\scriptscriptstyle 1}}{\mu_{\scriptscriptstyle 5}} 
    < \sqrt{\frac{ 0.46 }{m_{\nu_{\scriptscriptstyle 5}}}} \times 10^{-3} 
     \;\; \hbox{(with~} m_{\nu_{\scriptscriptstyle 5}} \hbox{~in MeV).}
     \label{znbb-1}
     \end{eqnarray} 
 An alternative approximation is to consider only the leading mixing
effect between the sole vev-bearing $Y = \frac{1}{2}$ superfield
basis state and the other $Y = \frac{1}{2}$ superfield basis states.
This yields the approximate expression [cf. Eqn.(\ref{Bln})]
\begin{eqnarray}
 \begin{array}{lc}
& \tilde{B}^{\scriptscriptstyle L}_{\ell_m\nu_{\scriptscriptstyle 5}}
 =  \frac{\sqrt{\mu_{\scriptscriptstyle 5}^2+\mu_{\scriptscriptstyle 0}^2}}
{\mu_{\scriptscriptstyle 0}} \frac{\mu_i}{\mu_{\scriptscriptstyle 5}} \, ,
\hbox{\phantom{aaaaaaaaaaaaaaaaaaaaa}}
 \\
\hbox{and thus} \;\;\;\;\;\;\;\; &
|\tilde{B}^{\scriptscriptstyle L}_{\ell_m\nu_{\scriptscriptstyle z1}}|^2
+ |\tilde{B}^{\scriptscriptstyle L}_{\ell_m\nu_{\scriptscriptstyle z2}}|^2
\simeq 1 -  \frac{ {\mu_i}^2
(\mu_{\scriptscriptstyle 5}^2 + \mu_{\scriptscriptstyle 0}^2)}
{\mu_{\scriptscriptstyle 5}^2 \mu_{\scriptscriptstyle 0}^2} \; .
\end{array}
\label{BLlgmu}
\end{eqnarray}
This approximation should hold even for large values
of ${\mu}_{\scriptscriptstyle 5}$. 
Using this  approximation
 along with Eqn.$\!$ (\ref{numass}) for
$m_{\nu_{\scriptscriptstyle 5}}$, the $0\nu\beta\beta$ constraint is
\begin{eqnarray}
\frac{\mu_{\scriptscriptstyle 1}}{\mu_{\scriptscriptstyle 0}} 
< \frac{4.29\times 10^{-5}} { {v} \cos\!\beta } 
 \sqrt{\frac{xM_{\scriptscriptstyle 2}}
{\left(x{g}_{\scriptscriptstyle 2}^{2} + 
{g}_{\scriptscriptstyle 1}^{2} \right)}}
 \simeq 2.15\times 10^{-7}
 \sqrt{(1+\tan^2\!\!\beta) M_{\scriptscriptstyle 2}}\;
\label{znbb-2}
\end{eqnarray}
(with mass parameters in GeV).
For the low neutrino mass region in which it is effective, the
$0\nu\beta\beta$ constraint on ${\mu}_{\scriptscriptstyle 1}$
(or $\frac{ {\mu}_{\scriptscriptstyle 1} }{ {\mu}_{\scriptscriptstyle 5} }$)
is much stronger than that from pion decay, as is seen in Fig.~9.  
At the generic
MSSM parameter point 
($M_{\scriptscriptstyle 2} = {\mu}_{\scriptscriptstyle 0} = 200\, \hbox{GeV}$),
Eqns.$\!$ (\ref{znbb-1}) and (\ref{znbb-2}) require
$\frac{\mu_{\scriptscriptstyle 1}}{\mu_{\scriptscriptstyle 5}} 
\raisebox{-.3em}{$\stackrel{\displaystyle <}{\sim}$} \frac{1}{100}$
for MeV scale neutrino masses and
${\mu}_{\scriptscriptstyle 1}:{\mu}_{\scriptscriptstyle 0}$ of around 
$1:150,000$ ($1:15,000$) for $\tan\!\beta = 2$ ($45$) --- setting 
$\frac{ {\mu}_{\scriptscriptstyle 1} }{ {\mu}_{\scriptscriptstyle 5} } 
\sim \frac{1}{100}$
allows ${\mu}_{\scriptscriptstyle 5}$ values of only tenths of a GeV for
$\tan\!\beta =2$ and a couple GeV for $\tan\!\beta = 45$.

The limitation from $0\nu\beta\beta$ is only applicable if 
$m_{\nu_{\scriptscriptstyle 5}} < 10\, \hbox{MeV}$.  Thus the
$0\nu\beta\beta$ injunction is turned off if ${\mu}_{\scriptscriptstyle 5}$
is large enough to push $m_{\nu_{\scriptscriptstyle 5}}$ above this threshold.
This means that at any given point in MSSM parameter space, and for 
${\mu}_{\scriptscriptstyle 1} \ne 0$, there will be at least two allowed
ranges for ${\mu}_{\scriptscriptstyle 5}$:
$\mu_{\scriptscriptstyle 5} \leq \mu_{\scriptscriptstyle 5}^{\beta, max}$ 
and
$\mu_{\scriptscriptstyle 5}^{(10)} \leq 
\mu_{\scriptscriptstyle 5} \leq 
\mu_{\scriptscriptstyle 5}^{max}$, where 
$\mu_{\scriptscriptstyle 5}^{\beta, max}$ is the upper bound from 
$0\nu\beta\beta$,
$\mu_{\scriptscriptstyle 5}^{(10)}$ is the ${\mu}_{\scriptscriptstyle 5}$ 
value at which $m_{\nu_{\scriptscriptstyle 5}}$ reaches
$10\, \hbox{MeV}$, and $\mu_{\scriptscriptstyle 5}^{max}$ is the cut-off 
value due to the strongest constraint aside from $0\nu\beta\beta$ 
(typically this is from $R^{\pi e}_{\pi \mu}$ for low to moderate 
$\tan\!\beta$ values and from $R^{\mu e}_{\tau e}$ for high $\tan\!\beta$).
This of course assumes
$\mu_{\scriptscriptstyle 5}^{\beta, max} < 
\mu_{\scriptscriptstyle 5}^{(10)} < \mu_{\scriptscriptstyle 5}^{max}$
as is almost universally true.

\section{Overall Combined Constraints}

Here we pull together all the constraints addressed individually
in the preceding sections.  These are combined numerically in a
comprehensive program to yield a maximum-allowed 
${\mu}_{\scriptscriptstyle 5}$
for any given point in MSSM parameter space for a specified
${\mu}_{\scriptscriptstyle 1}:
 {\mu}_{\scriptscriptstyle 2}:
 {\mu}_{\scriptscriptstyle 3}$ set.  Table I lists
all the experimental constraints applied.  Note that constraints
resulting from the WA66 and CHARM neutrino decay experiments as well as
those from neutrinoless double beta decay experiments are not implemented 
except to rule out potentially admissible large
$\mu_5$ regions beyond the first cutoff point for individual
constraints (a thorough implementation 
awaits a more complete analysis including scalar intermediates).

A few details associated with the numerical studies deserve mention.
First, (electroweak) gaugino unification is assumed; {\it i.e.},
$M_{\scriptscriptstyle 1} = xM_{\scriptscriptstyle 2}$ with 
$x = \frac{5}{3}{\tan}\!^2{\theta}_w$.
Second, running parameters are evaluated at the scale appropriate to
each particular process.  Thus for instance $Z^0$ decays use
${\sin}^2\!\!\;{\theta}_w (M_{\scriptscriptstyle Z})$ while $\tau$ decays 
have ${\sin}^2\!\!\;{\theta}_w (m_{\tau})$.  Specific
numerical inputs include (see \cite{LEP1}):
${\sin}^2\!\!\;{\theta}_w (M_{\scriptscriptstyle Z}) = 0.2315$,
$M_{\scriptscriptstyle Z} = 91.1867\, \hbox{GeV}$,
and
$M_{\scriptscriptstyle W} = 80.4\, \hbox{GeV}$.
 
Fig.$\!$ 12 scans the 
$M_{\scriptscriptstyle 2}$-${\mu}_{\scriptscriptstyle 0}$ plane and presents
contours of maximum-allowed ${\mu}_{\scriptscriptstyle 5}$ values in GeV. 
Figs. 12a and 12b
have $\tan\!\beta = 2$ while 12c and 12d have $\tan\!\beta =45$;
${\mu}_{\scriptscriptstyle 1}:
 {\mu}_{\scriptscriptstyle 2}:
 {\mu}_{\scriptscriptstyle 3} = 0:1:1$ in 12a and 12c
whereas
${\mu}_{\scriptscriptstyle 1}:
 {\mu}_{\scriptscriptstyle 2}:
 {\mu}_{\scriptscriptstyle 3} = 0:1:10$ in 12b and 12d.
With $\tan\!\beta = 2$,  the $R^{\mu e}_{\tau e}$ constraint dominates
for the $0:1:1$ ratio set; but $R^{\pi \mu}_{\pi e}$ is
stronger for the $0:1:10$ ratio set.  The charged current constraints
are slightly more restrictive than those from the $149\, \hbox{MeV}$
neutrino mass bound shown in Fig.$\!$ 7c --- the difference is greater
when
${\mu}_{\scriptscriptstyle 1}:
 {\mu}_{\scriptscriptstyle 2}:
 {\mu}_{\scriptscriptstyle 3} = 0:1:10$,
but the order of magnitude is still the same.  
If ${\mu}_{\scriptscriptstyle 2}$ is set to zero
(${\mu}_{\scriptscriptstyle 1}:
  {\mu}_{\scriptscriptstyle 2}:
  {\mu}_{\scriptscriptstyle 3} = 0:0:1$), then the 
$18.2\, \hbox{MeV}$ ${\nu}_{\tau}$ mass bound of Fig.$\!$ 7a dominates.

Another distinction between Figs.$\!$ 7 and 12 is in the central
region where $M_{\scriptscriptstyle 2}$ or 
$|{\mu}_{\scriptscriptstyle 0}|$ is small.
This region is ruled out in Fig.$\!$ 12 by limits from processes 
involving charginos and neutralinos.  Such restrictions, embodied in 
the Table I bounds on the chargino mass, on anomalous ``visible'' $Z^0$ 
decay modes
($Z^0 \rightarrow {\chi}^{\pm}{\ell}^{\mp}, {\chi}^0_c {\chi}^0_d,
{\chi}^0_d \nu;\; d \ne 1$), and on $Z^0$ full and invisible decay widths,
are in general rather conservative, reflecting present intangibles
concerning sparticle decays ---  a more complete study of these
is in progress.  As discussed in Section II.E. and seen in Fig.$\!$ 1,
${\mu}_i$ values sufficient to significantly affect the lighter chargino
mass are in the high-${\mu}_i$ realm and thus only depend on
${\mu}_{\scriptscriptstyle 5}$.  
Results for the combined constraints concur:  bounds in
the small $M_{\scriptscriptstyle 2}$, 
small $|{\mu}_{\scriptscriptstyle 0}|$ region are basically
independent of the ratios among the ${\mu}_i$'s.  
One additional point worth noting is that driving up 
${\mu}_{\scriptscriptstyle 5}$ to extremely high values will
not be able to push up a very small MSSM chargino mass.

Regarding $\tan\!\beta = 45$:
the general pattern of the contour lines is similar to that for the
$\tan\!\beta = 2$ plots; however, admissible 
${\mu}_{\scriptscriptstyle 5}$ values are much larger.  Upper bounds on 
${\mu}_{\scriptscriptstyle 5}$ are relaxed by a factor of ${\sim}\tan\!\beta$.
The limits in Figs.$\!$ 12c and 12d are much stronger than those in
Fig.$\!$ 7d which just uses the $149\, \hbox{MeV}$ neutrino mass bound.
Both of these changes are due to the dominant charged current
constraints ($R^{\mu e}_{\tau e}$ is strongest except for
${\mu}_{\scriptscriptstyle 1}:
 {\mu}_{\scriptscriptstyle 2}:
 {\mu}_{\scriptscriptstyle 3} = 0:1:10$
and $M_{\scriptscriptstyle 2} 
\raisebox{-.3em}{$\stackrel{\displaystyle <}{\sim}$} 160\, \hbox{GeV}$ 
in which case $R^{\pi \mu}_{\pi e}$ dominates). 
As ${\mu}_{\scriptscriptstyle 2} \rightarrow 0$, 
we again return to the $18.2\, \hbox{MeV}$ $m_{{\nu}_{\tau}}$
limits of Fig.$\!$ 7b (with the central region again excluded by chargino
and neutralino constraints).

\section{Conclusions and Outlook}

The degree of generality possible within the single-VEV parametrization
deserves special notice.  With only a trio of inputs
(the bilinear ${\mu}_i$ couplings in the superpotential) beyond those of
the MSSM\footnote{Of course the MSSM already has a fair
number of input parameters --- $M_{\scriptscriptstyle 2}$, 
${\mu}_{\scriptscriptstyle 0}$, and
$\tan\!\beta$ all enter into the mass matrices. This is unchanged and
precludes an exhaustive scan of the full beyond-the-SM parameter space.
}, 
the tree-level mass matrices of all color-singlet fermions
are completely determined, and
a broad range of leptonic phenomenology can be analyzed with
a reasonable level of sophistication.
Furthermore, the trio can often be collapsed into a single input,
${\mu}_{\scriptscriptstyle 5}$, which carries the full weight of 
$R$-parity violation.
This is to be contrasted with analyses which
 either contain a plethora of RPV parameters which preclude a
meaningful coverage of the parameter space within the model
or arbitrarily pick one or two RPV paramaters to be non-zero while the
others are all set to zero by hand.  As a result, 
the true freedom within an RPV model will be masked or
muddied by a less optimal choice of flavor basis (or by no choice at
all!).

We have presented tree-level coverage for the variety
of electroweak signals summarized in Table I.  In the single-VEV
parametrization analysis presented here, {\em the trilinear RPV
couplings have not been set to zero}.  Rather, they simply do not
contribute (at tree level with gauge boson propagators) to the
impressive spectrum of leptonic processes studied.  In principle, 
intermediate scalars (sleptons, Higgs bosons) can usher back in the
tree-level trilinear dependence; however, such contributions should be
suppressed by the larger masses of the scalars relative to the SM
intermediate vector bosons, and are nonnegligible only in special regions 
of the parameter space.  A more in depth look at the scalar sector
(along with the related issue of electroweak sparticle decays) is
now in progress.  An exception to ignoring the scalar intermediates
may be necessary with neutrinoless double beta decay due to the
presence of strongly-interacting sparticle intermediates.  This
question is also under study.

Also beyond the present study but slated for future work 
are loop effects, necessary to
describe interesting processes such as $\mu \rightarrow e \gamma$.
This process may place an additional significant restriction on the
${\mu}_i$'s.  Other loop processes will be important in constraining the
trilinear RPV couplings.
Loops will also lift the degeneracy of the two tree-level massless
neutrinos.  This extra degree of precision is certainly needed
to study the very low-mass neutrinos preferred by several
neutrino oscillation experiments.  Some models to describe such
experiments also suggest that one or more extra light (sterile) neutrinos
be added.
This could strongly affect the analysis of the charged current
constraints presented here, but not the $Z^0$-mediated neutral
current constraints for the charged leptons.  Thus, while the
charged current constraints are more restrictive, those for
neutral current processes are more robust against such possible
model extensions.

The single tree-level neutrino mass, $m_{{\nu}_{\scriptscriptstyle 5}}$, 
depends on the three ${\mu}_i$ only through ${\mu}_{\scriptscriptstyle 5}$.  
If MeV-scale neutrino masses are allowed, then 
${\mu}_{\scriptscriptstyle 5}$ may well be large enough to yield signal 
rates near, at, or above present experimental
bounds for the numerous other processes described herein.  Such
a neutrino mass is not ruled out by direct or indirect
machine (terrestrial) mass bounds.  Cosmological constraints favoring
light neutrinos may not be applicable;  this depends on the decay
properties of the massive neutrino and requires more study.  
An MeV-scale neutrino has been found to be consistent
with cosmology in at least one study (which did include fields beyond 
those in the MSSM) \cite{RomVal}.

The results of this analysis reinforce those of \cite{PapI}:
${\mu}_{\scriptscriptstyle 5}$ values of the same order of magnitude, or 
even much larger, than $M_{\scriptscriptstyle 2}$ and 
${\mu}_{\scriptscriptstyle 0}$ (the standard MSSM inputs) are allowed by 
the experimental bounds.
This is particularly true for high $\tan\!\beta$ 
($
\raisebox{-.3em}{$\stackrel{\displaystyle >}{\sim}$} 45$), where RPV
signals are strongly suppressed.  In this case even much tighter
bounds on the neutrino mass do not preclude large 
${\mu}_{\scriptscriptstyle 5}$
values.  This $\tan\!\beta$ dependence is almost universal among the
processes studied.  This plus the fact that both the neutrino mass and 
numerous (though by no means all) approximate expressions for other bounds
depend only on ${\mu}_{\scriptscriptstyle 5}$ lead to the interesting 
question of how much variation is possible among these other processes if
${\mu}_{\scriptscriptstyle 5}$ is fixed\footnote{
If both ${\mu}_{\scriptscriptstyle 5}$ and neutrino mass are fixed, then
restrictions are placed upon the MSSM input parameters.
How much $\tan\!\beta$ dependence and variation is then possible in the
remaining processes is currently being investigated.}.
To look at this various possible
${\mu}_{\scriptscriptstyle 1} : 
 {\mu}_{\scriptscriptstyle 2} : 
 {\mu}_{\scriptscriptstyle 3}$ combinations
were studied using the exact (at tree-level) numerical
expressions.\footnote{Naturally, studies that truncate the
number of generations cannot preform such analyses and miss
very significant and interesting effects.}
A rough hierarchy is seen in the constraints on the individual
${\mu}_i$'s:  ${\mu}_{\scriptscriptstyle 1}$ is strongly restricted,
${\mu}_{\scriptscriptstyle 2}$ is less restricted, and 
${\mu}_{\scriptscriptstyle 3}$ is still less restricted.  
This suggests ratio sets of the general
form $0:1:x$ ($x \ge 1$).  With only two free inputs, this further
suggests that, along with a dimensionless ratio, 
${\mu}_{\scriptscriptstyle 5}$ is the preferred  indicator
of RPV effects in leptonic phenomenology.  

\vskip 1.cm

\bigskip
\bigskip

The authors thank  P.Tipton 
for helpful discussions and A.M.~Cooper-Sarkar
for discussions regarding the BEBC data. We also
benefitted from questions and comments of colleagues,
particularly C.-C. Chen,
S. Davidson, M. Losada, E. Nardi, F. Vissani, and C. Wagner. 
K.L. Chan and Gad Eilam are greatly appreciated for reading over the
manuscript.
O.K. thanks P.H. Frampton for being a constant source of encouragement.
This work was supported in part by the U.S. Department of Energy,
under grant DE-FG02-91ER40685 and by the U.S. National Science Foundation,
under grants PHY-9600155 and INT-9804704.\\

\clearpage

{\centerline{{\bf Table Captions}}

\vskip 2.0cm

{\bf Table I}:

Summary of phenomenological constraints incorporated in the
overall parameter constraint plots. 
The invisible $Z^0$-width is assumed to include decays to end
states composed of neutrinos and the lightest neutralino.
Bounds imposed on $Z^0$ decay constraints involving charginos and
neutralinos are quite conservative, representing uncertainty in
signal detection.

\vskip 1.5cm

{\bf Table II}:

Experimental left-right asymmetry results: 
${\cal A}_i$ obtained from ${\cal A}_{\scriptscriptstyle FB}^\ell$
(using ${\cal A}_{\scriptscriptstyle FB}^\ell =
 \frac{3}{4}{\cal A}_e{\cal A}_\ell $)
measured at LEP and SLD, ${\cal P}_\tau$ measured at LEP,
and a direct measurement from SLD.  

\vskip 1.5cm

\clearpage

{\centerline{{\bf Figure Captions}}

\vskip 1.0cm

{\bf Figure 1}:

Branching ratio for $\mu^- \to e^-e^+e^-$ as a function of 
$\mu_{\scriptscriptstyle 5}$ (in GeV), with
$M_{\scriptscriptstyle 2} = \mu_{\scriptscriptstyle 0}=200\, \hbox{GeV}$
and $\tan\!\beta =2$ (left) or $\tan\!\beta = 45$ (right)
for ratios of
$\mu_{\scriptscriptstyle 1}:
 \mu_{\scriptscriptstyle 2}:
 \mu_{\scriptscriptstyle 3}$ as marked.
The solid horizontal line is the experimental bound.
If ${\mu}_{\scriptscriptstyle 3}$ --- here set to zero --- is varied,
it  affects the curves only via ${\mu}_{\scriptscriptstyle 5}$,
and hence stretches out the horizontal scale. 

\vskip 1cm

{\bf Figure 2}:

Branching ratio for $\tau^- \to e^-\mu^+\mu^-$ as a function of 
$\mu_{\scriptscriptstyle 5}$ (in GeV), with
$M_{\scriptscriptstyle 2}=\mu_{\scriptscriptstyle 0}=200\, \hbox{GeV}$
and $\tan\!\beta = 2$, for ratios of 
$\mu_{\scriptscriptstyle 1}:
 \mu_{\scriptscriptstyle 2}:
 \mu_{\scriptscriptstyle 3}$ as marked.
The solid horizontal line is the experimental bound.
Here, ${\mu}_{\scriptscriptstyle 2}$ is set to zero; again, varying this 
stretches the horizontal scale. 

\vskip 1cm

{\bf Figure 3}:

Leptonic partial decay widths of the $Z^0$ (in MeV) as a function of 
$\mu_{\scriptscriptstyle 5}$ (in GeV), with
$\mu_{\scriptscriptstyle 1}:
 \mu_{\scriptscriptstyle 2}:
 \mu_{\scriptscriptstyle 3}=1:1:1$
(so that $\mu_i = \frac{1}{\sqrt{3}}\mu_{\scriptscriptstyle 5}$); 
$M_{\scriptscriptstyle 2} = \mu_{\scriptscriptstyle 0} = 200\, \hbox{GeV}$,
and $\tan\!\beta =2$.  Horizontal lines at the left edge of the
plot are the $\pm 3\sigma$ 
experimental bounds \cite{NovPil} for the corresponding quantities;
they are independent of $\mu_{\scriptscriptstyle 5}$ and are truncated 
at right for clarity. 
Explicit values for the experimental bounds on these
partial decay widths are found in \cite{LEP1}.
\vskip 1cm

{\bf Figure 4}:

Leptonic L-R asymmetry: deviations from SM predictions as functions of 
$\mu_{\scriptscriptstyle 5}$ (in GeV), with
$\mu_{\scriptscriptstyle 1}:
 \mu_{\scriptscriptstyle 2}:
 \mu_{\scriptscriptstyle 3}=1:1:1$ and
$M_{\scriptscriptstyle 2} = \mu_{\scriptscriptstyle 0}=200\, \hbox{GeV}$
for $\tan\!\beta =2$ (top) and $\tan\!\beta =45$ (bottom).
The solid line approximates the experimental $3\sigma$ upper bounds
for each of the 3 asymmetries.  

\vskip 1cm

{\bf Figure 5}:

Chargino masses.
Contours show minimum values of ${\mu}_{\scriptscriptstyle 5}$
(in GeV) necessary to push the lighter chargino mass above
$90\, \hbox{GeV}$ for $\tan\!\beta =2$ (top) 
and $\tan\!\beta = 45$ (bottom).
The region above or outside a given 
contour has $\bar{M}_{\scriptscriptstyle c1} > 90\, \hbox{GeV}$ for
${\mu}_{\scriptscriptstyle 5}$ at or above the designated value.

\vskip 1cm

{\bf Figure 6}:

Neutrino mass $m_{\nu_{\scriptscriptstyle 5}}$
as a function of $\mu_{\scriptscriptstyle 5}$ (in GeV), with
$M_{\scriptscriptstyle 2} = \mu_{\scriptscriptstyle 0} = 200\, \hbox{GeV}$ 
and a) $\tan\!\beta =2$, b)  $\tan\!\beta =45$. The lower horizontal line
is the $18.2\,\hbox{MeV}$ machine bound for a pure $\nu_\tau$;
the upper horizontal line is the $149\,\hbox{MeV}$ bound for
a generic $\nu_{\scriptscriptstyle 5}$.
($M_{\scriptscriptstyle 1} = xM_{\scriptscriptstyle 2}$, with
$x = \frac{5}{3}\tan\!\!^2{\theta}_w$.)   Insets:  Low 
$\mu_{\scriptscriptstyle 5}$ portion of curve on log-log scale.
\vskip 1cm

{\bf Figure 7}:

Maximum allowed values of ${\mu}_{\scriptscriptstyle 5}$ (in GeV) consistent 
with neutrino mass bounds:
machine bound $m_{{\nu}_{\tau}} < 18.2\, \hbox{MeV}$ (applicable for
${\mu}_{\scriptscriptstyle 1}:
 {\mu}_{\scriptscriptstyle 2}:
 {\mu}_{\scriptscriptstyle 3} = 0:0:1$)
for a) $\tan\!\beta=2$ and b) $\tan\!\beta=45$;   
the absolute bound $m_{{\nu}_{\scriptscriptstyle 5}} < 149\, \hbox{MeV}$ for 
c) $\tan\!\beta=2$ and d) $\tan\!\beta = 45$.
The region below or inside of a given contour 
is excluded for ${\mu}_{\scriptscriptstyle 5}$'s above the indicated value.

\vskip 1cm

{\bf Figure 8}:

The invisible $Z^0$-width (in MeV) as a function of 
$\mu_{\scriptscriptstyle 5}$ (in GeV), with
$M_{\scriptscriptstyle 2} = \mu_{\scriptscriptstyle 0} = 200\, \hbox{GeV}$ 
and $\tan\!\beta =2$.
The solid horizontal lines are the upper and lower
experimental bounds.
${\Gamma}^{\scriptstyle R\!P\!V}_{
\!\scriptstyle Z_{\scriptscriptstyle i\!n\!v}}$
 is assumed to be
$\Gamma(Z^0 \rightarrow {\nu}_c {\nu}_d)$ (dotted curve) or
$\Gamma(Z^0 \rightarrow {\nu}_c {\nu}_d, 
{\nu}_c {\chi}^0_{\scriptscriptstyle 1},
{\chi}^0_{\scriptscriptstyle 1}{\chi}^0_{\scriptscriptstyle 1})$ 
(dashed curve)
where ${\chi}^0_{\scriptscriptstyle 1}$ is the lightest neutralino
($4^{th}$ lightest neutral color-singlet fermion).

\vskip 1cm

{\bf Figure 9}:
Constraints on $\mu_1/\mu_5$ and $\mu_2/\mu_5$ ratios.  Solid lines:
pion decay [cf. Eqn. (\ref{Rpiell})]; in the upper half of 
the figure, the lower solid line corresponds to $\mu_2/\mu_5=1/12$, while
the upper line to $\mu_2/\mu_5=0$; in the lower half of the figure, the
solid line corresponds to $\mu_1/\mu_5=0$.  
Dotted lines:  the WA66 experiment [cf. \cite{WA66}].
Dashed line:  neutrinoless double beta decay (cf. Eqn. (\ref{znbb-1})).

\vskip 1cm

{\bf Figure 10}:

$R^{\pi e}_{\pi \mu}$ as a function of 
$\mu_{\scriptscriptstyle 5}$ (in GeV)
 for various $\mu_i$ ratios, with
$M_{\scriptscriptstyle 2} = \mu_{\scriptscriptstyle 0} = 200\, \hbox{GeV}$ 
and a) $\tan\!\beta =2$, b) $\tan\!\beta =45$.   In a) the 
$\mu_i$ ratios are 1 : 1 : 1 (long-dashed line); 1 : 10 : 110 (dot-dashed
line; appears twice in the figure); 0 : 1 : 1 (open-spaced dotted line);
0 : 1 : 11 (closely-spaced dotted line); 1 : 0 : 110 (short-dashed line). 

\vskip 1cm

{\bf Figure 11}:

$R^{\mu e}_{\tau e}$ {\it vs}. ${\mu}_{\scriptscriptstyle 5}$ for
assorted
${\mu}_i$ ratios and a) $\tan\!\beta = 2$, b) $\tan\!\beta = 45$.
Horizontal lines denote $3\sigma$ deviations from the
measured central value as given in \cite{BFKM}.

\vskip 1cm

{\bf Figure 12}:

Maximum allowed values of ${\mu}_{\scriptscriptstyle 5}$ (in GeV) consistent
with all the constraints listed in Table I:
a) for $\tan\!\beta = 2$ and $\mu_i$ ratios $0:1:1$;
b) for $\tan\!\beta = 2$ and $0:1:10$;
c) for $\tan\!\beta = 45$ and $0:1:1$;
d) for $\tan\!\beta = 45$ and $0:1:10$.

\clearpage

\begin{table}[th]
\caption{}
\vspace*{0.15cm}
  
\begin{tabular}{lcl} 
       {\quad Quantity \quad} & 
\raisebox{-.3em}{$\stackrel{\displaystyle {\mu}_{i}\hbox{ combo.}}
{{ \scriptstyle { \hbox{constrained}}}}$} &
       {\quad Experimental bounds \cite{expts} \quad} \\ \hline
       & & \\[-.2in]
       \framebox{$Z^0$-coupling:} & & \\
    $\bullet$ $U_{br}^{e\mu}$ \hspace*{.2in} ($e$-$\mu$ universality)
      &  $\mu_{\scriptscriptstyle 1}^2-\mu_{\scriptscriptstyle 2}^2$ 
       &  $(0.596 \pm 4.37)\times 10^{-3}$ \\[-.15in]
       $\bullet$ $U_{br}^{e\tau}$ \hspace*{.2in} ($e$-$\tau$ universality)
       &  $\mu_{\scriptscriptstyle 1}^2-\mu_{\scriptscriptstyle 3}^2$ 
       &  $(0.955 \pm 4.98)\times 10^{-3}$ \\[-.15in]
       $\bullet$  $U_{br}^{\mu\tau}$ \hspace*{.2in}
       ($\mu$-$\tau$ universality)
       &  $\mu_{\scriptscriptstyle 2}^2-\mu_{\scriptscriptstyle 3}^2$ 
       &  $(1.55 \pm 5.60)\times 10^{-3}$ \\[-.15in]
       $\bullet$ $\Delta{\cal A}_{e\mu}$ \hspace*{.15in}
       ($e$-$\mu$ L-R asymmetry)
       &  $\mu_{\scriptscriptstyle 1}^2-\mu_{\scriptscriptstyle 2}^2$
       &  $(0.346\pm 2.54)\times 10^{-2}$
       (from $U_{br}^{e\mu}$) 
       \\[-.15in]
       $\bullet$ $\Delta{\cal A}_{\tau e}$ \hspace*{.1in}
       ($\tau$-$e$ L-R asymmetry)
       &  $\mu_{\scriptscriptstyle 3}^2-\mu_{\scriptscriptstyle 1}^2$
       + Rt. contrib.
       &  $0.0043\pm 0.104$ \\[-.15in]
       $\bullet$ $\Delta{\cal A}_{\tau\mu}$ \hspace*{.1in}
       ($\tau$-$\mu$ L-R asymmetry)
       &  $\mu_{\scriptscriptstyle 3}^2-\mu_{\scriptscriptstyle 2}^2$
       + Rt. contrib.
       &  $0.082\pm 0.25$ \\[-.15in]
     $\bullet$ $Br$($Z^0 \to e^{\pm} \mu^{\mp}$)
       &  $|\mu_{\scriptscriptstyle 1}\mu_{\scriptscriptstyle 2}|$ 
       &  $<1.7\times 10^{-6}$ \\[-.15in]
       $\bullet$ $Br$($Z^0 \to e^{\pm} \tau^{\mp}$)
       &  $|\mu_{\scriptscriptstyle 1}\mu_{\scriptscriptstyle 3}|$ 
       &  $<9.8\times 10^{-6}$ \\[-.15in]
       $\bullet$ $Br$($Z^0 \to \mu^{\pm} \tau^{\mp}$)
       &  $|\mu_{\scriptscriptstyle 2}\mu_{\scriptscriptstyle 3}|$ 
       &  $<1.2\times 10^{-5}$ \\[-.15in]  
      $\bullet$ $Br$($\mu^- \to e^- e^+e^-$)
       &  $|\mu_{\scriptscriptstyle 1}\mu_{\scriptscriptstyle 2}|$ 
       &  $<1.0\times 10^{-12}$ \\[-.15in]
        $\bullet$ $Br$($\tau^- \to e^- e^+e^-$)
       & $|\mu_{\scriptscriptstyle 1}\mu_{\scriptscriptstyle 3}|$ 
       & $<2.9\times 10^{-6}$\\[-.15in]
        $\bullet$ $Br$($\tau^{-} \to \mu^{-} e^+ e^-$)
       & $|\mu_{\scriptscriptstyle 2}\mu_{\scriptscriptstyle 3}|$ 
       &  $<1.7\times 10^{-6}$\\[-.15in]
       $\bullet$ $Br$($\tau^{-} \to \mu^{+} e^- e^-$)
       &  $|\mu_{\scriptscriptstyle 1}^2\mu_{\scriptscriptstyle 2}
       \mu_{\scriptscriptstyle 3}|$
       & $<1.5\times 10^{-6}$\\[-.15in]
       $\bullet$ $Br$($\tau^{-} \to e^{-} \mu^+ \mu^-$)
       &  $|\mu_{\scriptscriptstyle 1}\mu_{\scriptscriptstyle 3}|$ 
       & $<1.8\times 10^{-6}$\\[-.15in]
       $\bullet$ $Br$($\tau^{-} \to e^{+} \mu^- \mu^-$)
       &  $|\mu_{\scriptscriptstyle 1}\mu_{\scriptscriptstyle 2}^2
       \mu_{\scriptscriptstyle 3}|$
       & $<1.5\times 10^{-6}$\\[-.15in]
       $\bullet$ $Br$($\tau^- \to \mu^- \mu^+ \mu^-$)
       &  $|\mu_{\scriptscriptstyle 2}\mu_{\scriptscriptstyle 3}|$ 
       & $<1.9\times 10^{-6}$\\[-.15in]
    $\bullet$ $Br$($Z^0 \to \chi^{\pm} \ell^{\mp}$)
       &   $\mu_{\scriptscriptstyle 5}$
       &  $< 1.0\times 10^{-5}$ \\[-.15in]
       $\bullet$ $Br$($Z^0 \to \chi^{\pm} \chi^{\mp}$)
       &  $\mu_{\scriptscriptstyle 5}$ 
       &  $< 1.0\times 10^{-5}$ \\[-.15in]
      $\bullet$ $Br$($Z^0 \to \chi^{0}_i\chi^{0}_j,
                                               \chi^{0}_j\nu) ; \; j \ne 1$
       &  $\mu_{\scriptscriptstyle 5}$ 
       &  $< 1.0\times 10^{-5}$ \\[-.15in]
      $\bullet$ ${\Gamma}_{\!\scriptscriptstyle Z}$ \hspace*{.25in}
      (total $Z^0$-width)
       &  $\mu_{\scriptscriptstyle 5}$
       &  $2.4948 \pm 0.0075 \, \hbox{GeV}$ \\[-.15in]
       $\bullet$
       ${\Gamma}_{\!\scriptstyle Z_{\scriptscriptstyle i\!n\!v}}$
       \hspace*{.2in} (invisible $Z^0$ width:
       &  $\mu_{\scriptscriptstyle 5}$ 
       &  $500.1\pm5.4\, \hbox{MeV}$ \\[-.15in]
       \hspace*{.8in}  $Z^0 \rightarrow {\nu}_c{\nu}_d, {\nu}_c{\chi}^0_1,
       {\chi}^0_1{\chi}^0_1$)
      & & \\[-.1in]
       \framebox{$W^{\pm}$-coupling:}
        & & \\
       $\bullet$        $\overline{\Gamma}^{\mu e}$ \hspace*{.2in}
       ($\mu \to e \nu \nu$)
       &  $m_{\nu_{\scriptscriptstyle 5}}\,/\,\mu_i$ ratio  &
        $0.983\pm0.111$ \\[-.1in]
       $\bullet$        $\overline{\Gamma}^{\tau e}$ \hspace*{.2in}
       ($\tau \to e \nu \nu$)
       &  $m_{\nu_{\scriptscriptstyle 5}}\,/\,\mu_i$ ratio  &
        $0.979\pm0.111$ \\[-.1in]
       $\bullet$        $\overline{\Gamma}^{\tau \mu}$ \hspace*{.2in}
       ($\tau \to \mu \nu \nu$)
       &  $m_{\nu_{\scriptscriptstyle 5}}\,/\,\mu_i$ ratio  &
        $0.954\pm0.108$ \\[-.1in]
       $\bullet$        $R^{\pi e}_{\pi \mu}$ \hspace*{.2in} ($\pi$ decays)
       &  $m_{\nu_{\scriptscriptstyle 5}}\,/\,
       \frac{\mu_{\scriptscriptstyle 1}}{\mu_{\scriptscriptstyle 5}}$ and
       $\frac{\mu_{\scriptscriptstyle 2}}{\mu_{\scriptscriptstyle 5}}$   &
        $(1.230\pm0.012)\times 10^{-4}$ \\[-.1in]
           $\bullet$        $R^{\tau e}_{\tau \mu}$ \hspace*{.2in}
           ($\tau$ decays)
       &  $m_{\nu_{\scriptscriptstyle 5}}\,/\,\mu_i$ ratio   &
        $1.0265\pm0.0222$ \\[-.1in]
        $\bullet$        $R^{\mu e}_{\tau e}$ \hspace*{.2in}
        (decays to $e$'s)
       &  $m_{\nu_{\scriptscriptstyle 5}}\,/\,\mu_i$ ratio &
        $1.0038\pm0.0219$ \\[-.1in]
       $\bullet$ $m_{\nu_{\scriptscriptstyle 5}}
      |\tilde{B}^{\scriptscriptstyle L}_{e\nu_{\scriptscriptstyle 5}}|^2$
      \hspace*{.2in} [$(\beta\beta)_{0\nu}$]
       &  $m_{\nu_{\scriptscriptstyle 5}}\,/\,
       \frac{\mu_{\scriptscriptstyle 1}}{\mu_{\scriptscriptstyle 5}}$  &
        $< 0.46\, \hbox{eV}$
        (only for $m_{\nu_{\scriptscriptstyle 5}}\!<\!10\,\mbox{MeV}$)
        \\[-.1in]
           & & \\[-.2in]
      \framebox{mass constraints:}
       & & \\
        $\bullet$ $\nu_{\scriptscriptstyle 5}$ mass
       &  $\mu_{\scriptscriptstyle 3}$ 
       &  $<18.2\, \hbox{MeV}$ if $\nu_{\scriptscriptstyle 5} = \nu_{\tau}$ 
       \\[-.15in]
      &  $\mu_{\scriptscriptstyle 5}$  
      &  $<149\, \hbox{MeV}$ if $\nu_{\scriptscriptstyle 5} \ne \nu_{\tau}$
    \\[-.15in]
    $\bullet$ $\chi^\pm$ mass
         &  $\mu_{\scriptscriptstyle 5}$ & $ > 70\,\mbox{GeV}$\\
\end{tabular} \normalsize 
\end{table}

\clearpage

\noindent
\begin{table}
\begin{center}
\begin{minipage}{3.0in}
\caption{}
\vspace*{.2in}
\begin{tabular}{|l|c|c|c|} 
${\cal A}_{\ell}$ & Method & LEP combined\cite{LEP1} &
SLD\cite{SLD} \phantom{aaaa} \\ \hline
                          & $FB$ & $0.1461\pm .0110$ & $0.152\pm .012$
\\
\cline{2-4}
${\cal A}_e$              & ${\cal P}_{\tau}$
                                 & $0.1399\pm .0073$ & ---
\\
\cline{2-4}
                          & direct
                                 & ---               & $0.1543\pm .0039$
\\
\hline
${\cal A}_{\mu}$  & $FB$ & $0.1488\pm .0170$ & $0.102\pm .034$
\\
\hline
\smash{\lower 1.75ex \hbox{${\cal A}_{\tau}$}}
& $FB$ & $0.1753\pm .0210$  & $0.195\pm .034$
\\
\cline{2-4}
                          & ${\cal P}_{\tau}$
                                 & $0.1411 \pm .0064$  &  ---
\\
\end{tabular}
\end{minipage}
\end{center}
\end{table}


\bigskip
\bigskip

\clearpage


\includegraphics{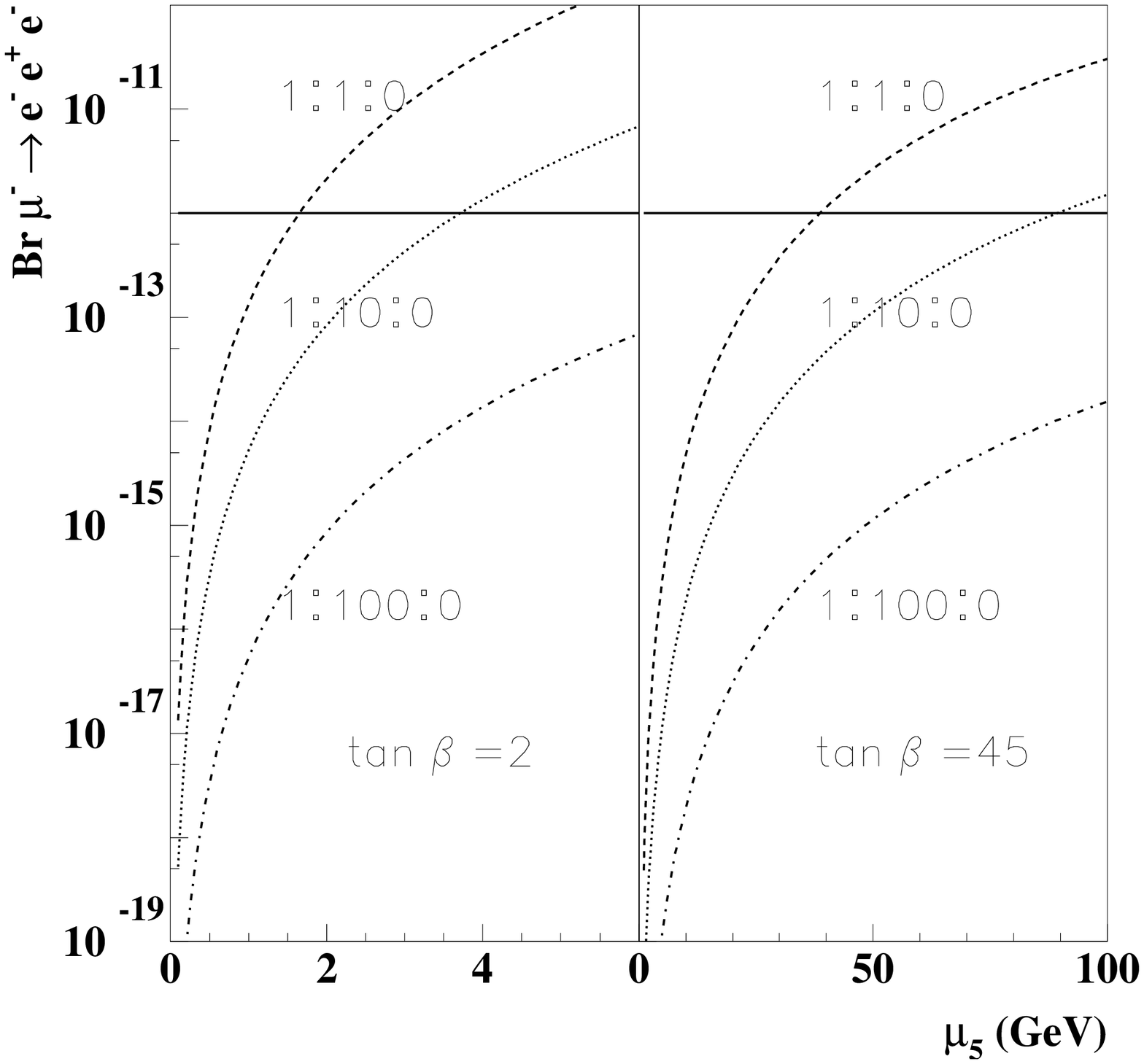}
Figure 1  
\clearpage

\includegraphics{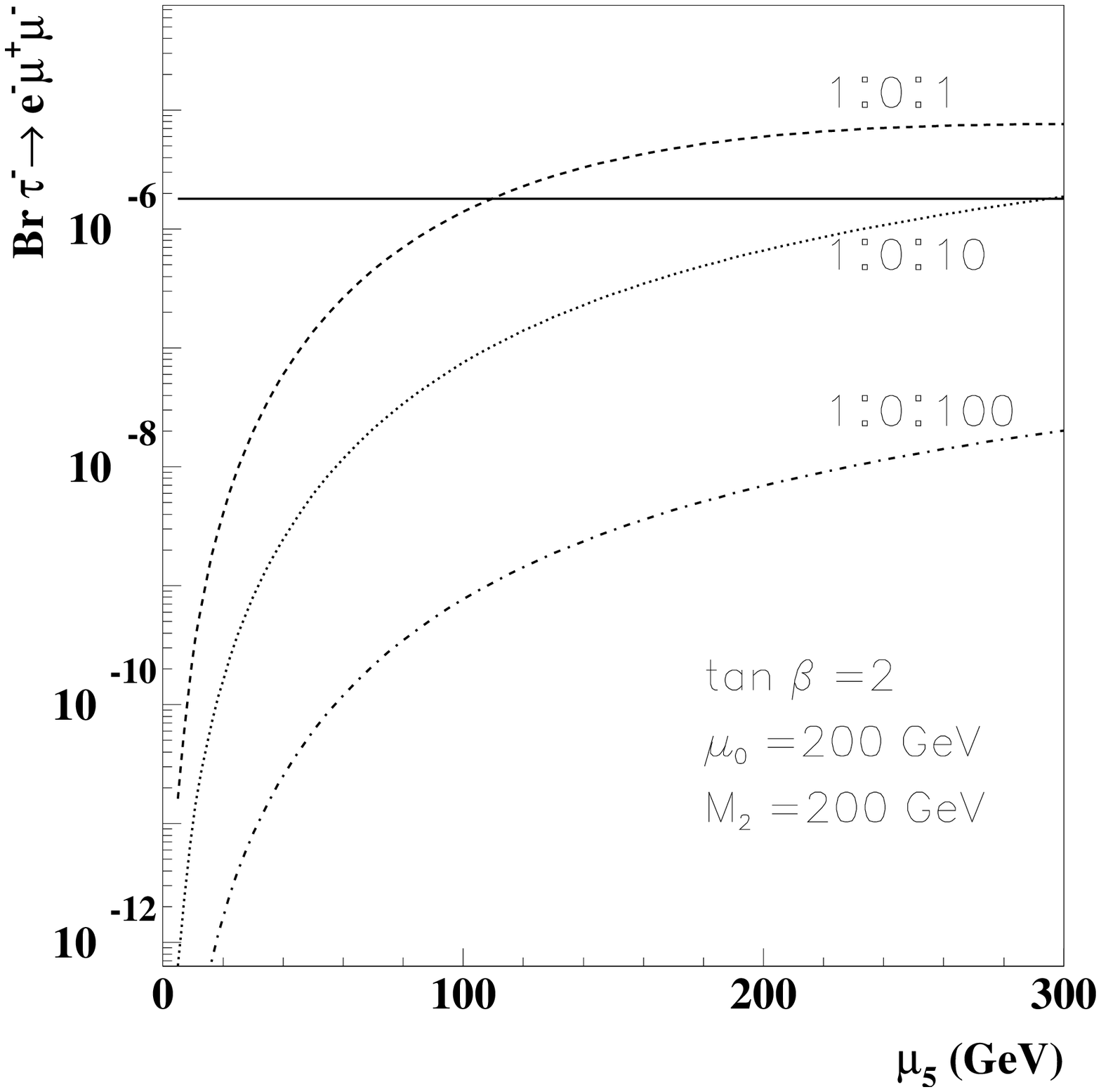}
Figure 2  
\clearpage

\includegraphics{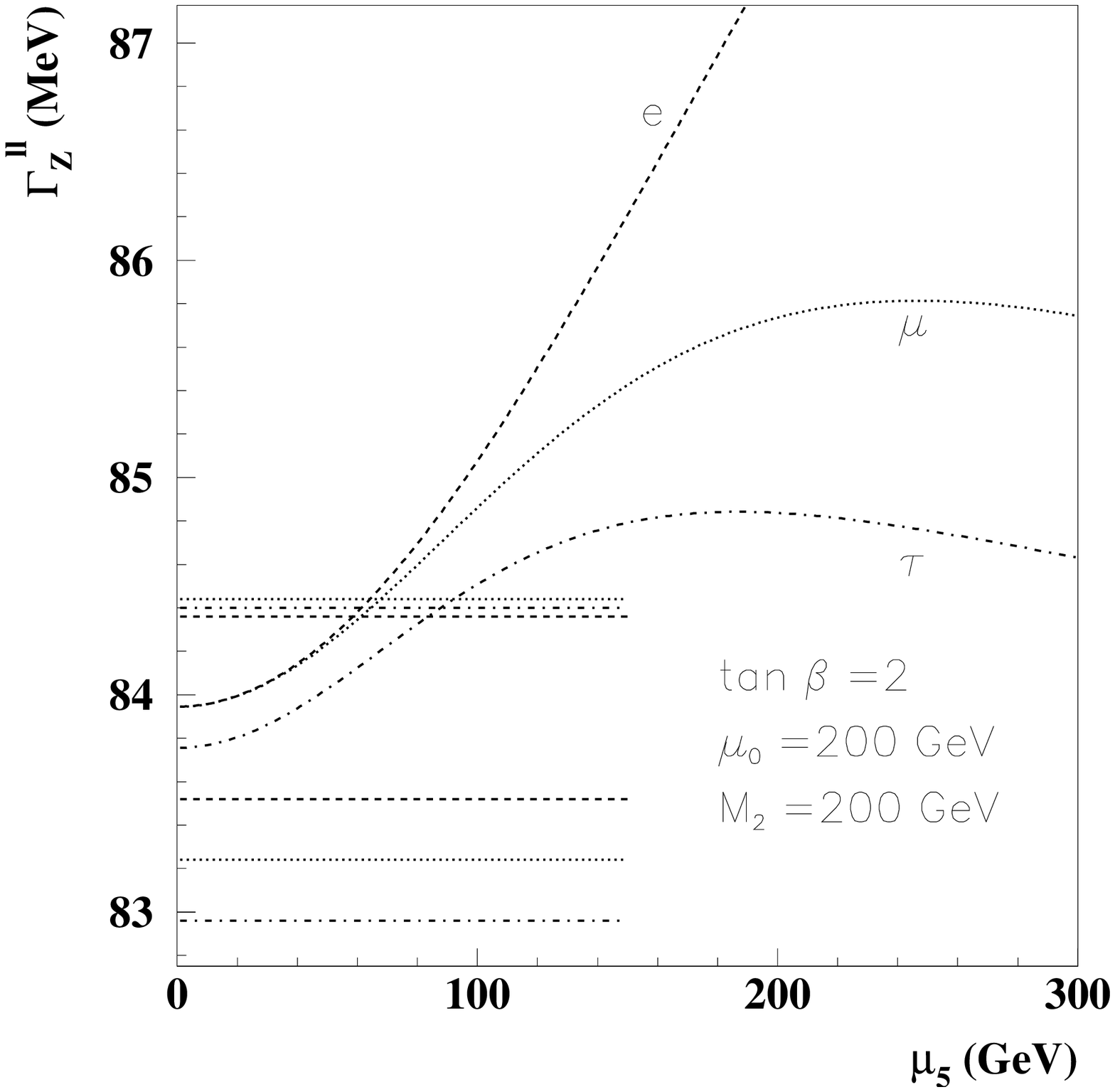}
Figure 3  
\clearpage

\includegraphics{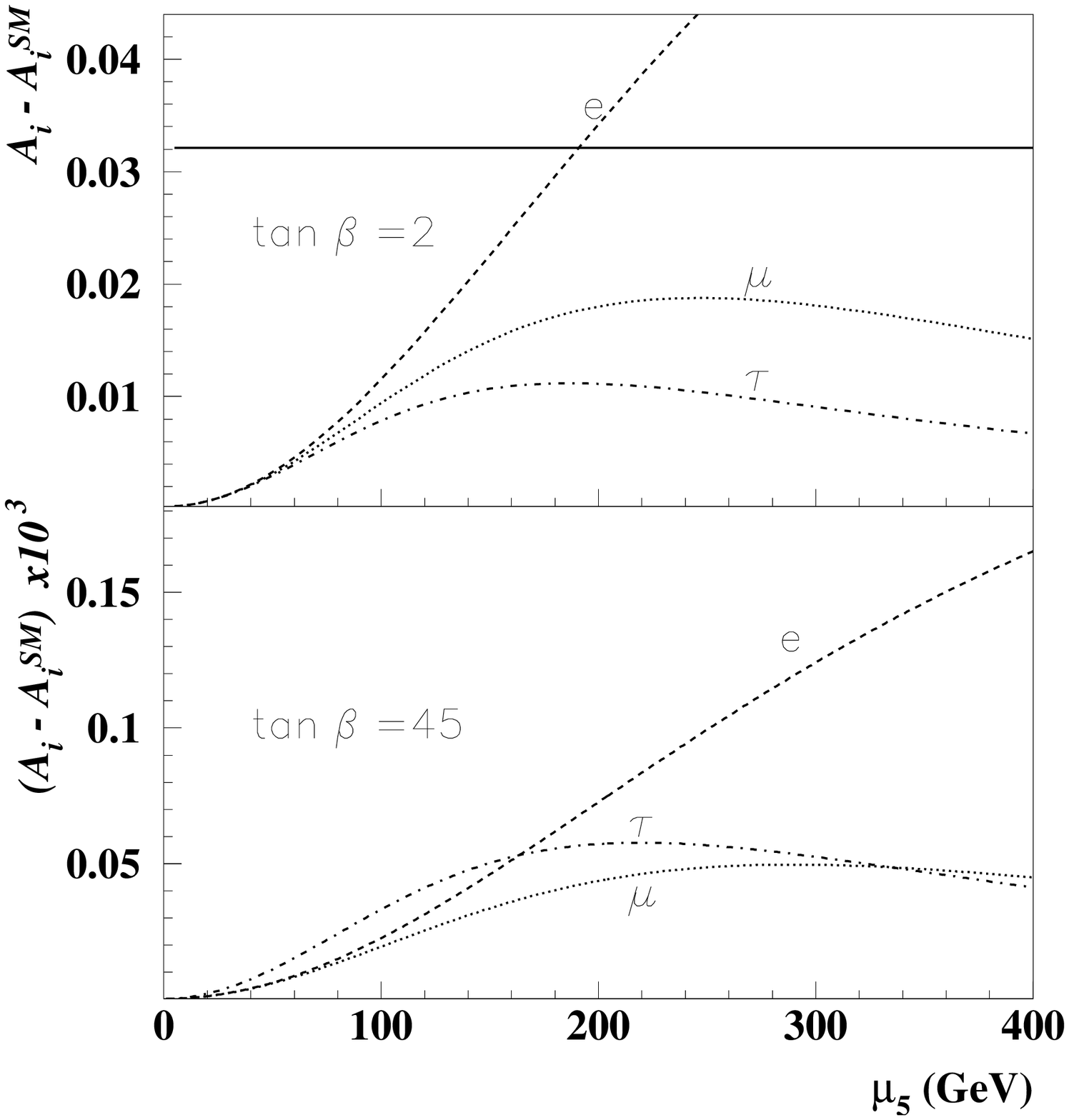}
Figure 4  
\clearpage

\includegraphics{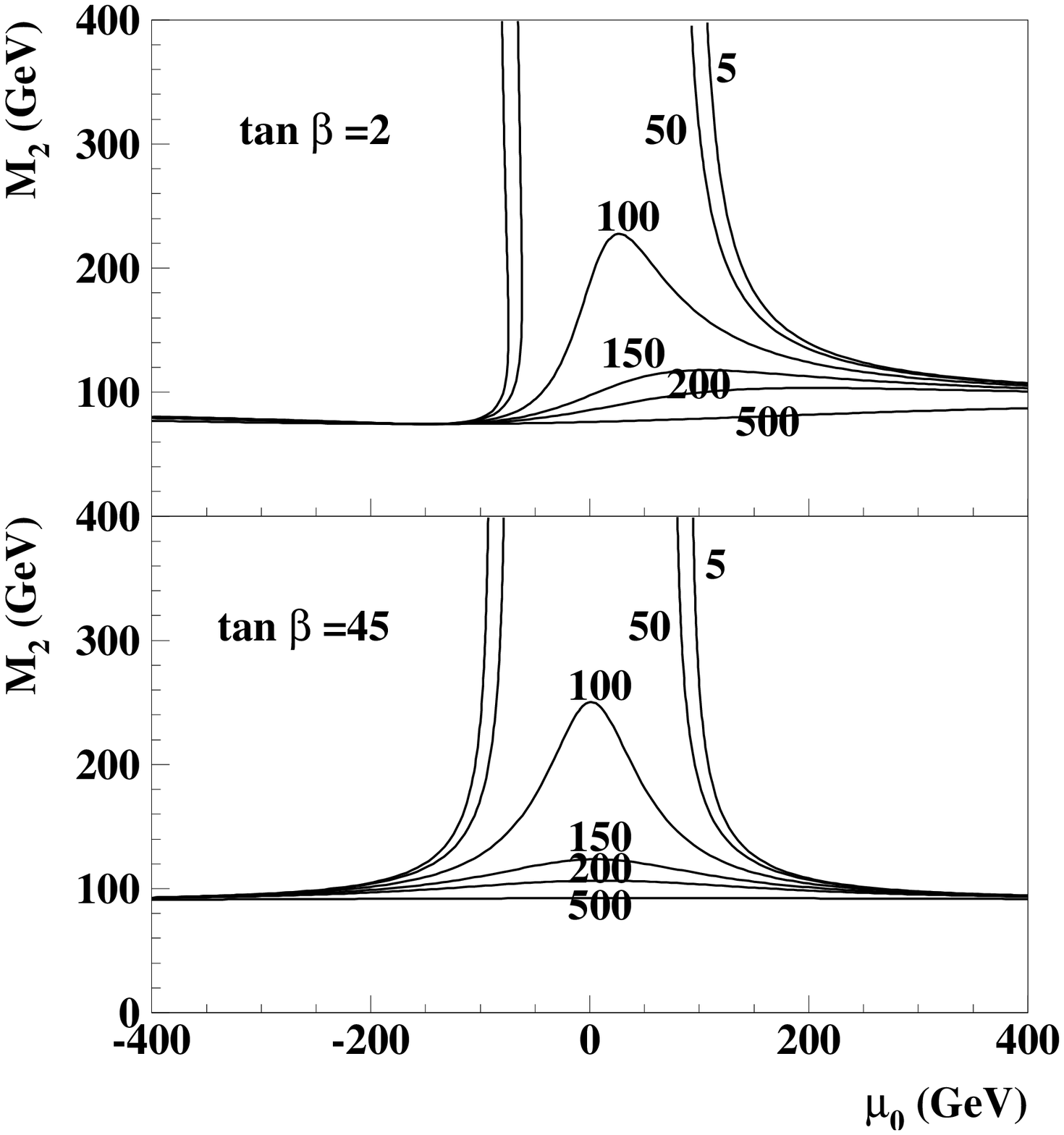}
Figure 5  
\clearpage

\includegraphics{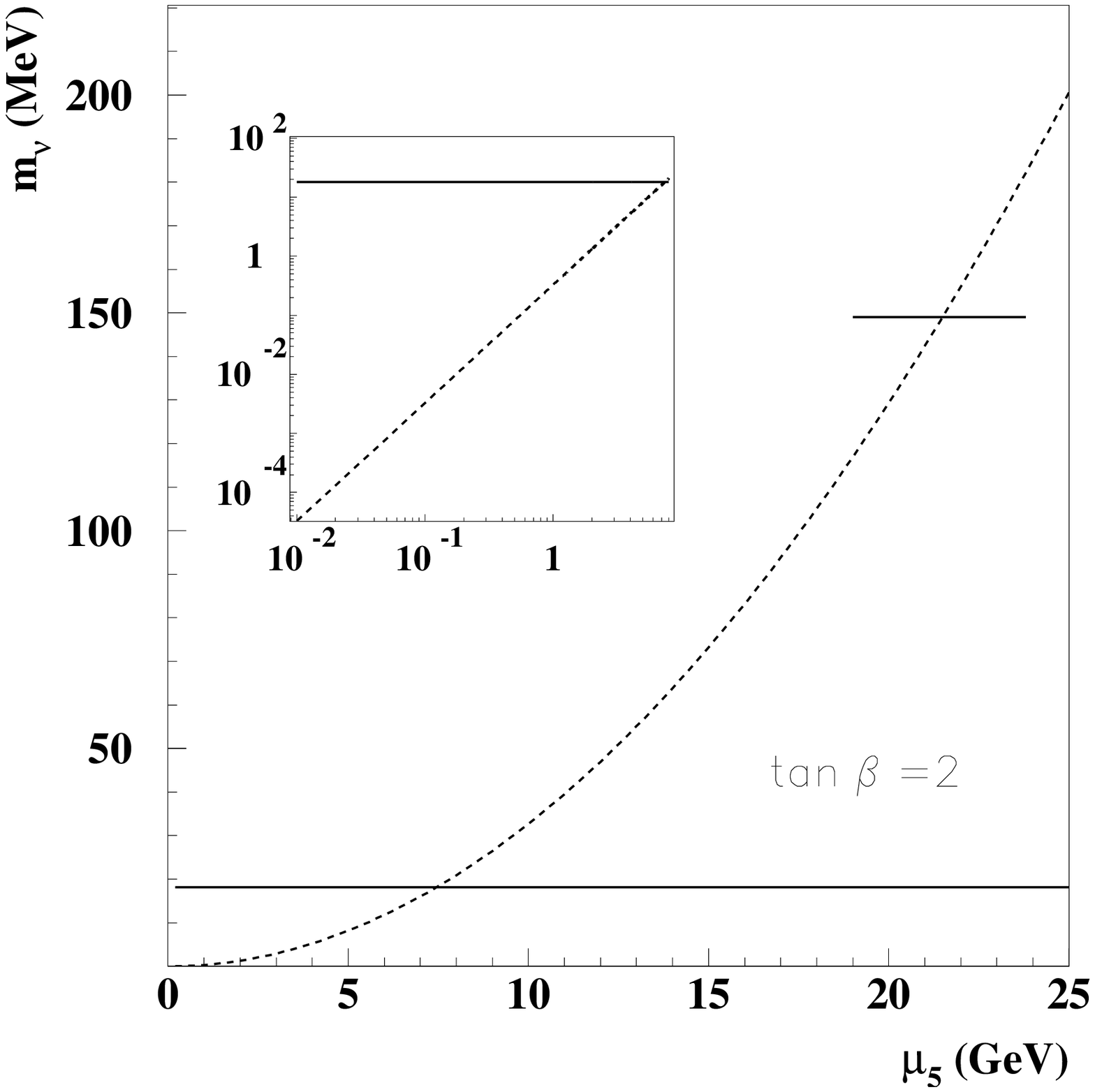}
Figure 6a
\clearpage

\includegraphics{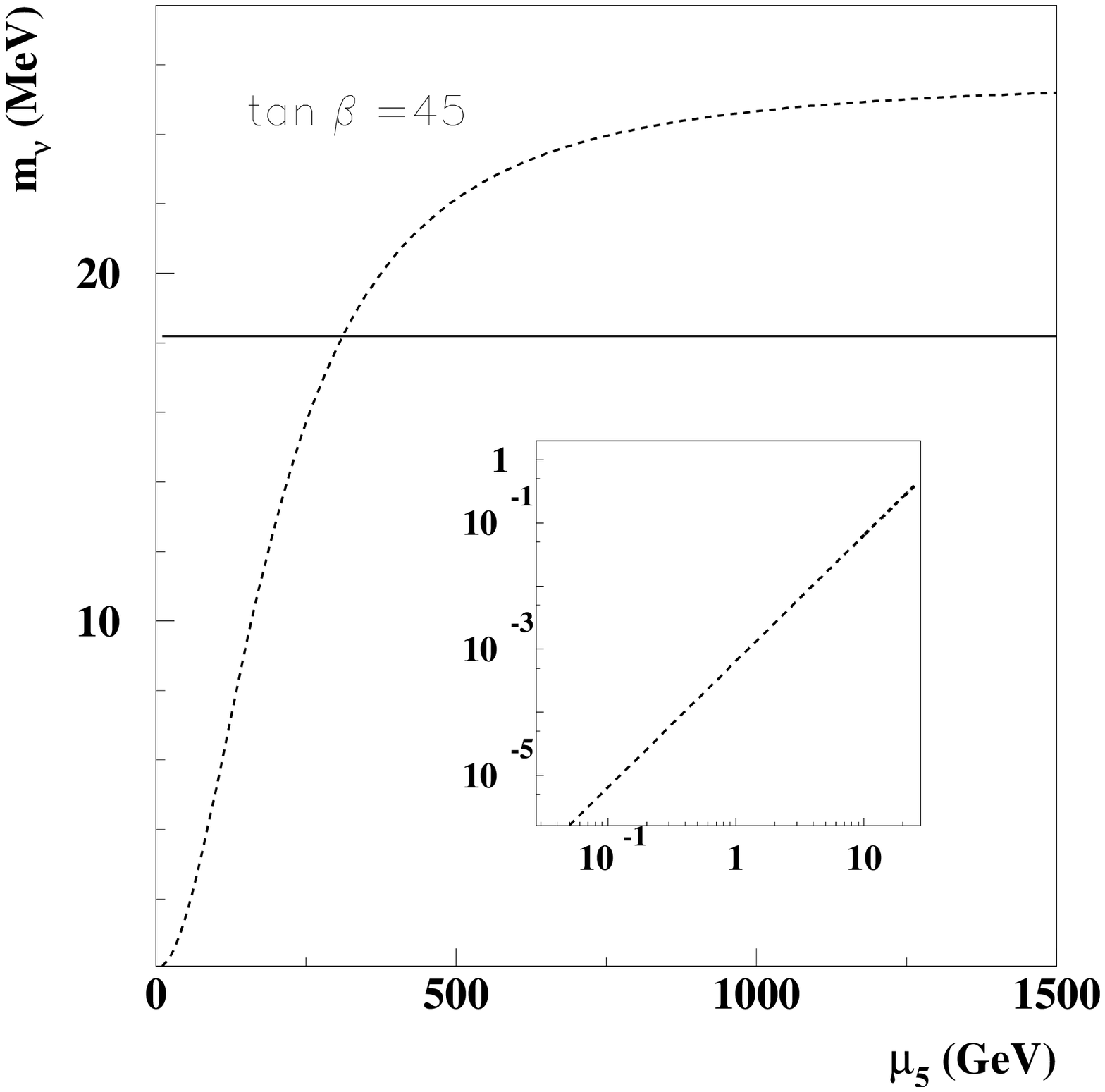}
Figure 6b  
\clearpage

\includegraphics{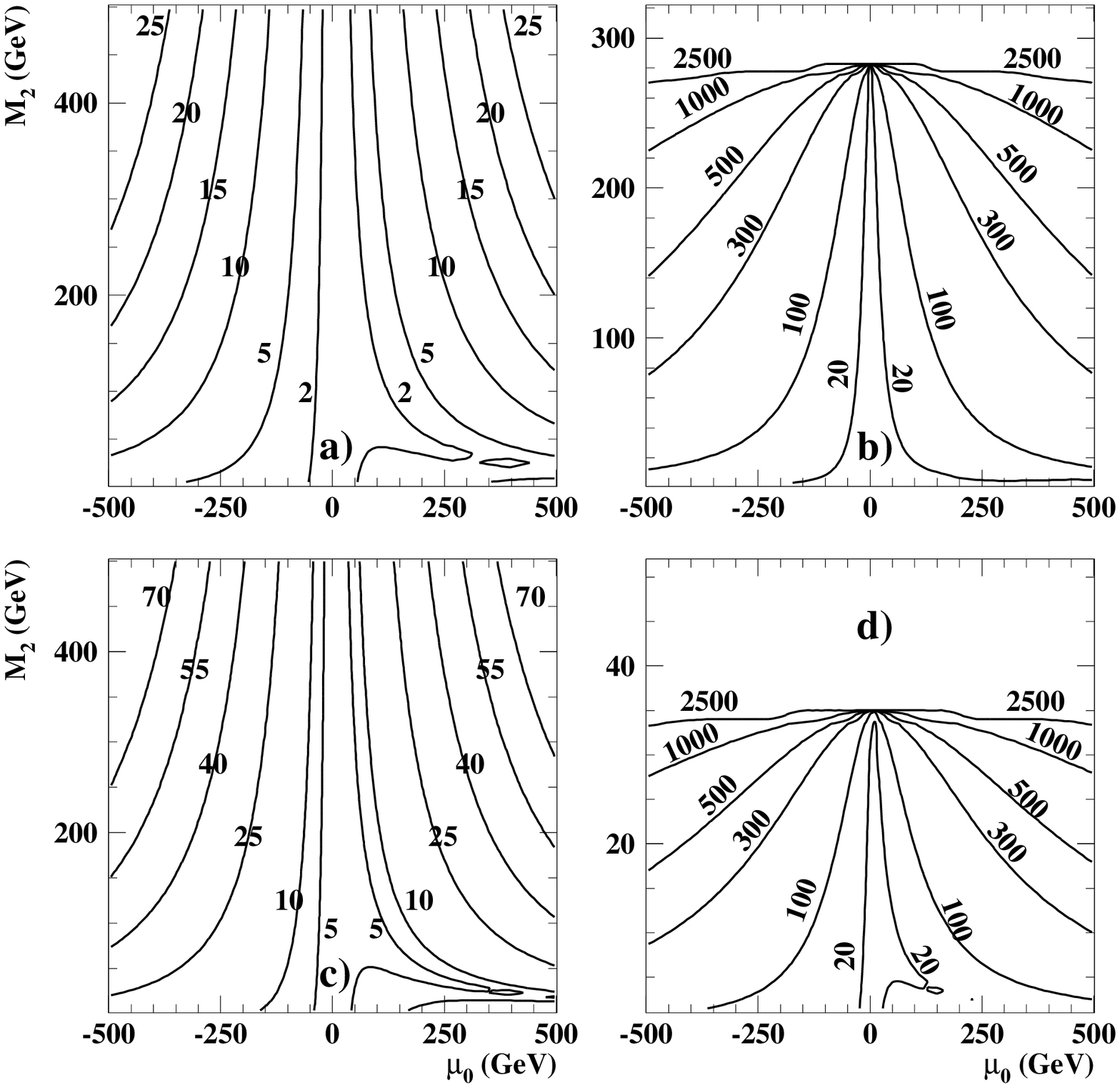}
Figure 7  
\clearpage

\includegraphics{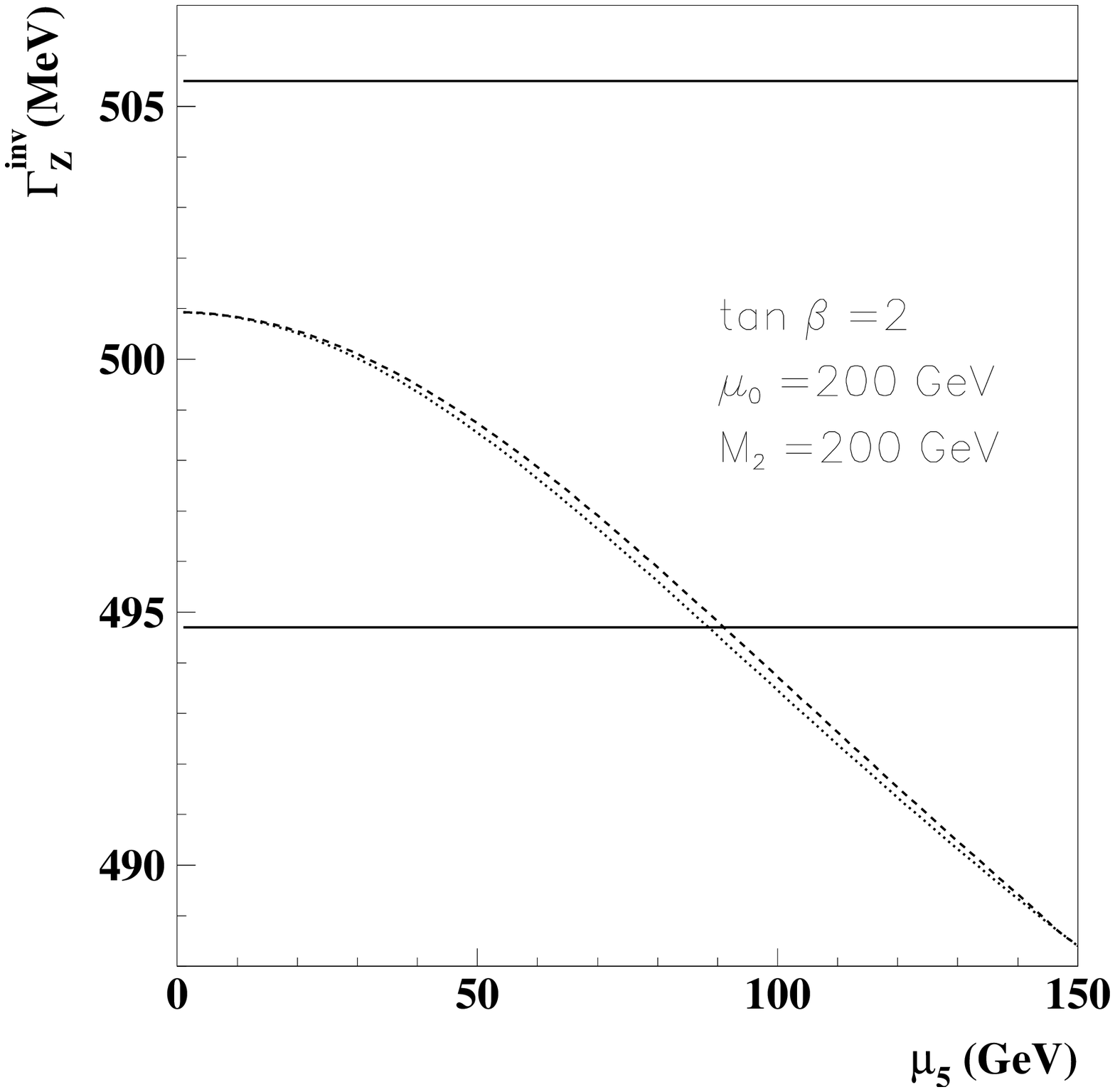}
Figure 8  
\clearpage

\includegraphics{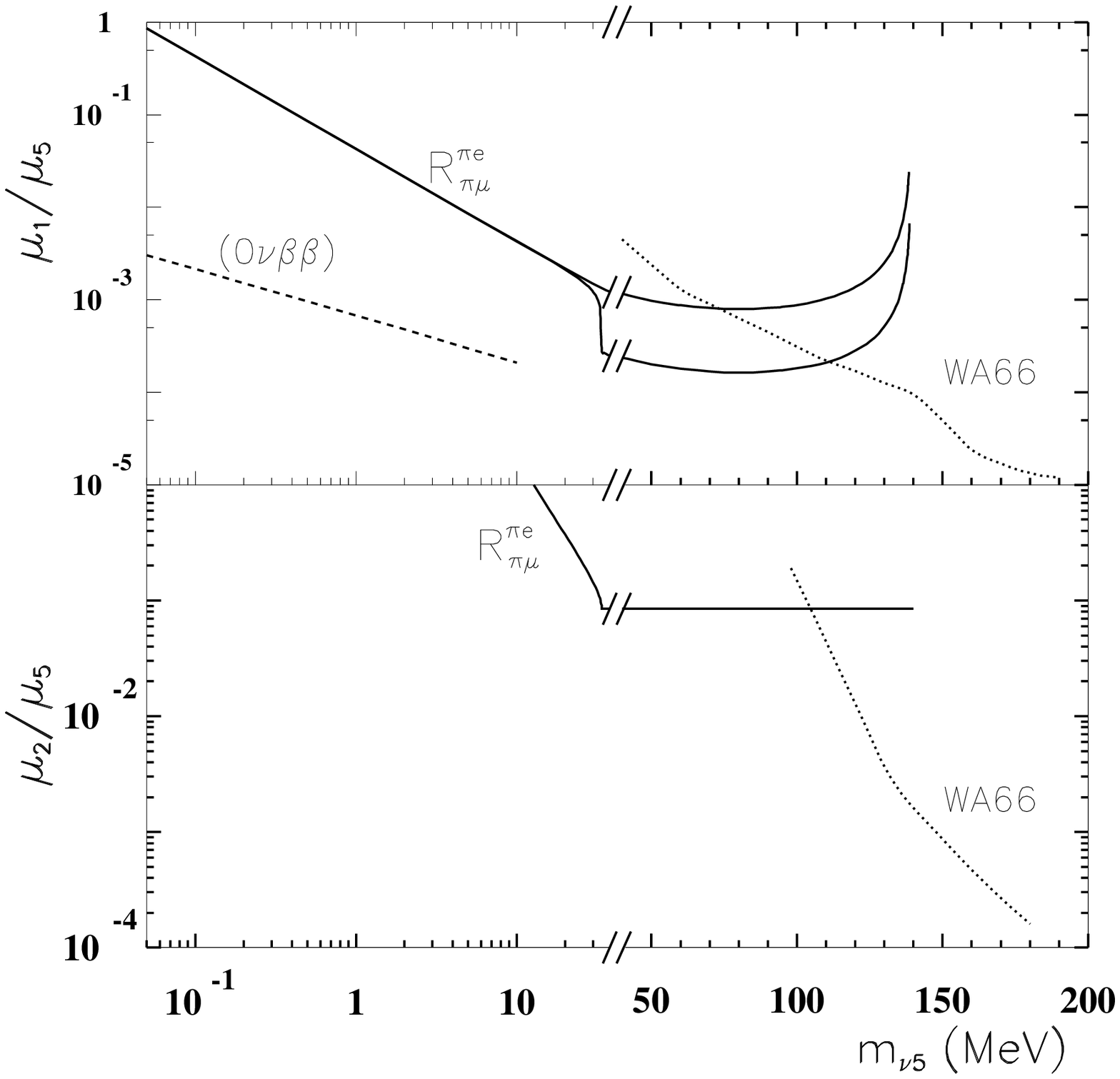}
Figure 9  
\clearpage

\includegraphics{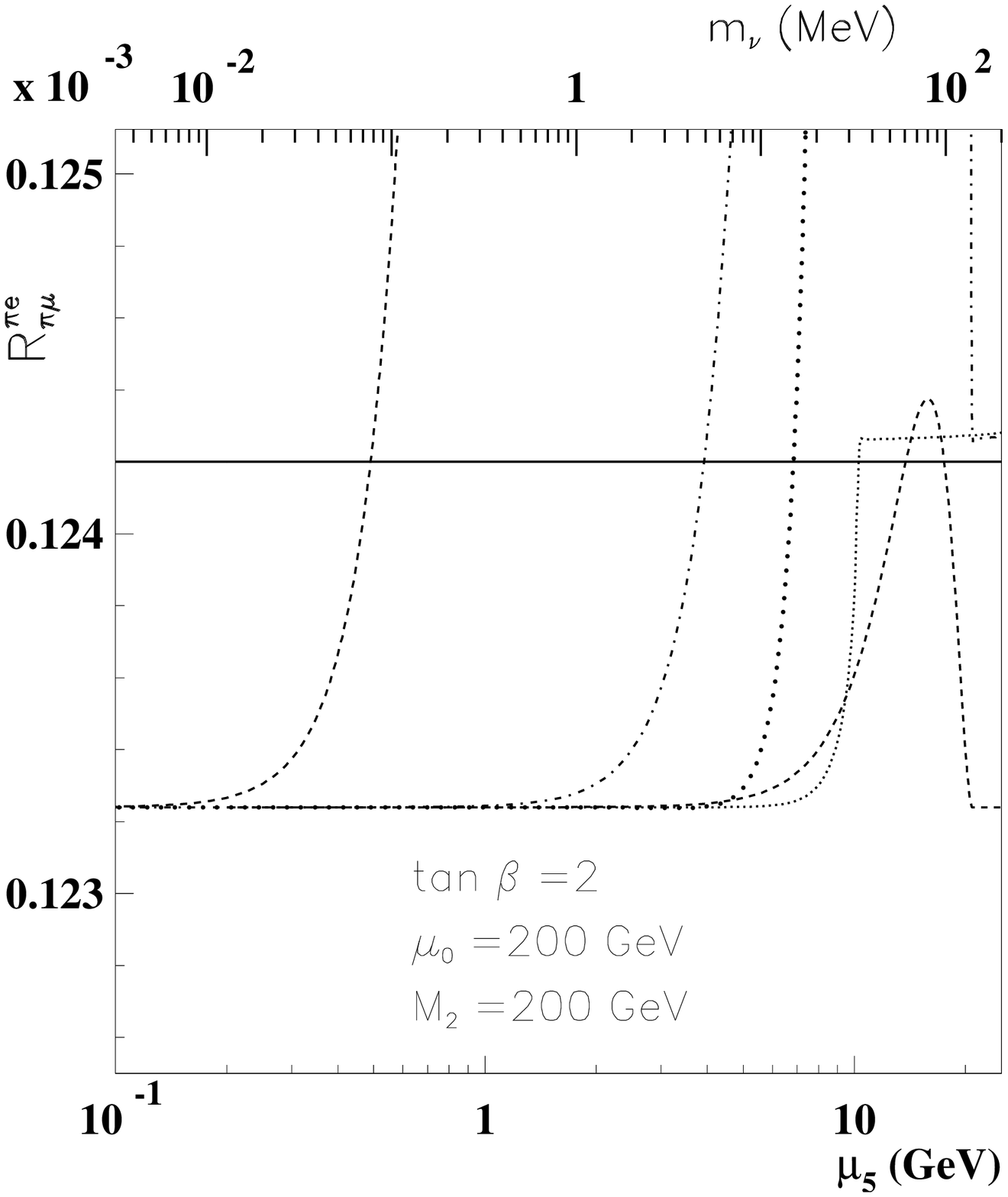}
Figure 10a  
\clearpage

\includegraphics{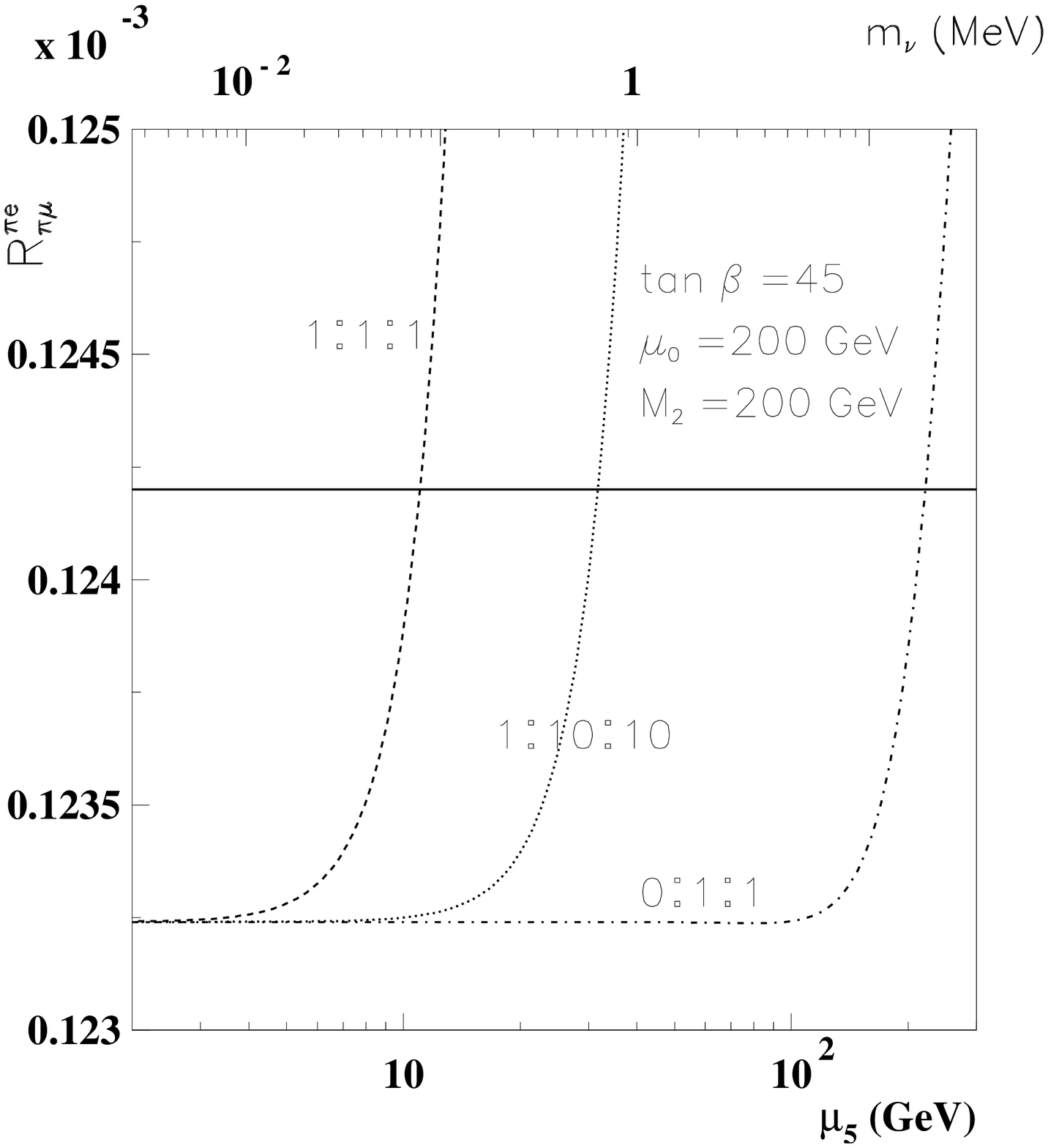}
Figure 10b  
\clearpage

\includegraphics{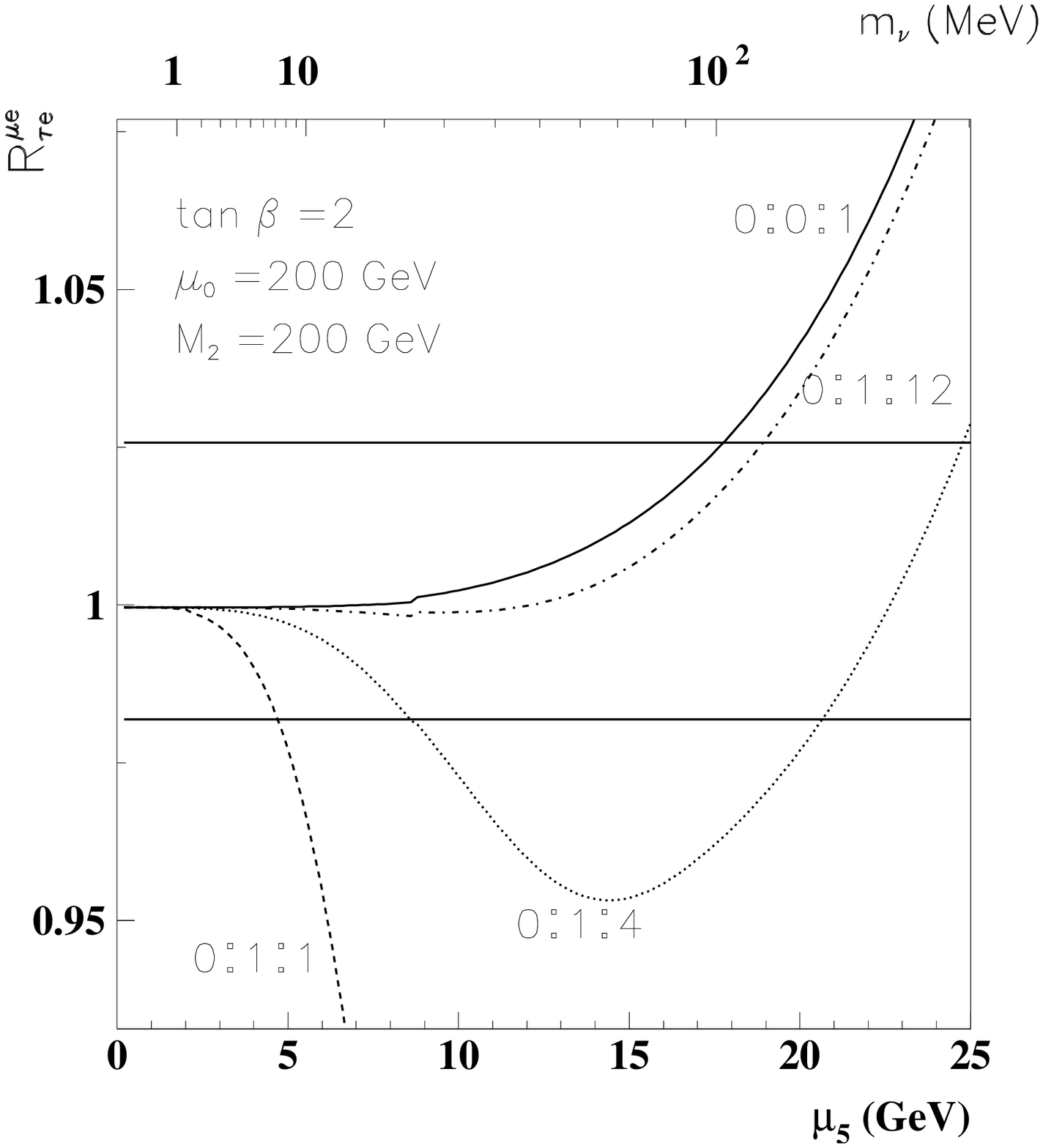}
Figure 11a  
\clearpage

\includegraphics{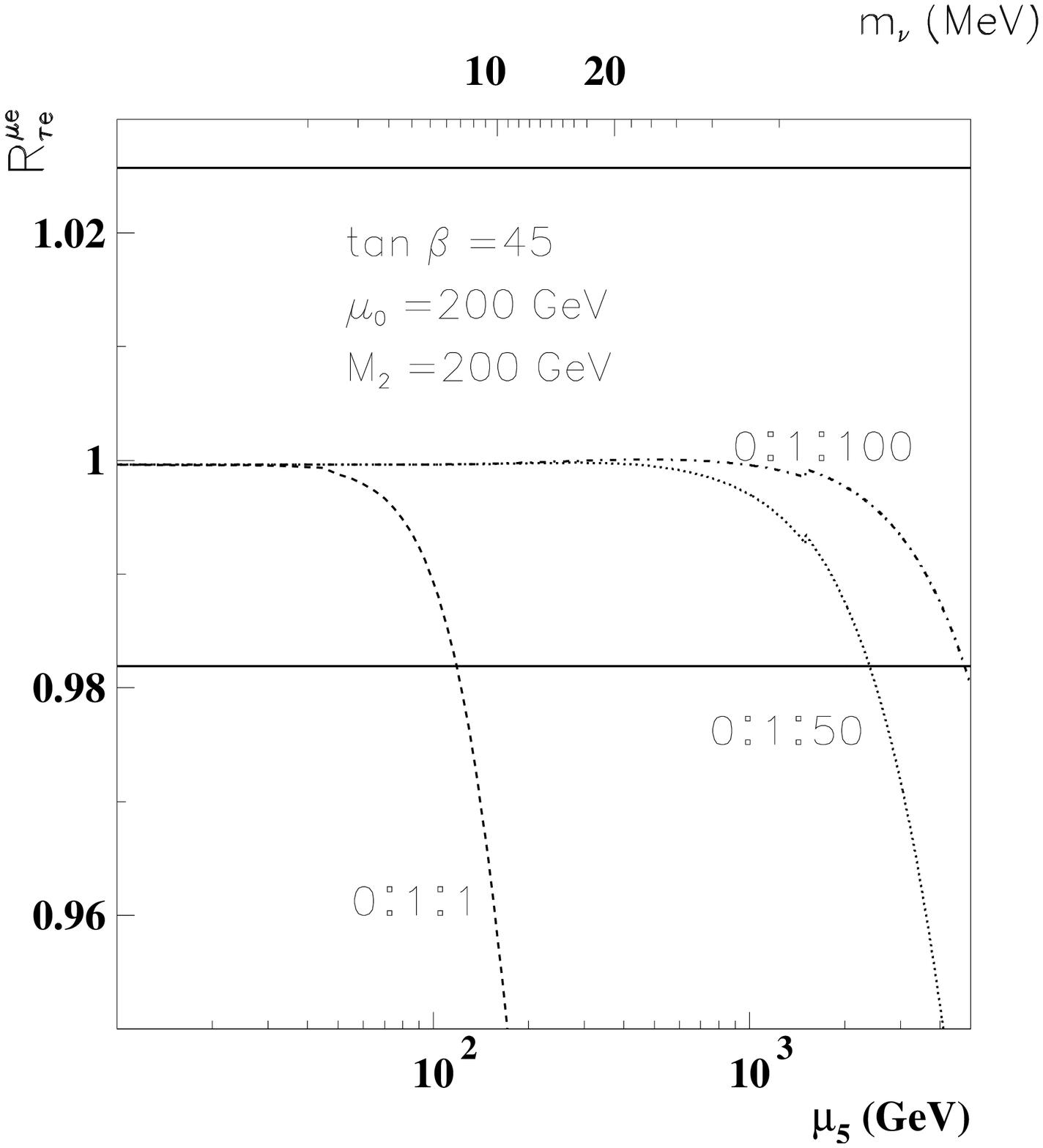}
Figure 11b  
\clearpage

\includegraphics{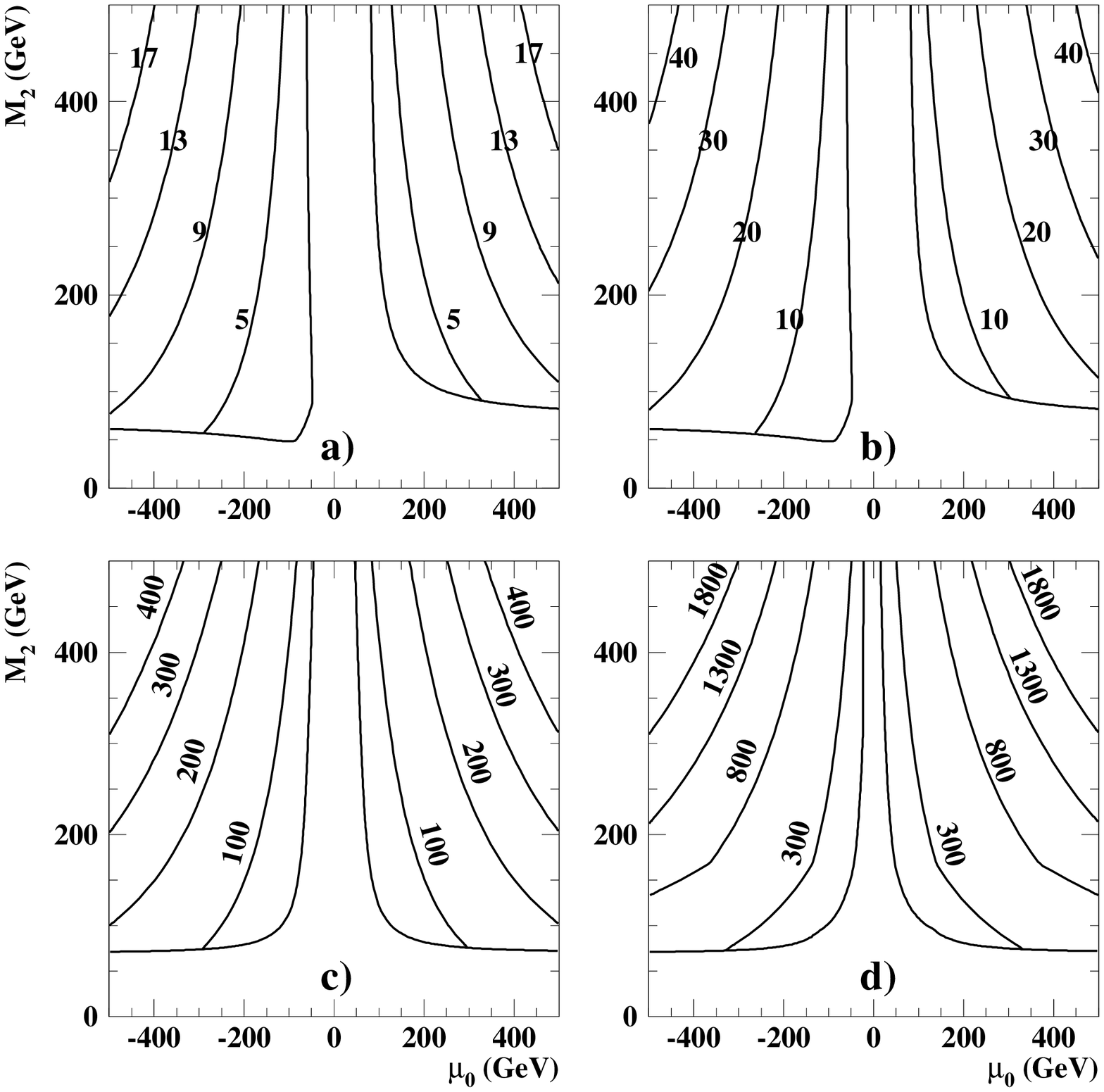}
Figure 12  
\clearpage

\end{document}